\documentclass[final,5p,times,twocolumn,numbers]{elsarticle}

\usepackage{amssymb,ulem}
\usepackage{braket}
\usepackage{lipsum,xcolor}
\usepackage{bbm}
\usepackage{placeins}
\usepackage{amsmath}
\usepackage{natbib}

\journal{Nuclear Physics B}

\begin{document}

\begin{frontmatter}

\title{Symmetry and Quantum Interference in the Attosecond Domain}

\author[first]{C. Figueira de Morisson Faria} 
\author[first]{S. Hashim}
\author[first]{J. Rivera-Dean}
\author[second]{T. Rook}
\affiliation[first]{organization={University College London, Department of Physics and Astronomy},
            addressline={Gower Street}, 
            city={London},
            postcode={WC1E 6BT}, 
            country={UK}}
\affiliation[second]{organization={Clarendon Laboratory, University of Oxford},
            addressline={Parks Road}, 
            city={Oxford},
            postcode={OX1 3PU}, 
            country={UK}}

\begin{abstract}

In this review, we examine the interplay of symmetry and quantum interference in the attosecond domain primarily through one- and two-electron photoelectron momentum distributions. Using our own work, we discuss how  tailored laser fields provide control over dynamical symmetries, and how quantum-orbit methods make it possible to connect the resulting momentum-space structures to specific ionisation, propagation, and rescattering pathways. Particular emphasis is placed on ultrafast photoelectron holography and nonsequential double ionisation through recollision excitation with subsequent ionisation, where several distinct forms of quantum interference can be identified and manipulated. We also discuss how target properties, including the binding potential, orbital symmetry and parity, modify this mapping. Finally, we consider emerging opportunities offered by non-classical states of light, for which photon statistics and quantum fluctuations provide additional degrees of freedom for controlling symmetry and strong-field electron dynamics.
\end{abstract}


\end{frontmatter}




\section{Introduction and Scope}
\label{sec:introduction}

Phenomena resulting from the interaction of matter with intense laser fields\footnote{In this context, intense means that the electric fields involved are comparable to the forces that bind matter. In gases, strong-field physics usually requires intensities around 
$10^{13}$--$10^{15}\,\mathrm{W/cm^2}$, because the laser field must compete with the Coulomb force binding an electron to an atom \cite{Krausz2009}. In solids and other condensed-matter  systems, the relevant binding scale is often a band gap or collective energy over an interatomic distance, so comparable electronic control can occur at fields of $0.01$--$1\,\mathrm{V/\AA}$, corresponding roughly to $10^{9}$--$10^{13}\,\mathrm{W/cm^2}$, and sometimes even lower in nanostructured, narrow-gap, or THz-driven materials \cite{Ghimire2014,Yue2022,Heide2024}. Liquids occupy an intermediate regime between gases and solids: strong-field and ultrafast phenomena are typically observed for intensities around $10^{11}$--$10^{14}\,\mathrm{W/cm^2}$, corresponding to microscopic fields of approximately $0.01$--$1\,\mathrm{V/\AA}$ \cite{Mondal2023}. As in solids, the relevant scales are set by intermolecular and electronic binding energies over nanometre or sub-nanometre distances, although optical breakdown and collective effects often lower the experimentally accessible thresholds compared to gases.}, such as high-order harmonic generation (HHG), above-threshold ionisation (ATI) and laser-induced nonsequential double and multiple ionisation (NSDI, NSMI), are associated with the recollision or recombination of an electron, typically freed by tunnelling or over-the-barrier ionisation, with its parent ion.  
This physical picture, known as the three-step model, allows a direct link with a time-domain picture, in which the field dictates very specific ionisation and recollision times \cite{Corkum1993}. Ionisation happens close to the field extrema, while recollision takes place near a field zero crossing. This is in stark contrast with the weak-field regime, in which the field does not substantially modify the target, acting merely as a perturbation. In the perturbative  framework, what matters is the spectral overlap between the field frequency and the energy gap associated with the excitation or ionisation into a continuum state. Thus, there is no well-defined ionisation time. 

Furthermore, in strong fields, the field enters the continuum dynamics. The  electron accumulates different phases depending on when it is freed and on its propagation in the continuum (up to the detector or its parent ion). These well defined times and phase differences  enable the use of strong-field phenomena as tools for steering ultrafast electron motion. They also allow quantum pathways to be associated with electron orbits.~For that reason, path-integral methods are widespread in strong-field physics~\cite{Salieres2001} and have been used and perfected since this research area's inception.~Their main advantage is to introduce an intuitive understanding of the electron's dynamics, which can be mapped to HHG spectra or photoelectron momentum distributions (PMDs), and, yet, keep features such as tunnelling and quantum interference.   

Different trajectories associated with the same final state may interfere, and this interference has enabled a myriad of applications~\cite{Lein2007,Faria2020}. Examples are ultrafast imaging using high-order harmonics~\cite{Lein2002HHG,Kanai2005,Baker2006} or photoelectron holography~\cite{Spanner2004,Bian2011,Faria2020}, reconstruction of ionisation dynamics~\cite{Dudovich2006,Pedatzur2015}, coherent control of electron motion~\cite{Goulielmakis2004,Goulielmakis2010}, HHG~\cite{Burnett1977,Mcpherson1987,Ferray1988}, precision metrology~\cite{Walker1994,Paulus2003}, and emerging quantum information studies based on electron--photon coherence and entanglement~\cite{RiveraDean2022}. 
In addition to proof of principle studies, in which quantum interference has been used to explain the high-harmonic or ATI peaks~\cite{Lewenstein1994,Lewenstein1995,Arbo2010}, or to early applications, such as attosecond pulses and bursts of photoelectrons~\cite{Antoine1996,Paul2001,Drescher2001,Paulus2001}, which depend on phase-locked, well-defined ionisation times, quantum interference has been widely explored in schemes to characterise the field or the target using time-delayed quantum pathways. 

Examples are metrology techniques to characterise attosecond pulses, such as FROG (Frequency-Resolved Optical Gating) \cite{Trebino1993,Kane1993}, FROG-CRAB (Frequency-Resolved Optical Gating for Complete Reconstruction of Attosecond Bursts) \cite{Mairesse2005}, SPIDER (Spectral Phase Interferometry for Direct Electric-field Reconstruction) \cite{Iaconis1998} or RABITT (Reconstruction of Attosecond Beating by Interference of Two-photon Transitions)~\cite{Paul2001,Muller2002,Dahlstrom2012,Agostini2024}, which employ pump-probe schemes combining XUV pulses (the pump) and a synchronised infrared field (the probe) to ionise a target gas, or ultrafast photoelectron holography \cite{HuismansScience2011,Faria2020}, which uses interference patterns from qualitatively different quantum electron pathways.~Recently, quantum interference was used to probe coherence and entanglement between electronic and nuclear degrees of freedom, uniting attosecond physics and quantum information science \cite{Koll2022,Koll2026}, and an interferometric pump-probe scheme known as KRAKEN (quantum-state tomography of attosecond electron wavepackets) has been used for tomographically reconstructing the photoelectron quantum state created by an XUV pulse \cite{Laurell2022,Laurell2025}.~Because it measures purity, coherence, and entanglement-relevant quantities, it provides a bridge between attosecond science and quantum metrology.~Furthermore, above-threshold ionisation interference rings have been linked to the quantum and classical Fisher information, highlighting their potential for quantum sensing applications \cite{Maxwell2021a}. 

Often, these schemes bring together quantum interference and synthesised fields.~The latter allows one to manipulate the electric field waveform directly in the time domain, essentially providing ``on-demand'' control of light-matter interactions.~Examples of synthesised fields in attosecond and
strong-field physics include bichromatic and multicolour wave-
forms \cite{Wirth2011,Chan2011,Skruszewicz2015}, orthogonally polarised two-colour fields
\cite{Brugnera2011,Xie2017Disentangling,Zhang2014}, bicircular and
polarisation-shaped pulses \cite{Baykusheva2016,Pisanty2014,Heinrich2021}, carrier-envelope-phase-stabilised
few-cycle pulses \cite{Apolonski2000,Baltuska2003,Paulus2003}, THz-assisted drivers
\cite{Balogh2011}, plasmonically enhanced near fields \cite{Stockman2007,Susmann2011}, and emerging combinations involving non-
classical light \cite{Gorlach2023,Rasputnyi2024BSVHHG,Heimerl2025NeedleTipsBSV}.

Powerful tools for understanding and manipulating light-matter interaction in synthesised fields are  provided by the dynamical-symmetry framework.~Symmetries are a major organizing principle in several areas of physics, such as quantum physics, particle physics, solid-state physics, gravity, chemistry,  and, as expected, strong-field physics and attosecond science. They are linked to conservation laws, as shown by Noether's theorem~\cite{Noether1971}, enable the determination of selection rules, allow the classification of states and phases, and, by constraining a problem's dynamics, simplify computations.~Symmetries also provide additional physical insights by shedding light on common patterns in apparently unconnected phenomena. Conversely, symmetry breaking may add additional structure to physical observables and be an important metrological tool. One way of breaking symmetries is using quantum interference. 

This review aims at uniting quantum interference and symmetry, and discussing how symmetry and its breaking can be mapped into key observables such as photoelectron momentum distributions.~It is by no means a complete discussion of how these properties are used in strong-field and attosecond physics.~For that purpose, we refer to Ref.~\cite{Neufeld2025}.~Rather, we provide examples within photoelectron holography, and laser-induced nonsequential double ionisation, focusing on our own work. Ultrafast photoelectron holography brings together high electron currents and sub femtosecond timescales, and uses the quantum interference patterns between qualitatively different quantum pathways to obtain temporal information about a target~\cite{HuismansScience2011}. Laser-induced nonsequential double ionisation (NSDI) is the archetypal process in which electron-electron correlation is present for strong-field laser-matter interaction \cite{Weber2000,Moshammer2000}\footnote{The first experimental signatures of non-sequential double ionisation, and later non-sequential multiple ionisation (NSMI), emerged in the early 1980s \cite{LHuillier1983}, first in Xenon and then for other noble gas atoms \cite{Walker1994, Kondo1993,Talebpour1997} and molecules \cite{Cornaggia1998, Eremina2004}, revealing behaviour that strongly deviated from the expectations of single-electron ionisation theory. For instance, with intensities above $1.5\times10^{12}$W/cm$^2$, the multiple ionisation yield of Xe atoms was consistent with predictions of sequential double ionisation (SDI), but below this threshold intensity, the ionisation yield showed an enhancement and differed by many orders of magnitude from sequential predictions, resulting in the famous `knee' structure \cite{Walker1994}. This implied the existence of another more dominant ionisation mechanism requiring an explicit energy-transfer mechanism between the two electrons, thereby ruling out independent-electron models.}.  It is described by the laser-induced inelastic recollision of an electron with its parent ion, whereupon it transfers enough kinetic energy to the core so that a second electron is released. This can either happen instantaneously, in what is known as electron-impact  (EI) ionisation, also sometimes known as recollision impact ionisation~\cite{Weber2000,Rudenko2007}, or the second electron can be promoted to an excited bound state, from where it tunnels with a time delay.~This latter scenario is known as recollision-excitation with subsequent ionisation (RESI). For reviews see \cite{FigueiraDeMorissonFaria2011,Becker2012}.

Until the 2010s, NSDI  was  viewed as a classical type of electron-electron correlation, perhaps due to the immense success of classical-trajectory models reproducing key features, such as the shapes of the electron-momentum distributions. However, if excitation too is involved, there has been theoretical~\cite{Hao2014,Maxwell2015} and experimental~\cite{Quan2017}  evidence that quantum interference is robust, and survives focal averaging and the integration over several degrees of freedom \cite{Maxwell2016}. Furthermore, several types of quantum interference in RESI have recently been classified by us for arbitrary field shapes~\cite{Hashim2024,Hashim2024b,Hashim2024c}, and we will draw from this recent work.

\begin{figure*}[h]
    \centering
\includegraphics[width=0.9\linewidth]{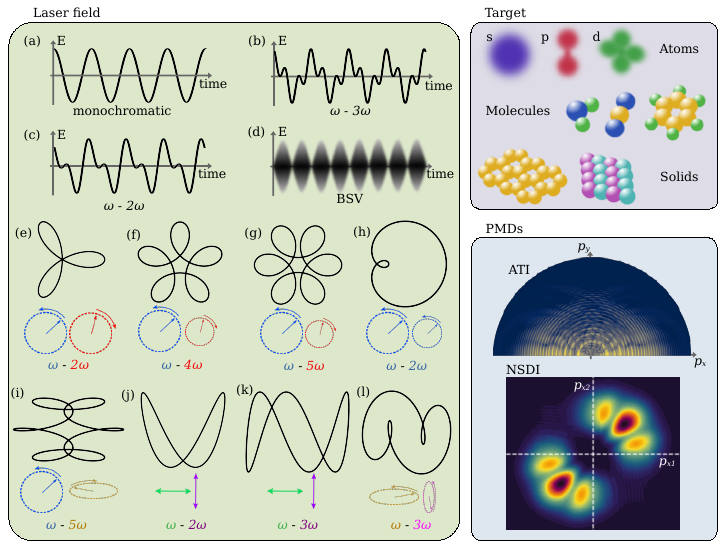}
    \caption{Schematic showing a variety of shaped fields, targets with different point-group symmetry, and examples of the electron observables we are considering. 
    In the left-hand green box, examples of electric fields, satisfying different dynamical symmetries are shown. In the first two rows, only linearly polarised fields are considered. (a) shows a monochromatic field which has more symmetry than the fields (b) and (c) where a second colour has been introduced to break symmetry. The exact relative phase between the constituent fields determines precisely which symmetries are retained. (d) provides an example of a field which differs from the ordinary coherent light; in particular, it shows the field associated with a monochromatic bright-squeezed vacuum which is even more symmetric than the monochromatic coherent light field.
    (e)-(g) are examples of counter-rotating bichromatic circularly-polarised fields with different combinations of frequency. Notably, these display symmetry under discrete rotations combined with time translations. (h) shows the field for the co-rotating polarisation counterpart.
    (i) is an example of a counter-rotating field with one of the components being elliptical.
    (j) and (k) show orthogonally-polarised two-colour fields, while (l) shows what happens when some ellipticity is given to the fields in (k), forming co-rotating elliptically polarised fields.
    The purple box in the top right, contains some illustrative examples of possible targets of strong-field processes which exhibit different geometrical symmetry. Firstly, atomic orbitals themselves have different symmetry. The s-orbitals have spherical symmetry but the higher angular momentum quantum number orbitals have more complicated symmetry structure.
    Examples of some molecules with different symmetry are also provided. $H_2O$ exhibits a reflection symmetry across an axis passing through its oxygen atom, while $CO_2$ also displays cylindrical symmetry. The benzene ring has a discrete rotational symmetry. 
    The solids have some point-group symmetry associated with their crystallographic structure.
    The bottom right blue box contains examples of the observables associated with the ATI and NSDI (in particular RESI) processes, which are considered throughout this review. Subsequently, we shall discuss how the symmetry of laser field and target will inform the symmetry and other properties of the photoelectron momentum distributions derived from these processes.
     }
    \label{fig:symmetries}
\end{figure*}

The review is organised as follows. Section~\ref{sec:context} provides a broader conceptual context, introducing  shaped fields and their applications and briefly tracing the development of symmetry concepts in strong-field and attosecond physics. Section~\ref{sec:quantumorbits} introduces the quantum-orbit framework used throughout the review, with emphasis on the strong-field approximation (SFA), the Coulomb quantum-orbit strong-field approximation (CQSFA), and the saddle-point equations that connect electron trajectories to measurable observables. Section~\ref{sec:symmetries} summarises the dynamical symmetries relevant to the driving fields considered here. In Sec.~\ref{sec:mapping}, we discuss how these symmetries are mapped onto photoelectron momentum distributions, including the kinematic constraints imposed by the saddle-point equations and biases introduced by the target. Sections \ref{sec:holography} and \ref{sec:NSDI} then develop two complementary case studies. Section~\ref{sec:holography} focuses on photoelectron holography, relating characteristic interference structures to specific quantum orbits and illustrating how target properties such as parity may be extracted from them. Section~\ref{sec:NSDI} turns to nonsequential double ionisation, with particular emphasis on RESI, and discusses how exchange, temporal, combined temporal--exchange, and interchannel interference shape correlated two-electron momentum distributions and may break or reveal field-induced symmetries. Finally, Sec.~8 summarises the main conclusions and discusses future opportunities, including the use of non-classical light as a new resource for controlling strong-field symmetries and quantum interference.

\section{Conceptual framework: Providing a broader context}
\label{sec:context}

\subsection{Shaped fields and some applications}
\label{sec:shaped}
Next, we discuss some widely used shaped fields, with emphasis on their metrology and coherent control applications.

\textit{Bi- or polychromatic linearly polarised fields} and \textit{few-cycle carrier-envelope phase (CEP)-stabilised pulses} are widely used to break symmetries and thus create controlled directional currents in ATI, HHG and NSDI. 
Typical knobs employed are the ratio between the frequencies and field amplitudes and time delays between the compounding waves. The most common bichromatic fields are ($\omega,2\omega$) and ($\omega, 3\omega$) fields, i.e., composed of a wave of frequency $\omega$ and its second and third harmonics. However, in many instances chirped waves or non-commensurate waves have also been used~\cite{Petrakis2021}. Generalizing this further, one may combine several phase-locked colours and sculpt nearly arbitrary wave forms~\cite{Wirth2011,Chan2011}. CEP-stabilised few-cycle pulses are essential for the generation of isolated attosecond pulses~\cite{Antoine1996,Baltuska2003}, and can be written as a three-color field~\cite{Haessler2015}.~Thus, many of the symmetries used in the two-color case can be generalised to few-cycle pulses.  A further example of a widely used bichromatic field is a strong THz field combined with near-IR driving. Instead of employing comparable frequencies in order to reshape the sub-cycle waveform, there is a huge separation of timescales between the THz and near-IR waves. Hence, the former acts as a quasi-static bias, or a current-driving background, being often applied to modify transport or solid-state currents \cite{Meineke2022,Fujimoto2025}. 

\textit{Orthogonally polarised fields} allow one to control electron motion in more than one dimension, as they introduce transverse components in the electron tunnelling and propagation. The ellipticity can be used to suppress rescattering, or to confine it to specific spatiotemporal windows, acting as a gate. 
Examples are elliptically or circularly polarised fields \cite{Bashkansky1987,Torlina2013}, orthogonally polarised two-color (OTC) fields \cite{Kim2005,Zhang2014,Xie2017Disentangling}, and bicircular fields \cite{Pisanty2014,Baykusheva2016}. Elliptically or circularly polarised fields are used in attosecond angular streaking (`attoclock') and tunnelling-time studies \cite{Torlina2013}. 
For high enough driving-field polarisations, one may perform a one-to-one mapping of a specific ionisation time to a single offset angle in a PMD \cite{Torlina2013}. Elliptically polarised fields find use in controlling recollision probability (as recollision gates for RESI, for instance \cite{Emmanouilidou2017}), and introduce transverse drift momentum, possibly altering the transverse momentum structure in one-electron distributions. 
Orthogonally polarised two-color (OTC) fields generate highly sensitive asymmetries used in phase-of-the-phase spectroscopy, chiral discrimination, and ultrafast metrology. Bicircular fields, composed of counter-rotating circularly polarised components of the same or commensurate frequencies, are commonly used for chiral HHG~\cite{Ayuso2019,Ayuso2022}, symmetry engineering~\cite{Skruszewicz2015,Tulsky2018,Demekhin2018}, and spin/angular momentum control~\cite{Pisanty2014,Fleischer2014,Pisanty2019}.

\textit{Plasmonic near fields} and \textit{nanostructure-enhanced fields} emerged from the convergence of ultrafast laser science and nanoplasmonics, exploiting the ability of metallic nanostructures to concentrate electromagnetic energy into nanoscale volumes. This produces local electric fields far stronger than the incident laser field, which are highly localised,
spatially inhomogeneous,
and often phase shifted relative to the driving laser. Their strong local field enhancement enables nanoscale strong-field physics~\cite{Ciappina2017}, ultrafast photoemission~\cite{Bormann2010,Schenk2010}, attosecond streaking at surfaces~\cite{Stockman2007,Susmann2011}, enhanced nonlinear optics~\cite{Bouhelier2003,Neacsu2005}, and petahertz-scale control of electronic motion~\cite{Kruger2011,Kruger2012}.

\textit{Vortex beams} possess helical phase fronts and carry orbital angular momentum (OAM) in addition to energy and, possibly, spin angular momentum~\cite{Bliokh2015}. They enabled structured transfer of angular momentum, topology, and spatial phase to light-matter interactions.~Their applications span optical manipulation~\cite{Padgett2011}, imaging~\cite{Furhapter2005}, communications~\cite{Wang2012}, quantum information~\cite{Magana-loaiza2019}, and emerging strong-field and attosecond phenomena involving structured electron and high-harmonic dynamics~\cite{Habibovic2024}. In the context of strong-field and attosecond physics, vortex beams are increasingly important for 
transferring angular momentum to photoelectrons~\cite{Maxwell2021}, probing chiral~\cite{Planas2022} and topological matter~\cite{Granados2025}, and generating vortex harmonics~\cite{Dorney2019}.

\textit{Valleytronics} is also an emerging direction that exploits the valley degree of freedom as a carrier of information, in analogy to spin in spintronics~\cite{Zutic2004}.~Tailored fields, in particular, provide a powerful route to selectively control electronic populations in specific valleys of a material's band structure, i.e., local minima (maxima) in the conduction (valence) band located at inequivalent points in the Brillouin zone, typically high-symmetry ones~\cite{Schaibley2016}. In materials with multiple such valleys, electrons acquire an additional label, the valley index, alongside their charge and spin, making such systems promising platforms for valley-based optoelectronic applications~\cite{Vitale2018}.~In systems with broken inversion symmetry, the valley index acquires a well-defined pseudospin character~\cite{Schaibley2016}, resulting in valley-dependent optical selection rules allowing different valleys to be selectively excited through the polarisation of light. In this context, strong-field and shaped-light techniques offer a route to drive and probe electron dynamics on subfemtosecond timescales, opening the possibility of extending valley control in the petahertz regime~\cite{Ossiander2022}. Furthermore, strong field physics offers a versatile toolbox of shaped light fields (such as elliptically polarised, bichromatic or bicircular drivers) that can be readily adapted to implement valley-selective control across a wide range of material platforms~\cite{Langer2018,Jimenez-galan2020,Mrudul2021,Tyulnev2024}.

Last but not least, emerging \textit{quantum-light synthesised fields}, including combinations of coherent and squeezed-vacuum states~\cite{Lemieux2025PhotonBunching,tzur_attosecond-resolved_2025}, cat-state driving fields~\cite{Lewenstein2020,Lamprou2025}, and non-classical bichromatic waveforms, are opening new directions in attosecond quantum technologies by enabling ultrafast control through quantum fluctuations, coherence, and electron-photon entanglement~\cite{RiveraDean2022,RiveraDean2025}. For instance, recent studies have shown that the light statistics may modify the electron propagation in the continuum~\cite{EvenTzur2023,RiveraDean2025}, leading to strong HHG in circularly polarised fields~\cite{RiveraDean2025StructuredSqueezed,Stammer2026FluctuationSymmetry}, or substantially change tunnel ionisation in bichromatic fields, amplifying asymmetries in photoelectron momentum distributions~\cite{Singh2026BSVAsymmetry}. 

Fig.~\ref{fig:symmetries} illustrates the range of symmetries that can arise from both the driving field and the target, and how these may be reflected in strong-field observables. Shaping the laser field provides a particularly direct means of controlling its dynamical symmetries. For example, while a monochromatic linearly polarised field has several symmetries, introducing a second colour can selectively break them, with the surviving symmetries depending on parameters such as the relative phase between the two driving waves. Furthermore, non-classical light may have statistical properties with symmetries beyond those of a monochromatic coherent field.  Conversely, bichromatic circularly polarised fields may exhibit discrete dynamical symmetries combining spatial rotations with time translations. The target contributes its own geometrical symmetries, ranging from those of atomic orbitals and molecules to the crystallographic point-group symmetries of solids. 

\FloatBarrier

\subsection{Symmetries in attoscience: a brief historical summary}
\label{sec:historicover}
\begin{figure*}[h]
    \centering
    \includegraphics[width=\linewidth]{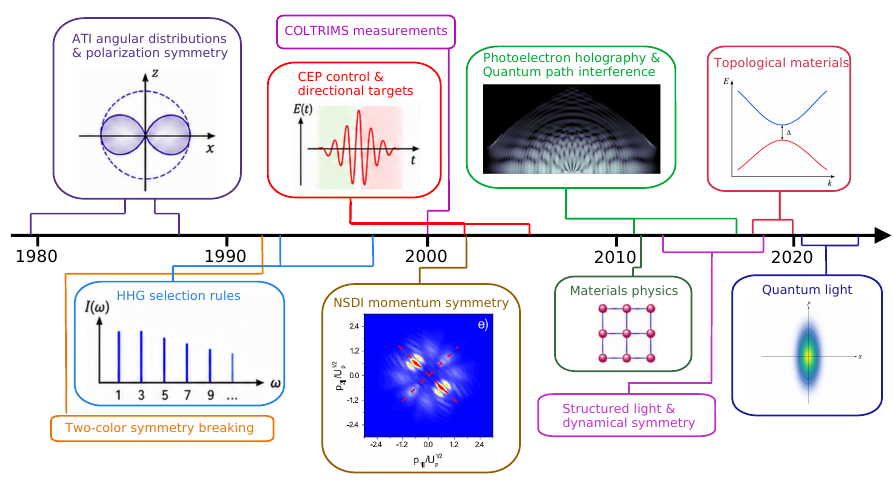}
    \caption{Schematic timeline showing a brief historical summary of symmetries in attosecond science. The first experiments on strong-field ionisation took place in the late 1970s~\cite{Agostini1979}, with subsequent studies in the late 1980s and early 1990s focusing on the photoelectron angular distributions in ATI~\cite{Bashkansky1987}
    and HHG~\cite{Alon1998}. The evolution of attosecond physics techniques~\cite{Dorner2000} and the gained control over light sources~\cite{Baltuska2003} in the early 2000s allowed for a more complete experimental understanding of the symmetries under more complex processes. This was extended to NSDI~\cite{Rudenko2007}, which required correlated measurements between electrons. These discoveries also prompted a conceptual shift in how symmetries in strong-field physics are interpreted, requiring theory models to be fine-tuned to include the effect of the Coulomb potential and the wide variety of structures it could generate on photoelectron spectra~\cite{HuismansScience2011,Maxwell2017}. Within the last fifteen years, the continued evolution of attosecond physics and the increasing control over driving field sources has made it possible to extend strong-field driving to a myriad of materials, from semiconductors~\cite{Ghimire2011} to topological materials~\cite{Luu2018B,Chacon2020}. More recently, this progress has also enabled the engineering of spatiotemporal waveforms in the emitted harmonics~\cite{Rego2019}, as well as the investigation of quantum light in strong-field processes~\cite{Gorlach2023,Rasputnyi2024}.
    }
    \label{fig:timeline}
\end{figure*}

These are examples of fields whose symmetries have been studied systematically in a myriad of ways, in addition to the symmetry of the target.~Their interplay influences the symmetry of the electron trajectories, and of the quantum amplitudes and observables, such as high-order harmonic or photoelectron spectra.~The research area evolved from relatively simple symmetry arguments toward increasingly sophisticated dynamical and group-theoretical analyses.~Next, we provide a brief historical overview intended to illustrate this evolution. A timeline is provided in Fig.~\ref{fig:timeline}, which we will match with the discussion below. 

Early studies of symmetry within strong-field and attosecond physics, in the 1980s and early 1990s, focused on photoelectron angular distributions in ATI. Seminal work on linearly polarised fields found that these distributions are predominantly concentrated along the polarisation axis, becoming narrower with increasing photoelectron energy \cite{Humpert1985}, and, in some cases, their symmetry could be explained by laser-induced scattering modified by ponderomotive potentials \cite{Freeman1986,McIlrath1987}. For circularly polarised light, it was shown that the light angular momentum was transferred to the emitted photoelectrons \cite{McIlrath1987}, and, for elliptically polarised fields, the angular distributions were predicted to exhibit a fourfold symmetry about the minor and major polarisation axes \cite{Bashkansky1987}. Further experiments have shown that these reflection symmetries were broken, but the angular distributions were reflection-symmetric with regard to the field helicity \cite{Bashkansky1988}. Theoretical studies have shown that the symmetry breaking could be attributed to the residual Coulomb potential in the continuum \cite{Basile1988,Jaron1999}. Decades later, path-integral methods incorporating this potential provided more rigorous evidence that this was indeed the case \cite{Goreslavski2004,Popruzhenko2008}. Around this period, two-color fields, combining $\omega$ and $2\omega$ fields, have been identified as tools for breaking inversion symmetry and controlling angular distributions.

Furthermore, selection rules were investigated in the foundational HHG theory articles \cite{Becker1994,Lewenstein1994}, in the 1990s. Examples are the absence of even harmonics for systems with inversion symmetry and linearly polarised monochromatic fields, or the prediction that elliptically polarised light will lead to elliptically polarised harmonics.~A more formal approach based on dynamical symmetries and general group-theoretical arguments has been used in \cite{Alon1998}, in which HHG selection rules were derived for time-periodic Hamiltonians.~Around the same time, intense, bichromatic two-color fields, both collinearly and orthogonally polarised, were theoretically \cite{Long1995} and experimentally studied \cite{Perry1993,Watanabe1994,Eichmann1995}. Specifically, for $(\omega, 2\omega)$ fields, there was the appearance of even harmonics, resulting from the temporal inversion symmetry being broken \cite{Long1995}. It was also demonstrated theoretically that high-order harmonics are phase locked and can be used to generate attosecond pulse trains \cite{Antoine1996}, although attosecond pulses were realised experimentally only a few years later \cite{Paul2001,Hentschel2001} (for reviews see, e.g.,  \cite{Krausz2016,LHuillier2023}). 

In the 2000s, few-cycle pulses \cite{Nisoli1997,Sartania1997} and carrier-envelope phase (CEP) stabilisation \cite{Telle1999,Jones2000,Apolonski2000} enabled direct manipulation of temporal symmetry, and turned symmetry breaking into a diagnostic tool \cite{Baltuska2003,Paulus2003} (for reviews see \cite{Brabec2000,Milos2006}). For example, in \cite{Paulus2003} ultrafast electron emission was directed to the right or to the left by varying the CEP, serving as a tool for its diagnosis,  building upon former evidence that the CEP influenced ionisation \cite{Paulus2001}. Additionally, aligned molecules and directional targets extended symmetry studies in strong-field physics beyond the laser field itself, allowing molecular orbital symmetry, nodal structures, and orientation-dependent quantum interference to be directly mapped through photoelectron and high-harmonic spectra \cite{Kanai2005,Kang2020,Meckel2014}. Specifically, aligned and oriented molecules in strong fields brought directional targets with preferred molecular axes, breaking the effective cylindrical symmetry characteristic of atomic systems \cite{Kanai2005,Habibovic2021}.~Consequently, strong-field ionisation and high-harmonic generation became highly sensitive to orbital parity, nodal-plane structure, and molecular orientation \cite{Kang2020,Kanai2005,Lein2002HHG}. Different alignments produced distinct rescattering and interference conditions, including two-centre interference patterns analogous to double-slit interference \cite{Lein2002HHG,Kanai2005,Spanner2004}. Ionisation probabilities were found to depend strongly on the angle between the laser field and the molecular orbitals, while molecular orientation could break inversion symmetry and enable even-harmonic generation \cite{Kanai2005,Habibovic2021}.
These studies paved the way towards other areas of research, such as molecular imaging, orbital tomography, laser-induced electron diffraction (LIED), chiral strong-field physics, and HHG spectroscopy \cite{Lein2007,Spanner2004,Meckel2014,Ayuso2022,Heinrich2021}.

In the early 2000s, symmetry was also present in NSDI momentum distributions, first for the ion and then for the correlated two-electron momentum distribution. This was made possible by the development of the COLTRIMS (Cold Target Recoil Ion Momentum Spectroscopy) technique \cite{Dorner2000}, which allowed the measurement of the recoiling target ion, and thus shifted NSDI investigations from the total yield to momentum distributions. This experimental technique was instrumental in confirming that NSDI resulted from a laser-induced inelastic recollision mechanism \cite{Corkum1993}, and not shake-off ionisation, in which the first electron, upon ionisation, would trigger a very fast rearrangement of the core  \cite{Fittinghoff1992, Kheifets2001, Kheifets2022}, or collective tunnelling, where both electrons are freed together and reach the continuum simultaneously \cite{Eichmann2000}. Those alternative mechanisms would require peaks around vanishing momentum in the COLTRIMS distribution \cite{Moshammer2000,Weber2000}, which would signal that both electrons tunneled near the maxima of the laser field. In contrast, recollision takes place near the field zero crossings, ensuring non-vanishing momentum transfer from the first electron to the core \cite{Weber2000,Rudenko2007}. In addition to that, symmetries in correlated two-electron momentum distributions were also observed experimentally \cite{Weber2000,Rudenko2007} and modelled theoretically, both for EI \cite{Faria2004,Chen2010} and RESI \cite{Shaaran2010a,Shaaran2012,Maxwell2015}. Similarly to what happened to ATI decades earlier, the emphasis was on the shapes of the momentum distributions, which were modelled incoherently \cite{Faria2004b,FigueiraDeMorissonFaria2011}. Several studies indicated that, for EI, both electrons were released without time delay, occupying the first and third quadrants of the plane spanned by the momentum components parallel to the driving-field polarisation \cite{Weber2000,Rudenko2007,Faria2004,Chen2010}. Anticorrelated distributions, measured experimentally \cite{Rudenko2007} and modelled by classical-trajectory methods \cite{Chen2017}, were associated with RESI.

From the late 2000s onwards, attention shifted to photoelectron holography. This shift was caused by the necessity of determining phase differences between different quantum pathways, which, in the context of molecular imaging with high-order harmonics has proven to be highly non-trivial, and by the ultrashort time scales and high electron currents involved. For a detailed discussion, we refer to our previous review \cite{Faria2020}. 

One should note, in addition, that using photoelectron holography as a tool also meant a conceptual shift on how symmetry was interpreted in strong-field physics. Previously, symmetry studies focused on the properties of the field, and of the target, while photoelectron holography tied symmetry to electron trajectories and phase differences between quantum pathways. This period also coincided with the rise of orbit-based methods in which the Coulomb potential was incorporated in the electron continuum propagation, and, from the experimental side, increasingly detailed momentum-resolved methods.  The myriad structures that arise in this context, and their symmetry, can be manipulated by an appropriate choice of the field or the target. Examples are  the fan-shaped fringes that occur near the ionisation threshold \cite{Rudenko2004,Maharjan2006}, the spider-like fringes near the polarisation axis \cite{Huismans2012} and the spiral-like fringes perpendicular to the polarisation axis \cite{Maxwell2020}.

In the 2010s, structured light, with fully spatiotemporal engineered waveforms, became possible  through the convergence of few-cycle pulse compression  \cite{Nisoli1997,Sartania1997}, carrier-envelope-phase stabilisation \cite{Telle1999,Jones2000,Apolonski2000}, phase-locked multicolour waveform synthesis \cite{Watanabe1994,Kim2005,Shafir2009,Brugnera2011,Wirth2011}, programmable femtosecond pulse shaping \cite{Weiner1990,Assion1998,Weiner2000}, and advanced polarisation and wavefront control \cite{Sansone2006,Fleischer2014}. Further developments involve structured light, such as vortex beams \cite{Allen1992,Beijersbergen1993,Marrucci2006,Gauthier2017,Dorney2019} (for a review see \cite{Wang2025}). Structured light required more systematic theoretical approaches, uniting space and time when interpreting and classifying the symmetries of the field. This led to an upsurge in group-theoretical approaches describing strong-field laser-matter interaction. Although there were seminal publications since the late 1990s and early 2000s applying these methods to periodic  fields, predominantly focusing on selection rules \cite{Alon1998,Ceccherini2001,ALon2002}, they have gained traction with the advent of shaped fields. A key issue is that the symmetries must account for the spatial and temporal dependence of the field, and for the symmetry of the target \cite{Baykusheva2016,Neufeld2019,Habibovic2021}. These group-theoretic considerations are crucial to characterise, for instance, chirality, which is one important application of vortex beams \cite{Habibovic2024}.

That decade also brought quantum-mechanical simulations of NSDI, which, supported by more sophisticated classical models, have shown that RESI may lead to distributions of a myriad of shapes~\cite{Ye2008,Shaaran2010,Maxwell2015}. In particular, SFA computations have highlighted a fourfold symmetry for RESI in monochromatic fields or long enough pulses~\cite{Faria2004,Shaaran2010,Shaaran2010a}, which was broken by quantum interference~\cite{Maxwell2015,Maxwell2016}. Another important development in the 2010s brought together electron-electron correlation and synthesised light, in a series of investigations of NSDI in shaped light, such as few-cycle pulses \cite{Liu2004,Bergues2012,Huang2016,Kubel2016,Chen2017}, circularly polarised fields \cite{Fu2012,Huang2018}, polarisation gated fields \cite{Quan2009}, OTC fields \cite{Zhang2014,Mancuso2016,Song2018}, or few-cycle counter-rotating two-colour circularly or elliptically polarised (TCCP, TCEP) laser fields \cite{Pang2020,Ge2023,Liu2024}. These studies were mainly performed using classical-trajectory models, with emphasis on how features such as final-state electron-electron repulsion \cite{Huang2016}, qualitatively different types of recollisions \cite{Chen2017} and temporal windows for recollision dynamics \cite{Fu2012,Zhang2014,Mancuso2016} influenced the shape of the electron-momentum distributions. This does not mean, however, that quantum mechanical approaches were not employed.  For example, the full solution of the time-dependent Schr\"odinger equation (TDSE) was used to determine how the type of electron-electron interaction \cite{Lein2000,Parker2006} , or the field shape \cite{Baier2006,Baier2007}, influences the symmetry and the maxima of correlated electron momentum distributions. Similar questions were asked in early work using the strong-field approximation (SFA) \cite{Faria2004,Faria2004b,Faria2005,Faria2008}, or  the quantitative rescattering theory (QRS) \cite{Chen2010,Chen2019,Chen2021,Chen2022}.

More recently, symmetry has become both a design principle for emerging attosecond light sources and a diagnostic tool for quantum materials, enabling circularly polarised and vortex high harmonics \cite{Saito2017}, symmetry-controlled HHG emission from crystals \cite{Langer2017}, using HHG to examine crystal anisotropy \cite{You2016,Luu2018}, and ultrafast probes of Berry curvature \cite{Silva2019,Luu2018B}, topology \cite{Chacon2020,Baykusheva2021,Heide2022}, chirality \cite{Heinrich2021}, and correlated phases \cite{Murakami2018,Uchida2022,Shao2022}.

Finally, recently, HHG from non-classical Bright-Squeezed Vacuum (BSV) light sources whose average intensities are $10^{12}\mathrm{W/cm}^2$ and with broad photon-number distributions, covering states from 0 to $2\times 10^{13} $ photons per pulse, has been reported \cite{Rasputnyi2024BSVHHG}. These experiments are evidence that quantum light can drive strong-field processes and provide a new, genuinely quantum route for symmetry control in attosecond physics. For classical drivers, the field symmetry is a property of the deterministic laser waveform and leads to well-defined selection rules for strong-field observables. With non-classical light, however, the relevant symmetry is no longer fixed solely by the mean field: photon statistics and quantum fluctuations can break, soften, or statistically sample otherwise forbidden configurations. Further experimental results indicate that high-order harmonics or photoelectrons inherit field properties \cite{Heimerl2025NeedleTipsBSV,Lemieux2025PhotonBunching,tzur_attosecond-resolved_2025}. 
Moreover, theoretical studies show that the HHG \cite{Gorlach2023} and ATI \cite{Wang2023QuantumATI,Lyu2025PhotonStatisticsATI} features, such as the plateau and the cutoff, are sensitive to the quantum state of light. This establishes the field statistics as another degree of freedom. It is even so that the quantum noise may lead to an effective photon statistics force, which will influence the electron trajectories in the continuum \cite{EvenTzur2023}. A striking example of how these dynamics are altered is given in \cite{RiveraDean2025StructuredSqueezed}, in which is shown that the fluctuations associated with squeezing may lead to strong harmonics in circularly polarised light, while for classical fields the absence of rescattering precludes their existence. Further work has shown that quantum fluctuations can break symmetries in a bicircular squeezed driver, leading to harmonics that are forbidden for its classical counterpart \cite{Stammer2026FluctuationSymmetry}. Recently, we have found that squeezing may amplify asymmetries in above-threshold ionisation by collinear $(\omega, 2\omega)$ bichromatic fields \cite{Singh2026BSVAsymmetry}. This effect comes from tunnelling ionisation being boosted by squeezing for specific time delays between the two field waves. Recently, quantum-boosted tunnelling was reported experimentally for BSV light \cite{Jiang2026BSVTunnelling}, albeit in a different context (elliptically polarised light).

\section{Theoretical Framework: Quantum-Orbit Methods}
\label{sec:quantumorbits}

The starting point, from which the quantum-orbit methods considered in this review are derived, is the time-dependent Schr\"odinger equation
\begin{equation}
    i \frac{\partial}{\partial t}  |\psi (t) \rangle = \hat{H}(t) |\psi (t) \rangle ,
\end{equation}
where the Hamiltonian $\hat{H}(t) $ includes the target and the interaction with the field, and is chosen according to the problem at hand. Throughout, we consider the light field to be classical, and refer to \cite{Stammer2023} for a full quantum-electrodynamical (QED) treatment. We work within the length gauge and dipole approximation, and, unless otherwise stated, employ atomic units. 

 For a two electron atomic system, the Hamiltonian reads as
\begin{equation}
    \hat{H}(t) = \sum_{j=1}^2 \left[\frac{\hat{\mathbf{p}}_j^2}{2} + V_{Z_\mathrm{eff}}(\hat{\mathbf{r}}_j)+\hat{\mathbf{r}}_j\cdot \mathbf{E}(t)\right] + V_{12}(\hat{\mathbf{r}}_1-\hat{\mathbf{r}}_2) ,
\end{equation}
where the hats denote operators, $\hat{\mathbf{p}}_j$ and $\hat{\mathbf{r}}_j$, with $j=1,2$, represent the electron momenta and coordinates, respectively, $\mathbf{E}(t)$ is the time-dependent field, assumed to be spatially homogeneous, $V_{12}(\hat{\mathbf{r}}_1-\hat{\mathbf{r}}_2)$ is the electron-electron interaction and $V_Z(\mathbf{r}) = -Z/|\mathbf{r}|$ is the Coulomb potential of arbitrary strength. The interaction of the two active electrons with the residual ion is via a potential with effective charge $Z_\mathrm{eff}$. The interaction of each electron with the field is given by the dipole terms $\hat{\mathbf{r}}_j\cdot E(t)$. This is an appropriate starting point for a treatment of NSDI, in which electron-electron correlation plays a huge role. 

 For ATI, the main features are captured within the single-active-electron approximation. Thus, it suffices to consider the single electron Hamiltonian
\begin{equation}\label{eq:single_e_ham}
    \hat{H}(t) =  \frac{\hat{\mathbf{p}}^2}{2} + V_{Z_\mathrm{eff}}(\hat{\mathbf{r}})+\hat{\mathbf{r}}\cdot E(t),
\end{equation}
where, similarly, the residual ion is represented by the effective potential $ V_{Z_\mathrm{eff}}(\hat{\mathbf{r}})$ and the coupling with the field reads $\hat{\mathbf{r}}\cdot E(t)$. For simplicity, we will refer to  $ V_{Z_\mathrm{eff}}(\hat{\mathbf{r}})$ as  $\hat{V}$  in the equations that follow.

In this review, the observable of interest is the photoelectron momentum distribution (PMD), upon which the symmetries of the electric field and of the target may be imprinted. This photoelectron momentum distribution is built by taking the system to an initial state $\ket{\Psi_i}$, at an initial time $t_i$ to a final continuum state $\ket{\Psi_f}$,  at a final time $t_f$, so that $\ket{\Psi_f}=U(t_f,t_i)\ket{\Psi_i}$ and $t_f \rightarrow \infty$, where $U(t_f,t_i)$ is the time evolution operator associated with the full Hamiltonian. 

One may solve the TDSE for the PMD, however, the effect of these symmetries will remain entangled, unless one uses a temporal filter or an orbit-based model to separate them. Such methods are often in the spirit of path-integral formulations of quantum mechanics, or in some cases approximations of this theory. Generally, the transition amplitude may be written as a path integral over the ionisation time and the photoelectrons' path in the continuum.  This integral is made tractable by making use of the steepest-descent approximation, selecting only the ``stationary'' paths, which may be interpreted as quantum orbits.  Crucially, these path-integral based methods also allow us to disentangle the quantum interference present in PMDs. The ability to additively decompose the overall quantum mechanical amplitude into distinct quantum orbits allows us to pinpoint exactly where different interference features of the overall PMD originate by simply considering pairwise interference between different trajectory types. These key ideas are present in whole families of strong-field approaches, which have been developed over many decades. For details, together with a timeline and a detailed discussion, see the review article \cite{Faria2020}.

In addition to the path-integral methods highlighted above, other orbit-based approaches exist, although they are less widespread in strong-field and attosecond physics. Examples are Bohmian trajectories, which arise from a hydrodynamic formulation for the TDSE~\cite{Cruz-rodriguez2019}, or initial-value representations, which often explore real trajectories in phase space~\cite{Spanner2003}, or expansions of the time-dependent Schr\"odinger equation in a coupled-coherent state basis~\cite{Shalashilin2000,Shalashilin2008}. 

In the following section, we will review two path-integral methods which will each be particularly useful in understanding photoelectron holography of PMDs as well as the correlated PMDs of NSDI.
Firstly, the strong-field approximation (SFA), and specifically its extension to two-electron systems, is discussed. The SFA is the most employed method in strong-field physics, and relies on a series of simplifications: (a) neglect of the binding potential when the electrons are in the continuum; (b) neglect of the field when the electrons are bound; (c) neglect of the internal structure of the target. We will consider the SFA both in the context of ATI and NSDI, considering a single and two active electrons, respectively. Assumptions (b) and (c) mean that the state of the system is approximated by the coherent superposition of a field-free bound state and continuum states, while assumption (a) means that the continuum is approximated by Volkov states, i.e., by field-dressed plane waves. For NSDI processes that incorporate excitation to be computed within this formalism, it is necessary to relax assumption (c) and consider bound-state population transfer for the second electron.

Second, we will discuss the Coulomb quantum-orbit strong field approximation (CQSFA), which incorporates both the residual binding potential and the driving laser field within a path integral representation and approximates this via the steepest descent approach. This will only be done for the one-electron case in the present work.

We are interested in computing the ATI or NSDI photoelectron momentum distributions. Without loss of generality, we can assume that the final state is a field-dressed plane wave $\ket{\mathbf{p}+\mathbf{A}(t)}$, $t \rightarrow \infty$ or a tensor product $\ket{\mathbf{p}_1+\mathbf{A}(t)}\ket{\mathbf{p}_2+\mathbf{A}(t)} $ thereof if more than one electron is involved. The dressing is incorporated via the vector potential $\mathbf{A}(t)$. This expression has been verified to be an excellent asymptotic limit to the actual continuum states \cite{Rodriguez2023}.

The photoelectron momentum distributions will then be 
\begin{equation}
\mathcal{P}(\mathbf{p})=|\bra{\mathbf{p}+\mathbf{A}(t)}U(t,t_0)\ket{\psi(t_0)}|^2,
\label{eq:distroneelectron}
\end{equation}
or
\begin{equation}
\mathcal{P}(\mathbf{p}_1,\mathbf{p}_2)=|\bra{\mathbf{p}_1+\mathbf{A}(t),\mathbf{p}_2+\mathbf{A}(t)}U(t,t_0)\ket{\Psi(t_0)}|^2,
\label{eq:distr2electrons}
\end{equation}
for one or two-electron systems, respectively, where evolution took place from an initial time $t_0$ to a final time $t$. The one- and two-electron initial states were specified as $\ket{\psi(t_0)}$ and$\ket{\Psi(t_0)}$, respectively.

Next, we will bring the key ideas behind the path-integral methods used in this work. For simplicity, we discuss the one-electron case. We start by writing the time-evolution operator as the Duhamel integral equation
\begin{equation}\label{eq:timeEv1}
    \hat{U}(t,t_0) =  \hat{U}_{0}(t,t_0) - i \int_{t_0}^t \hat{U}(t,t') \hat{H}_I(t') \hat{U}_{0}(t',t_0)dt',
\end{equation}
where $\hat{U_0}$ is the time evolution operator for the field free part of the Hamiltonian and $H_I(t) = \mathbf{r}\cdot \mathbf{E}(t)$ is the dipole interaction term. Eq.~\eqref{eq:timeEv1} is exact. If it is iterated, this leads to perturbative series. This is the procedure followed, for instance, when building the low-intensity perturbation theory, which does not converge for strong fields. Eq.~\eqref{eq:timeEv1} is the starting point of both the SFA and the CQSFA. Throughout, we consider the limit $t_0 \rightarrow -\infty$. 

\subsection{The strong-field approximation}

The strong-field approximation (SFA) exhibits the formal structure of a perturbative scattering series, but with a field-dressed, modified basis. It was first developed by Keldysh \cite{Keldysh1965}, Faisal \cite{Faisal1973} and Reiss \cite{Reiss1980}, thus being also known as the Keldysh-Faisal-Reiss theory (KFR) and Perelomov, Popov and Terent'ev (PPT) \cite{Perelomov1966} to describe strong-field ionisation. In the 1990s, the SFA was adapted to describe HHG \cite{Lewenstein1994,Becker1994,Becker1997} and rescattered ATI \cite{Lewenstein1995,Becker1994b,Lohr1997}, and, eventually, to multielectron processes such as NSDI \cite{BeckerFaisal1996,Kopold2000,Goreslavskii2001}.  Since its early days, it has become the most widespread semi-analytical theory employed in strong-field and attosecond physics. 

Physically, the SFA effectively reframes the problem such that electrons in the continuum are dressed by the strong electromagnetic field and may interact perturbatively with the binding potential, while electrons in the bound state can only interact perturbatively with the electromagnetic field. 

\subsubsection{Single active electron: direct and rescattered ATI}
For a single electron, this is achieved by combining Eq.~\eqref{eq:timeEv1} with iterations of 
\begin{equation}\label{eq:timeEv2}
    \hat{U}(t,t') =  \hat{U}_{V}(t,t') - i \int_{t'}^t \hat{U}(t,\tau) \hat{V}(\mathbf{r}) \hat{U}_V(\tau,t')d\tau,
\end{equation}
where $\hat{U}_V$ is the Volkov time evolution operator. That is, the time evolution operator for a system which evolves under the sole influence of a time varying, spatially uniform electric field, which, in the length gauge, has the exact solution
\begin{equation}
    \hat{U}_{V}(t,t') = \exp\left\{-\frac{i}{2} \int_{t'}^t(\mathbf{p}+\mathbf{A}(\tau))^2d\tau\right\}|\mathbf{p}+\mathbf{A}(t) \rangle \langle \mathbf{p}+\mathbf{A}(t')|.
    \label{eq:volkovprojector}
\end{equation}
Once more, Eq.~\eqref{eq:timeEv2} is exact, but it can be repeatedly iterated around $\hat{U}_{V}$. To first order, this will give \begin{equation}\label{eq:timeEv3}
    \hat{U}(t,t') \simeq   \hat{U}_{V}(t,t') - i \int_{t'}^t \hat{U}_V(t,\tau) \hat{V}(\hat{\mathbf{r}}) \hat{U}_V(\tau,t')d\tau,
\end{equation}
which, if inserted in Eq.~\eqref{eq:timeEv1} will lead to the time evolution associated with direct and rescattered ATI, respectively \cite{Lohr1997}. After some manipulation, this leads to the direct ATI transition amplitude
\begin{equation}\label{Eq:ATI:prob:amp}
    M_{\mathrm{dir}}({\mathbf{p}}) = -i \int_{-\infty}^\infty \langle \mathbf{p}+\mathbf{A}(t')| \mathbf{r}\cdot \mathbf{E}(t') |\psi_0\rangle e^{iS_d(\mathbf{p};t')}dt',
\end{equation}
 where 
\begin{equation}
    S_d(\mathbf{p};t') = I_p t' - \frac{1}{2}\int_{t'}^\infty (\mathbf{p}_f+\mathbf{A}(t))^2 dt,
    \label{eq:directaction}
\end{equation}
is the semiclassical action. For direct ATI, the electron tunnels from an initial state $\ket{\psi_0}$ of energy $-I_p$ to a continuum state $\ket{\mathbf{p}+\mathbf{A}(t')}$ at $t'$ and propagates in the continuum up to the detector without further interaction. An equivalent way of writing Eq.~\eqref{Eq:ATI:prob:amp}, shown in Ref. \cite{Lohr1997}, is 
\begin{equation}\label{Eq:ATI:prob:ampV}
    M_{\mathbf{p}}^{(0)} = -i \int_{-\infty}^\infty \langle \mathbf{p}+\mathbf{A}(t')| \hat{V}|\psi_0\rangle e^{iS_d(\mathbf{p};t')}dt',
\end{equation}
which is widely used and given in terms of the binding potential. The prefactor 
\begin{equation}
    V_{\mathbf{p}0}=\langle \mathbf{p}+\mathbf{A}(t')| \hat{V}|\psi_0\rangle =\frac{1}{(2\pi)^{3/2}}\int d^3r e^{-i [\mathbf{p}+\mathbf{A}(t')]\cdot \mathbf{r}} \psi_0(\mathbf{r})V(\mathbf{r})
    \label{eq:directprefactor}
\end{equation}
contains all the information about the target via its bound-state wave function $\psi_0(\mathbf{r})$. In Eqs.~\eqref{Eq:ATI:prob:amp} and \eqref{Eq:ATI:prob:ampV}, a single integral over $t'$ must be solved. 

The second term of Eq.~\eqref{eq:timeEv3} will give the ATI transition amplitude in which a single act of rescattering is incorporated. After some rearranging, this rescattered ATI transition amplitude can be written as 
\begin{equation}
M_{\mathrm{resc}}
= - \int_{-\infty}^{\infty} dt
\int_{-\infty}^{t} dt'\,
\left\langle
\psi_{\mathbf{p}}^{(V)}(t)
\left|
\hat{V} U^{(V)}(t,t') \hat{V}
\right|
\psi_0(t')
\right\rangle .
\label{eq:ATI_resc}
\end{equation}
Strictly speaking, Eq.~\eqref{eq:ATI_resc} contains the direct and rescattered ATI expressions, as shown in \cite{Lohr1997}. However, once saddle-point equations are derived from this transition amplitude, rescattered ATI is the process one focuses on unless one considers $t\rightarrow t'$. 

If one considers the explicit expression for the time evolution operator and closure relations in momentum space, it is convenient to write $M_{\mathrm{resc}}$ as 
\begin{equation}
    M_{\mathrm{resc}}
=
-\int_{-\infty}^{\infty} dt
\int_{-\infty}^{t} dt'
\int d^{3}\mathbf{k}\,
e^{iS_{\mathbf{p}}(t,t',\mathbf{k})}
V_{\mathbf{p}\mathbf{k}}V_{\mathbf{k}0},
\label{eq:ATIrescpspace}
\end{equation}
where
\begin{equation}
    S_{\mathbf{p}}(t,t',\mathbf{k})
=
-\frac{1}{2}\int_{t}^{\infty} d\tau\,
\left[\mathbf{p}+\mathbf{A}(\tau)\right]^2
-\frac{1}{2}\int_{t'}^{t} d\tau\,
\left[\mathbf{k}+\mathbf{A}(\tau)\right]^2
+
I_pt' \label{eq:actionATIresc}
\end{equation}
is the semiclassical action associated with rescattered ATI, and the prefactors read
\begin{equation}
    V_{\mathbf{p}\mathbf{k}}
=
\left\langle \mathbf{p}+\mathbf{A}(t) \left| \hat{V} \right|
\mathbf{k}+\mathbf{A}(t) \right\rangle
=
\frac{1}{(2\pi)^3}
\int d^3\mathbf{r}\,
e^{-i(\mathbf{p}-\mathbf{k})\cdot\mathbf{r}}
V(\mathbf{r}) 
\end{equation}
and 
\begin{eqnarray}
    V_{\mathbf{k}0}
&=&
\left\langle \mathbf{k}+\mathbf{A}(t') \left| \hat{V} \right| 0 \right\rangle \nonumber \\
&=&
\frac{1}{(2\pi)^{3/2}}
\int d^3\mathbf{r}\,
e^{-i\{\mathbf{k}+\mathbf{A}(t')\}\cdot\mathbf{r}}
V(\mathbf{r})\psi_0(\mathbf{r}),
\label{eq:atipref1}
\end{eqnarray}
respectively. In the above-stated equations $\mathbf{k}$ is the intermediate momentum of the electron, $V_{\mathbf{p}\mathbf{k}}$ is the prefactor associated with rescattering and the prefactor $V_{\mathbf{k}0}$ links the initial bound state with a continuum state. The ionisation prefactor is formally equivalent to that for direct ATI, with the difference that Eq.~\eqref{eq:directprefactor} considers the final instead of the intermediate momentum.  The action [Eq.~\eqref{eq:actionATIresc}] also contains the intermediate dressed momentum $\mathbf{k}+\mathbf{A}(\tau)$, for times $t'\leq \tau < t$, that is, between ionisation and rescattering, and the final momentum  $\mathbf{p}+\mathbf{A}(\tau)$ for later times. 

One also sees that Eq.~\eqref{eq:ATIrescpspace} has the form of nested integrals in the rescattering time $t$, the ionisation time $t'$ and the intermediate momentum $\mathbf{k}$, thus making it a 5-dimensional problem. In contrast, the direct-ATI transition amplitude contains a single integral over $t'$. Thus, a single act of rescattering has added significant complexity to the problem.  

\subsubsection{Two-electron systems: below-threshold nonsequential double ionisation}
Next, we will discuss the SFA transition amplitude for NSDI, focusing exclusively on the recollision-excitation with subsequent ionisation (RESI) mechanism. This choice is based on quantum-interference being robust for RESI, while being washed out for electron-impact ionisation (EI). One should note, however, that it is possible to derive a two-electron SFA from first principles, which contains both mechanisms (see, e.g., \cite{Symphony} for a review).

RESI prevails in the so-called ``below-threshold regime'', when the driving field intensities are not sufficient to provide the first electron with enough kinetic energy to trigger an immediate release of the second electron upon recollision. Instead, the returning electron excites a second electron, which reaches the continuum with a time delay. 

The strong-field approximation allows a well-defined, albeit simplified, description of RESI. This description is given in \cite{Shaaran2010,Shaaran2010a} and briefly outlined below. This is enabled by the clear scattering structure embedded in the SFA framework. The Feynman diagram employed to describe RESI is presented in Fig.~\ref{fig:FeynmanDiagram}. It corresponds to the following process: 
\begin{figure}
    \centering
    \includegraphics[width=\linewidth]{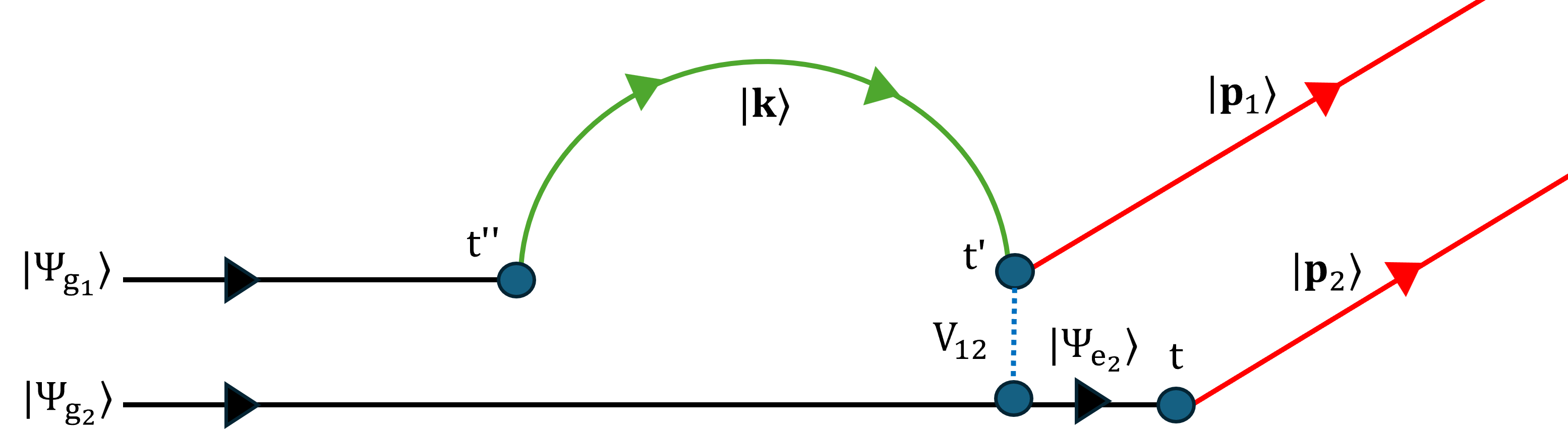}
    \caption{Feynman diagram associated with RESI, in which the initial, intermediate and final states of both electrons and the electron-electron interaction are illustrated. The arrows indicate the direction of the temporal evolution, the black lines indicate  the bound-state field-free propagation, and the coloured lines indicate the continuum propagation, with intermediate and final states signposted as green and red, respectively. The vector potential has been omitted in the Volkov states for simplicity, but the continuum states should be understood as laser-dressed.  }
    \label{fig:FeynmanDiagram}
\end{figure}
First, both electrons are in the initial state $\ket{\Psi_i}=\ket{\psi_{g_1}}\ket{\psi_{g_2}}$, where the subscripts $g$ indicate that both electrons are in the ground state and we take $t_0 \rightarrow -\infty$. The two-electron system evolves up to the time $t''$ guided by a field-free compound time evolution operator $U_0(t'',t_0)=U_{0_1}(t'', t_0)\otimes U_{0_2}(t'', t_0)$. At an initial time $t''$, the first electron tunnels into the continuum, which is approximated by Volkov states, while the second electron remains bound.~Thus, the intermediate state characterizing both electrons is $\ket{\mathbf{k}+\mathbf{A}(\tau)}\ket{\psi_{g_2}}$, $t''<\tau<t'$. The two-electron system then evolves in the continuum up to a rescattering time $t'$ according to the compound time-evolution operator $U(t',t'')=U_{V_1}(t',t'') \otimes U_{0_2}(t',t'')$, so that the first (second) electron evolves following the Volkov (field-free) time-evolution operator. At a time $t'$, the electron-electron interaction $V_{12}$ takes place, exciting the bound electron.~This means that the first electron will acquire its final momentum $\mathbf{p}_1$, and the second electron will transition to an excited state. Thus, the intermediate two-electron compound state at the rescattering time will be $\ket{\mathbf{p}_1+\mathbf{A}(t')}\ket{\psi_{e_2}}$, where excitation is indicated by the subscript $e_2$. The system then evolves from $t'$ to $t$ according to $U(t, t') = U_{V_1}(t, t') \otimes U_{0_2}(t, t').$ Finally, at a time $t$, the second electron reaches the continuum with field-dressed momentum  $\mathbf{p}_1+\mathbf{A}(t')$, so that the final state describing both electrons is $\ket{\mathbf{p}_1+\mathbf{A}(t)}\ket{\mathbf{p}_2+\mathbf{A}(t)}$. 

The transition amplitude describing the process outlined in Feynman diagram (see Fig.~\ref{fig:FeynmanDiagram}) is
\begin{eqnarray}
&&M^{(\mathcal{C})}_{RESI}(\mathbf{p}_{1},\mathbf{p}_{2})=\hspace*{-0.2cm}\int_{-\infty }^{\infty
}dt\int_{-\infty }^{t}dt^{^{\prime }}\int_{-\infty }^{t^{\prime
}}dt^{^{\prime \prime }}\int d^{3}k  \notag \\
&&\times V^{(\mathcal{C})}_{\mathbf{p}_{2}e}V^{(\mathcal{C})}_{\mathbf{p}_{1}e,\mathbf{k}g}V^{(\mathcal{C})}_{\mathbf{k}%
	g}\exp [iS^{(\mathcal{C})}(\mathbf{p}_{1},\mathbf{p}_{2},\mathbf{k},t,t^{\prime },t^{\prime
	\prime })],  \label{eq:MRESI}
\end{eqnarray}
where
\begin{eqnarray}
&&S^{(\mathcal{C})}(\mathbf{p}_{1},\mathbf{p}_{2},\mathbf{k},t,t^{\prime },t^{\prime \prime
})=  \notag \\
&&\quad E^{(\mathcal{C})}_{\mathrm{1g}}t^{\prime \prime }+E^{(\mathcal{C})}_{\mathrm{2g}}t^{\prime
}+E^{(\mathcal{C})}_{\mathrm{2e}}(t-t^{\prime })-\int_{t^{\prime \prime }}^{t^{\prime }}%
\hspace{-0.1cm}\frac{[\mathbf{k}+\mathbf{A}(\tau )]^{2}}{2}d\tau  \notag \\
&&\quad -\int_{t^{\prime }}^{\infty }\hspace{-0.1cm}\frac{[\mathbf{p}_{1}+%
	\mathbf{A}(\tau )]^{2}}{2}d\tau -\int_{t}^{\infty }\hspace{-0.1cm}\frac{[%
	\mathbf{p}_{2}+\mathbf{A}(\tau )]^{2}}{2}d\tau  \label{eq:singlecS}
\end{eqnarray} 
is the semiclassical action, and we have used the expansion of the Volkov time-evolution operators $U_{V_1}$ and $U_{V_2}$ into a field-dressed momentum basis Eq.~\eqref{eq:volkovprojector}. Here, we consider an arbitrary excitation channel $(\mathcal{C})$, and write the ground- and excited-state energies for each electron accordingly. The first electron is initially bound in a state of energy $-E^{(\mathcal{C})}_{1g}$, and, upon return, it  excites a second electron from a bound state with energy  $-E^{(\mathcal{C})}_{2g}$ to a state with energy  $-E^{(\mathcal{C})}_{2e}$. One should note that the RESI action is factorizable, although the transition amplitude is not. Thus, electron-electron correlation is accounted for via the time ordering: recollision of the first electron must happen before ionisation of the second electron \cite{Shaaran2012}. For the very same reason, the Feynman diagram in Fig.~\ref{fig:FeynmanDiagram} is not factorizable. 

Similarly to the single-electron case, the 
prefactors  $V^{(\mathcal{C})}_{\mathbf{k}g}$, $V^{(\mathcal{C})}_{\mathbf{p}_1e,\mathbf{k}g}$ and $V^{(\mathcal{C})}_{\mathbf{p}_2e}$ contain all information about the bound-state geometry, the binding potentials, and, in addition, the form of the electron-electron interaction \cite{Shaaran2010}. The superscript $(\mathcal{C})$ renders these expressions easily adaptable to coherent superpositions of channels and bound states.

The prefactor associated with the ionisation of the first electron reads
\begin{eqnarray}\label{eq:pre3}
V^{(\mathcal{C})}_{\mathbf{k}g} &=& \bra{\mathbf{k}+\mathbf{A}(t'')}V\ket{\psi_{1g}^{(\mathcal{C})}}\\ &= &\frac{1}{(2\pi)^{3/2}}\int d^3r_1e^{-i[\mathbf{k} +\mathbf{A}(t'')]\cdot\mathbf{r}_1}V(\mathbf{r}_1)\psi_{1g}^{(\mathcal{C})}(\mathbf{r}_1), \notag
\end{eqnarray}
where $V(\mathbf{r}_1)$ is the binding potential of the neutral atom, and $\braket{\mathbf{r}_1|\psi_{1g}^{(\mathcal{C})}}=\psi_{1g}^{(\mathcal{C})}(\mathbf{r}_1)$ is the ground-state wave function for the first electron.

The prefactor related to the recollision-excitation process is given by 
\begin{eqnarray}
\label{eq:Vp1ekg}V^{(\mathcal{C})}_{\mathbf{p}_1e,\mathbf{k}g}\hspace*{-0.1cm}&=& \hspace*{-0.1cm} \bra{\mathbf{p}_1,\psi_{2e}^{(\mathcal{C})}}V_{12}
\ket{\mathbf{k},\psi_{2g}^{(\mathcal{C})}}
    \\
    &=&\hspace*{-0.1cm} \frac{V_{12}(\mathbf{p}_1-\mathbf{k})}{(2\pi)^{3/2}}\hspace*{-0.1cm}\int \hspace*{-0.1cm}d^3r_2e^{-i(\mathbf{p}_1-\mathbf{k})\cdot \mathbf{r}_2}\psi_{2e}^{*(\mathcal{C})}(\mathbf{r}_2)\psi_{2g}^{(\mathcal{C})}(\mathbf{r}_2),  \notag 
\end{eqnarray}
where
\begin{equation}
    V_{12}(\mathbf{p}_1 - \mathbf{k}) =  \frac{1}{(2\pi)^{3/2}}\int d^3rV_{12}(\mathbf{r})\exp[-i(\mathbf{p}_1-\mathbf{k})\cdot\mathbf{r}]
\end{equation}
 gives the interaction by which the second electron is excited.~Throughout this review, this will be taken to be contact-type, i.e., $V_{12}=\delta(\mathbf{r}_1-\mathbf{r}_2)$, which is justified by assuming that the remaining core electrons screen the long range of the electron-electron interaction. The wave functions $\braket{\mathbf{r}_2|\psi_{2e}^{(\mathcal{C})}}=\psi_{2e}^{(\mathcal{C})}(\mathbf{r}_2)$ and $\braket{\mathbf{r}_2|\psi_{2g}^{(\mathcal{C})}}=\psi_{2g}^{(\mathcal{C})}(\mathbf{r}_2)$ are associated with the excited and ground states of the second electron, respectively.

Finally, 
\begin{eqnarray}
    V^{(\mathcal{C})}_{\mathbf{p}_2e} &=& \bra{\mathbf{p}_2+\mathbf{A}(t)}V_{\mathrm{ion}}\ket{\psi_{2e}^{(\mathcal{C})}}  \\ &=& \frac{1}{(2\pi)^{3/2}}\int d^3r_2V_{\mathrm{ion}}(\mathbf{r}_2)e^{-i[\mathbf{p_2}+\mathbf{A}(t)]\cdot\mathbf{r}_2}\psi_{2e}^{(\mathcal{C})}(\mathbf{r}_2),\notag
    \label{eq:Vp2e}
\end{eqnarray}
where $V_{\mathrm{ion}}(\mathbf{r}_2)$ is the binding potential of the singly ionised target, describes the ionisation of the second electron. One should note that prefactors \eqref{eq:pre3} and \eqref{eq:Vp2e} link a bound state to a continuum state, similarly to those obtained for ATI. In particular, Eq. \eqref{eq:pre3}, connecting the electronic ground state with an intermediate momentum, is formally equal to Eq. \eqref{eq:atipref1} given as the first step of rescattered ATI, while Eq. \eqref{eq:Vp2e}, linking the final momentum of the second electron with an excited bound state, formally resembles that obtained for direct ATI [Eq.~\eqref{eq:directprefactor}]. The key difference is in the recollision-excitation process, in which the electron-electron interaction takes the second electron from the ground to an excited state.  

Furthermore, Eq.~\eqref{eq:MRESI} has three temporal nested integrals and a three-dimensional integral over the intermediate momentum, resulting in a six-dimensional integral in total for a \textit{single} excitation channel. This adds complexity to the problem, although the SFA is much less demanding than solving the two-electron problem using ab-initio methods. Whenever dealing with a single excitation channel, for simplicity, we omit the superscript $\mathcal{(C)}$.

\subsection{The Coulomb Quantum-Orbit Strong-Field Approximation}
If, instead of iterating the time evolution operator \eqref{eq:timeEv1} or combining perturbative series in different approximations when calculating the transition probability $\mathcal{F}(\mathbf{p})$ we apply time-slicing techniques \cite{Kleinert2009,Milosevic2013JMP}, it is possible to construct path-integral methods which incorporate the residual binding potential and the electric field on equal footing.~One such method is the Coulomb Quantum-Orbit Strong-Field Approximation (CQSFA).  

The CQSFA transition amplitude reads as
\begin{align}\label{eq:CQSFATamp}
    M_{\mathrm{CQSFA}}(\mathbf{p}_f)&=-i\lim_{t\rightarrow \infty
}\int_{-\infty}^{t}dt' \int d\mathbf{\tilde{p}}_0
\int_{\mathbf{\tilde{p}}_0}^{\mathbf{\tilde{p}}_f(t)} \mathcal {D}'
\mathbf{\tilde{p}}  \int
\frac{\mathcal {D}\mathbf{r}}{(2\pi)^3}  \nonumber \\
&\quad  \times  e^{i S(\mathbf{\tilde{p}},\mathbf{r},t,t')}
\langle
\mathbf{\tilde{p}}_0 | H_I(t')| \psi _0  \rangle \,,
\end{align}
where $\mathcal{D}'\tilde{\mathbf{p}}$ and $\mathcal{D}\mathbf{r}$ are the integration measures for the path integrals, and the prime indicates a restriction. These represent a sum over all possible paths in position and momentum, that the electron can take, between its start and end points. The tildes over the initial and intermediate momenta indicate field dressing, i.e., $\mathbf{\tilde{p}}_0=\mathbf{p}_0+\mathbf{A}(t')$ and $\mathbf{\tilde{p}}=\mathbf{p}+\mathbf{A}(\tau)$, with $t \le \tau \le t'$. For more complete discussions and full derivations, together with time-slicing techniques, we refer to our original publications  \cite{Lai2015a,Lai2017,Maxwell2017,Rodriguez2023}.

The action in Eq.~\eqref{eq:CQSFATamp} is given by
\begin{equation}\label{stilde}
S(\mathbf{\tilde{p}},\mathbf{r},t,t')=I_pt'-\int^{t}_{t'}[
\dot{\mathbf{p}}(\tau)\cdot \mathbf{r}(\tau)
+H(\mathbf{r}(\tau),\mathbf{p}(\tau),\tau)]d\tau,
\end{equation}
and the Hamiltonian by 
\begin{equation}
H(\mathbf{r}(\tau),\mathbf{p}(\tau),\tau)=\frac{1}{2}\left[\mathbf{p}(\tau)+\mathbf{A}(\tau)\right]^2
+V(\mathbf{r}(\tau)).
\label{Hamiltonianpath}
\end{equation}

In Eq.~\eqref{eq:CQSFATamp}, both the residual binding potential and the external laser field are fully incorporated in the electron dynamics. This implies that the CQSFA considers rescattering in its full extent, and not only to zeroth or first order \cite{Lai2015a}. In the limit of vanishing Coulomb potential, Eq.~\eqref{eq:CQSFATamp} reduces to the Strong-Field Approximation (SFA) transition amplitude associated with direct ATI.  Eq.~\eqref{eq:CQSFATamp} contains several nested integrals, over the ionisation time $t'$, the initial and intermediate field-dressed momenta $\tilde{\mathbf{p}}_0$ and $\mathbf{\tilde{p}}$ and the intermediate coordinate $\mathbf{r}$. This gives ten nested integrals if the problem is solved in three dimensions, and seven nested integrals if we restrict momenta and coordinates to the driving-field polarisation plane. Thus, the level of complexity has increased substantially if compared to approximate methods in which the continuum is approximated by Volkov states. 

\subsection{Calculating saddle-point equations}\label{sec:spe}
\label{sec:saddlepoints}

The transition amplitudes in the previous sections have the general form of nested temporal integrals, and sometimes integrals over intermediate momenta and/or coordinates. These integrals contain a slowly varying prefactor or the product thereof, and a highly oscillating semiclassical action. One may either solve these integrals numerically, or resort to saddle-point methods, in which one seeks the values of these variables such that the action is stationary, i.e, so that its derivative vanishes. This leads to saddle-point equations, which can be directly related to the classical trajectories of an electron in the continuum. Furthermore, the nested integrals can be approximated by sums over the saddle-point solutions, weighted by the prefactors and additional stability factors. Thus, in addition to simplifying the actual computations of transition amplitudes, they provide valuable insight into the phenomenon in question, and allow a direct assessment of the interfering pathways from the parent ion to the detector.

Here, we will focus on the saddle-point equations for scenarios of one, two and more dimensions.~An example of a one-dimensional problem is direct ATI within the SFA. Rescattered ATI can be reduced to a two-dimensional problem, and more dimensions will be involved if the SFA is applied to more than one electron and there is a time delay, such as in RESI, or if the Coulomb potential is incorporated in the path-integral method.
\subsubsection{One-dimensional problem: direct ATI}

Seeking values of $t'$ in Eq.~\eqref{Eq:ATI:prob:ampV} for which the action is stationary leads to the saddle point equation (SPE) 
\begin{equation}\label{Eq:SPE}
    \frac{\partial S_d(\mathbf{p};t')}{\partial t'} = \frac{1}{2}\left[\mathbf{p}+\mathbf{A}(t')\right]^2 + I_p = 0,
\end{equation}
whose solutions $t'_s$ will be associated with ionisation times. Assuming that these solutions are all sufficiently well separated, we can then approximate the amplitude by
\begin{equation}\label{Eq:SPA:ATI:prob:amp}
    M_{\mathrm{dir}}(\mathbf{p}) \approx \sum_s \sqrt{\frac{2\pi i}{\partial^2 S(\mathbf{p};t'_s)/\partial t_s^{'2}}} \langle \mathbf{p}+\mathbf{A}(t'_s)| V(\mathbf{r})|\psi_0\rangle e^{iS_d(\mathbf{p};t'_s)}.
\end{equation}
Depending on the choice of electric field, Eq.~\eqref{Eq:SPE} may be solved analytically, but is often solved by numerical root-finding. 

\begin{figure*}
    \centering
    \includegraphics[width=0.7\linewidth]{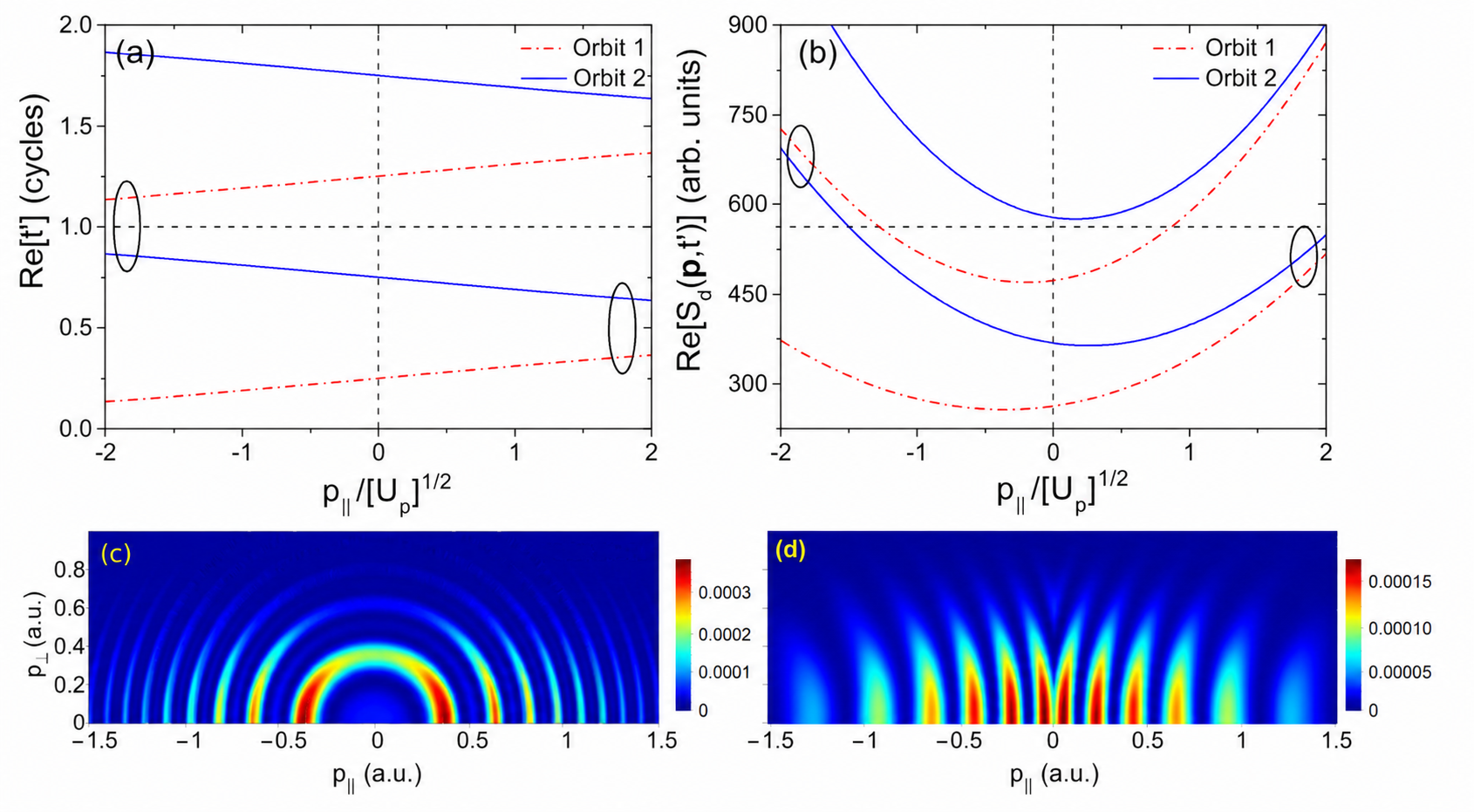}
\caption{SFA quantum orbits and the resulting inter- and intra-cycle interference patterns. (a) Real parts of the saddle-point ionisation times and (b) the corresponding semiclassical actions as functions of the electron momentum $p_{\parallel}$ parallel to the laser-field polarisation. The red dash-dotted and blue solid curves denote orbit 1 (short orbit) and orbit 2 (long orbit), respectively; solutions belonging to consecutive optical cycles are also shown and plotted using the same style. The ellipses identify the pairs of saddle-point solutions subsequently combined coherently to construct the intra-cycle interference pattern in panel (d). (c) Inter-cycle interference obtained by coherently superposing orbit-1 contributions from three consecutive optical cycles, giving the characteristic ATI-ring structure. (d) Type-A intra-cycle interference arising from the coherent superposition of orbit-1 and orbit-2 contributions within the same optical cycle, for which the relevant ionisation times are separated by less than half a cycle. The calculations are for hydrogen in its $1s$ ground state, with $I_p=0.5$ a.u., driven by a linearly polarised monochromatic field of intensity $I=2\times10^{14}\mathrm{W/cm^2}$ and frequency $\omega=0.057$ a.u. (corresponding to $\lambda\simeq800\mathrm{nm}$). Momenta in panels (a) and (b) are scaled by $\sqrt{U_p}$, while panels (c) and (d) show the parallel and perpendicular momentum components in atomic units. Adapted from Figs.~1 and 2 of Ref.~\cite{Maxwell2017}.}
    \label{fig:sfa_inter_intra}
\end{figure*}

At first sight, Eq.~\eqref{Eq:SPE} seems visually simple although, in practice, its evaluation can become as complex as the driving field itself.~Most of such complexity lies in the fact that their solutions require complex-valued $t$, with their real part denoting the time at which electrons are born in the continuum, while their imaginary component represents the time electrons spend within the tunnel~\cite{Smirnova2014}:~the greater it is, the less probable it is for the electron to ionise.~For a simple monochromatic field $E_0\sin(\omega t)$, Eq.~\eqref{Eq:SPE} can be solved analytically, yielding a total number of four solutions per field cycle, two of which are eliminated as they provide unphysical results~\cite{Popruzhenko2014a}.~These physically valid solutions read
\begin{equation}
    t_{\pm}
        = \frac{2\pi k}{\omega}
            \pm \frac{1}{\omega}
                \arccos\Bigg[\frac{-p_{\parallel} \mp i \sqrt{p_{\perp}^2 +2I_p}}{2\sqrt{U_p}}\Bigg],
    \ k \in \mathbbm{Z}
    \label{eq:saddlemonochromatic}
\end{equation}
corresponding to ionisation close to the two field extrema within the cycle.  The direct ATI spectra computed with the SFA for a monochromatic field using these expressions are given in Fig.~\ref{fig:sfa_inter_intra}. The figure shows rings that are associated with inter-cycle interference, on the left-hand side, and additional structure related to the two solutions considered within a field cycle, on the right-hand side. 

However, scenarios where analytical solutions are possible are a \textit{rara avis}, and the addition of an envelope or more complex multichromatic fields with potentially arbitrary polarisations can very easily complicate the form of the saddle-point equations as well as increase the number of saddle-point solutions, making the use of numerical approaches unavoidable.~Yet, the specific methodology used for solving the corresponding equations is a world of its own, since the driving field peculiarities can make the saddle-points move, merge and bifurcate~\cite{Faria2000,Faria2002,Jasarevic2020,Rook2022} when the field parameters transition between different regimes. This therefore makes the definition of a ``universal solver'' difficult to establish. To this end, the use of machine learning algorithms can help in finding solutions to Eq.~\eqref{Eq:SPE} for arbitrary shaped fields, as we have recently shown for direct ATI in Ref.~\cite{Jiakang2026} for symmetric fields, since unlike local-optimisation-based methods~\cite{Nocedal2006,Murray2011}, machine learning techniques are naturally built to adapt a broad range of conditions and parameters, thereby reducing the sensitivity of an algorithm to the problem's initial conditions.

Even further, in some cases the approximations leading to Eqs.~\eqref{Eq:SPE} and \eqref{Eq:SPA:ATI:prob:amp} are not valid, especially when saddle-points become close to each other, i.e., coalesce~\cite{Berry1989,Koch2018,Koch2018b}.~In these cases, one could retrace the underlying steps culminating in Eq.~\eqref{Eq:SPA:ATI:prob:amp} and incorporate the proper correction terms~\cite{Faria2002,Milosevic2002}, leading to the so-called Uniform Approximation.~This approach, however, requires a collective treatment of orbits as well as evaluating special functions ~\cite{Connor1976}. Although the numerical cost can be reduced by using the fact that, often, the orbits that appear in strong-field phenomena occur in pairs, the actual implementation requires some problem specific, sometimes ad-hoc classifications.  Alternatively, and as recently demonstrated, one can trace back even further to explicitly find adequate integration contours to Eq.~\eqref{Eq:ATI:prob:amp} allowing exact evaluation and identification of the relevant saddle-point contributions~\cite{Weber2025}.

\subsubsection{Two-dimensional problem: rescattered ATI}

Next, we will discuss rescattered ATI, which is an excellent case study for solving saddle-point equations in two dimensions.~Seeking solutions so that $ \partial S_{\mathbf{p}}(t,t',\mathbf{k})/\partial t'=\partial S_{\mathbf{p}}(t,t',\mathbf{k})/\partial t=0$, $\partial S_{\mathbf{p}}(t,t',\mathbf{k})/\partial \mathbf{k=\mathbf{0}}$ gives 
\begin{equation}
    \left[\mathbf{k}+\mathbf{A}(t')\right]^2 = -2I_p,
    \label{eq:rescSP1}
\end{equation}
\begin{equation}
    \left[\mathbf{p}+\mathbf{A}(t)\right]^2
=
\left[\mathbf{k}+\mathbf{A}(t)\right]^2,
\label{eq:rescSP2}
\end{equation}
\begin{equation}
    \int_{t'}^{t} d\tau\,
\left[\mathbf{k}+\mathbf{A}(\tau)\right]
=\mathbf{0}.
\label{eq:rescSP3}
\end{equation}
Their solutions determine the ionisation time \(t'\) [Eq.~\eqref{eq:rescSP1}], the rescattering time \(t\) [Eq.~\eqref{eq:rescSP2}],
and the electron intermediate momentum \(\mathbf{k}\) [Eq.~\eqref{eq:rescSP3}] in between those
two times.~Upon rescattering, the electron acquires the asymptotic momentum
\(\mathbf{p}\). Eq.~\eqref{eq:rescSP1} is associated with the conservation of energy upon tunnelling and has no real solution. Note that its main difference from its counterpart for direct ATI, is that it is written in terms of the intermediate momentum $\mathbf{k}$, not the final momentum $\mathbf{p}$.  Eq.~\eqref{eq:rescSP3} constrains the intermediate momentum so that the freed electron returns to the origin. This equation tells us that, in the saddle-point approximation, the intermediate momentum is along the driving-field polarisation direction, and also makes it possible to write it as a function of the remaining variables, i.e., $ \mathbf{k}=\mathbf{k}(t,t') $. Thus, the five-dimensional integral in Eq.~\eqref{eq:ATIrescpspace} can be reduced to 
\begin{equation}
M_{\mathrm{resc}}
=
-\int_{-\infty}^{\infty} dt
\int_{-\infty}^{t} dt'\,
e^{iS_{\mathbf{p}}(t,t')}
V_{\mathbf{p}\mathbf{k}(t,t')}
V_{\mathbf{k}(t,t')0},
\label{eq:ATIrescM2}
\end{equation}
where
\begin{equation}
    \mathbf{k}(t,t')
=
-\frac{1}{t-t'}
\int_{t'}^{t} d\tau\,\mathbf{A}(\tau)
\end{equation}
and \(S_{\mathbf{p}}(t,t') \equiv
S_{\mathbf{p}}(t,t',\mathbf{k}(t,t'))\). For well-separated saddles, this gives 
\begin{equation}
    M_{\mathrm{resc}}^{(\mathrm{SPA})}
=
\sum_s A_s \exp(iS_s),
\label{eq:SPAMresc}
\end{equation}
where 
\begin{equation}
  S_s = S_{\mathbf{p}}(t_s,t'_s),  
\end{equation}
and 
\begin{equation}
    A_s
=
(2\pi i)^{5/2}
\frac{
V_{\mathbf{p}\mathbf{k}(t_s,t'_s)}
V_{\mathbf{k}(t_s,t'_s)0}
}{
\sqrt{
(t'_s-t_s)^3
\det S''_{\mathbf{p}}(t,t')\big|_s
}
},
\end{equation}
where the $(t'_s-t_s)^{-3/2}$ term is associated with wave-packet spreading in the continuum. Physically, it is linked to the suppression of the transition amplitude for  larger excursion times. 

Moving from direct to rescattered ATI further accentuates the difficulties mentioned in the previous section. While, for direct ATI, the saddles tend to be well separated, unless there is a dramatic change in the field shape that will force them to merge (for examples see \cite{Weber2025}), for rescattered ATI they almost always coalesce or nearly coalesce. This implies that Eq.~\eqref{eq:SPAMresc} will not be a valid approximation for all photoelectron energy ranges, even if the driving field is linearly polarised.~Thus, in some instances, it will be necessary to treat more than one saddle collectively. In some particular cases, such as linearly polarised monochromatic fields, the saddles will always occur in pairs that nearly coalesce near the ATI cutoff energy. Thus, specific uniform asymptotic expansions can be applied \cite{Faria2002}.

 This is a consequence of the fact that, in addition to the ionisation condition [Eq.~\eqref{eq:rescSP1}, one must simultaneously satisfy the energy conservation condition at rescattering [Eq.~\eqref{eq:rescSP2}] and the return condition [Eq.~\eqref{eq:rescSP3}], which has been implicitly embedded in the other two saddle-point equations according to the above procedure~\cite{Lewenstein1995}. The problem therefore ceases to involve a single ionisation time and instead requires the determination of a coupled set of complex variables.~These quantities are linked through a highly nonlinear system of saddle-point equations whose solution often lie far from those associated with direct ATI trajectories \cite{Faria2002}, 
 rendering standard root-finding procedures increasingly sensitive to the choice of initial guesses. This challenge is further exacerbated in the presence of polychromatic or arbitrarily polarised driving fields~\cite{Hashim2025,Neufeld2025}, for which merging and bifurcation phenomena become more frequent as the field parameters are varied~\cite{Habibovic2025i,Habibovic2025,Habibovic2025iii,Jiakang2026}.
Another typical two-dimensional problem in the context of saddle-point equations and the SFA is high-order harmonic generation (HHG). It is handled in a very similar way as rescattered ATI and brings comparable technical challenges \cite{Weber2025}. This is not surprising, as the key difference between both processes is the third step: Instead of undergoing an elastic collision with its parent ion, for HHG to occur the active electron recombines with a bound state, releasing the kinetic energy it acquired in the continuum as high-frequency radiation. 

\subsubsection{Multidimensional problems: RESI and Coulomb distorted single electron ionisation }

Finally, we will illustrate two scenarios in which multiple nested integrals must be solved: the specific two-electron problem associated with NSDI RESI [Eq.~\eqref{eq:MRESI}] and the CQSFA transition amplitude [Eq.~\eqref{eq:CQSFATamp}]. 
Applying the steepest descent method to Eq.~\eqref{eq:MRESI} gives two saddle-point equations identical to the rescattered ATI equations \eqref{eq:rescSP1} and \eqref{eq:rescSP3} for the first electron, namely
\begin{equation}\label{eq:RESIsp1}
[\mathbf{k}+\mathbf{A}(t'')]^2 = -2E^{(\mathcal{C})}_{1g},
\end{equation}
\begin{equation}\label{eq:RESIsp2}
\mathbf{k} = -\frac{1}{t'-t''}\int_{t''}^{t'}d\tau \mathbf{A}(\tau). 
\end{equation}
These equations describe the same physics and the difference is mainly associated with notation: In RESI, we imply the convention $t''$ and $t'$ for its ionisation and rescattering time, instead of $t'$ and $t$, changing the arguments and the limits in the above integrals accordingly. We also consider the bound-state energy to be $E^{(\mathcal{C})}_{1g}$ to allow the possibility for different channels, although, for the first electron, this will be the ionisation potential of the neutral target.~This is expected, as the first electron behaves as in rescattered ATI: it is freed in the continuum by tunnelling, and returns to the site of its release at a later time.~The difference is that, upon return, an inelastic collision takes place and a second electron is excited.~Therefore, $\partial S^{(\mathcal{C})}(\mathbf{p}_{1},\mathbf{p}_{2},\mathbf{k},t,t^{\prime },t^{\prime \prime
})/\partial t'$ gives 
\begin{equation}\label{eq:RESIsp3}
[\mathbf{p}_{1}+\mathbf{A}(t')]^2 = [\mathbf{k}+\mathbf{A}(t')]^2 - 2(E^{(\mathcal{C})}_{2g}-E^{(\mathcal{C})}_{2e}),
\end{equation}
which gives the conservation of energy in this context.~This equation states that, upon return, the first electron gives a fraction $\Delta E=E^{(\mathcal{C})}_{2g}-E^{(\mathcal{C})}_{2e} $ of its kinetic energy to the core, whereupon it acquires the final momentum $\mathbf{p}_1$. One should also note that Eqs.~\eqref{eq:RESIsp1}-\eqref{eq:RESIsp3} only contain momenta (intermediate and final) associated with the first electron. 

Seeking the stationary values of the ionisation times $t$ for the second electron gives 
\begin{equation}\label{eq:RESIsp4}
[\mathbf{p}_{2} + \mathbf{A}(t)]^2 = -2E^{(\mathcal{C})}_{2e},
\end{equation}
which describes direct ATI from an excited bound state. Note that this equation only depends on the second electron. 

Thus, essentially, RESI results from two time-ordered ATI-like processes for the first and second electron.~This means that one must solve two coupled saddle-point equations for the first electron [Eqs.~\eqref{eq:RESIsp1} and \eqref{eq:RESIsp3}], with the intermediate momentum given by the return condition \eqref{eq:RESIsp2}, and an additional, independent saddle-point equation [Eq.~\eqref{eq:RESIsp4}] for the second electron. This stems from the RESI action being factorizable, i.e., $S^{(\mathcal{C})}(\mathbf{p}_{1},\mathbf{p}_{2},\mathbf{k},t,t^{\prime },t^{\prime \prime
})=S_1^{(\mathcal{C})}(\mathbf{p}_{1},\mathbf{k},t^{\prime },t^{\prime \prime
})+S_2^{(\mathcal{C})}(\mathbf{p}_{2},t)$.~Thus, solving the saddle-point equation for the first and the second electron will lead to essentially the same challenges as for rescattered and direct ATI, respectively. Nonetheless, there are differences. First, the inelastic rescattering process means that there will be a minimal energy for which rescattering will have a classical counterpart, which is  $\Delta E=E^{(\mathcal{C})}_{2g}-E^{(\mathcal{C})}_{2e} $. Second, ionisation of the second electron cannot happen before the first electron recollides, i.e., one must enforce causality. Because the variables are complex, this will require modifying the steepest descent contours to account for that \cite{Shaaran2012}, although, in practice, for most parameter ranges it suffices to assume that the real part of $t'$  is smaller than the real part of $t$. 

The next example, namely solving the transition amplitude \eqref{eq:CQSFATamp} obtained with the CQSFA for strong-field ionisation, is more convoluted, although Eq.~\eqref{eq:CQSFATamp} describes a process within the single-active-electron approximation. Apart from the nested integrals that should be taken into consideration, there will be singularities associated with the presence of the Coulomb potential. Physically, it is illustrative to compare the CQSFA transition amplitude with their counterparts for direct and rescattered ATI, although the CQSFA transition amplitude ~\eqref{eq:CQSFATamp} contains rescattering to all orders as it is fully Coulomb distorted. The key message is that the residual potential adds much more complexity, which is also reflected in the saddle-point equations. 

It is convenient to consider a two-pronged contour in time, which is widespread in approaches incorporating the Coulomb potential \cite{Popruzhenko2008,Yan2012,Torlina2012,Torlina2013}. The first arm of the contour goes from $t'$ to $\mathrm{Re}[t']$, and the second arm from  $\mathrm{Re}[t']$ to  $\mathrm{Re}[t]$, with $t \rightarrow \infty$. This allows splitting the action neatly into 
\begin{equation}\label{ssplit}
S(\mathbf{\tilde{p}},\mathbf{r},t,t')=S^{\mathrm{tun}}(\mathbf{\tilde{p}},\mathbf{r},t'_r,t')+S^{\mathrm{prop}}(\mathbf{\tilde{p}},\mathbf{r},t,t'_r),
\end{equation}
where 
$S^{\mathrm{tun}}(\mathbf{\tilde{p}},\mathbf{r},t'_r,t')$ and $S^{\mathrm{prop}}(\mathbf{\tilde{p}},\mathbf{r},t,t'_r)$ are associated with the sub-barrier and continuum dynamics, respectively. Keeping the real part of $t'$ constant in the first part of the contour is used as an intuitive argument to keep the initial momentum $\mathbf{p}_0$ constant. This simplifies the problem as it allows an easier expression for the coordinate 
\begin{equation}
\mathbf{r}_0(\tau) = \int^{\tau}_{t'}(\mathbf{p}_0+\mathbf{A}(t''))dt'',
\label{subBarrierPosition}
\end{equation}
in the first part of the contour, typically known as ``the tunnel trajectory''. A constant momentum can also be employed to justify neglecting the binding potential in the first part of the contour. Within these approximations, the saddle point equations read 
\begin{eqnarray}
[\mathbf{p}_0+\mathbf{A}(t')]^2 = -2I_p \label{SPEt}\\
\mathbf{\dot{r}}(\tau) = \mathbf{p}(\tau) + \mathbf{A}(\tau) \label{SPEp}\\
\mathbf{\dot{p}}(\tau) = -\nabla_rV(\mathbf{r}(\tau)). \label{SPEr}
\end{eqnarray}
Equation (\ref{SPEt}) gives the tunnelling time $t'$, and the coupled equations (\ref{SPEp}) and (\ref{SPEr}) are given in terms of the electron's intermediate momentum and position and determine its continuum dynamics for $\mathrm{Re}[t']<\tau\leq t$. A further approximation is to take tunnel exit to be real and along the laser's polarisation axis. This gives
\begin{equation}
z_0 = Re[r_{0||}(Re[t'])].
\label{tunnelExitGeneral}
\end{equation}
This approximation has been made in order to avoid branch cuts upon acts of rescattering \cite{Popruzhenko2014b,Pisanty2016}. For details in the context of the CQSFA  see \cite{Maxwell2018b,Faria2020}.

For well-separated saddles, this then gives
\begin{equation}
\label{eq:MpPathSaddle}
M(\mathbf{p}_f)\propto-i \lim_{t\rightarrow \infty } \sum_{s}\bigg\{\det \bigg[  \frac{\partial\mathbf{p}_s(t)}{\partial \mathbf{p}_s(t'_s)} \bigg] \bigg\}^{-1/2} \hspace*{-0.6cm}
\mathcal{C}(t'_s) e^{i
	S(\mathbf{\tilde{p}}_s,\mathbf{r}_s,t,t'_s)} ,
\end{equation}
where $t'_s$, $\mathbf{r}_s$ and $\mathbf{p}_s$ are the solutions of Eqs.~\eqref{SPEt}, \eqref{SPEp} and \eqref{SPEr} respectively, and the sum is over the distinct saddle point trajectories with  final momentum $\mathbf{p}_f$. The prefactor $\det[  {\partial\mathbf{p}_s(t)}/{\partial \mathbf{p}_s(t'_s)} ]$ comes from the quadratic fluctuations around the saddle points and 
\begin{equation}
\label{eq:Prefactor}
\mathcal{C}(t'_s)=\sqrt{\frac{2 \pi i}{\partial^{2}	S(\mathbf{\tilde{p}}_s,\textbf{r}_s,t,t'_s) / \partial t'^{2}_{s}}}\langle \mathbf{p}_s+\mathbf{A}(t'_s)|\mathbf{r}\cdot \mathbf{E}(t'_s)|\psi_{0}\rangle ,
\end{equation}
where  The stability factor $\det[  {\partial\mathbf{p}_s(t)}/{\partial \mathbf{p}_s(t'_s)} ]$ arises from a Legendre transformation via the $\mathbf{p}_0$ integral in Eq.~\eqref{eq:CQSFATamp} [see Ref.~\cite{Lai2015}].

The above-stated equations exhibit a series of key differences from the SFA transition amplitudes discussed in this work. First, it is clear that Eqs.~\eqref{SPEr} and \eqref{SPEp} are coupled, so that, in contrast to the previous example, the problem cannot be reduced to a simpler task. Second, Eq.~\eqref{SPEp} shows that the electron momentum from the time of ionisation up to the detector is no longer conserved, and the system is accelerated by the residual binding potential in addition to the driving field. In contrast, in the SFA, the field-dressed electron momentum only changes when rescattering takes place [see Eqs.~\eqref{eq:rescSP2} and \eqref{eq:RESIsp3}]. Thus, in case of direct ionisation [see Eq.~\eqref{Eq:SPE}], the momentum obtained at the ionisation time will be the final momentum at the detector.  

This brings the next challenge in the CQSFA, which also makes the problem richer:~it is no longer possible to neatly distinguish between direct and rescattered electrons, as the theory is not built as a field-dressed Born series. The electron orbit may behave like a direct or rescattered SFA orbit, or like neither, as the residual potential may only decelerate the electron without altering the orbit, lightly deflect an orbit, or trigger a soft or hard collision of the active electron with the core. Often, there are orbits which exhibit a hybrid character, with no SFA counterpart. This issue has been briefly discussed in a seminal publication \cite{Yan2010} in which hybrid orbits and caustics not present in the SFA have been identified. These and other Coulomb distorted orbits have been studied by us in several publications, in which we have compared Coulomb-free and Coulomb-distorted orbits \cite{Maxwell2018}, have shown that, in general, there are no clear cut ``direct'' or ``rescattered'' criteria \cite{Rodriguez2023}. However, among the many possibilities and types of orbits, it is still possible to single out the subset of orbits that are well mimicked by those of the rescattered SFA \cite{Rook2024a}. 

Furthermore, the interplay of the residual binding potential and the external field brings with it focal points, caustics and singularities, which are associated with singularities in the Jacobian matrix $J=\partial\mathbf{p}_s(t)/{\partial \mathbf{p}_s(t'_s)}$. These singularities give rise to, for instance, low-energy structures \cite{Blaga2009,Quan2009} which can be associated with laser-driven distant collisions of an electron with its parent ion \cite{Kaestner2012,Kaestner2012b} and Coulomb focusing \cite{Kelvich2016}. Furthermore, depending on the type of binding potential, broken symmetries may influence the range of rescattering angles considerably \cite{Rook2024}. For simplicity, the CQSFA is solved on the polarisation plane. This may require additional phases to
be inserted by hand, in order to account for focal points which exist if the full three-dimensional ensemble of trajectories is considered \cite{Brennecke2020,Werby2021}. 

In practice, the saddle-point equations obtained with the CQSFA are solved in three different ways: (a) The standard, boundary-type CQSFA, which matches the asymptotic final momentum $\mathbf{p}_f$ with the initial momentum $\mathbf{p}_0$ at the tunnel exit using some pre-knowledge of the orbits' dynamics, imposing that Eqs.~\eqref{SPEt}-\eqref{SPEr} be satisfied; (b) the rate-based CQSFA (R-CQSFA), in which, in order to account for the sub-barrier contributions, we construct a nonadiabatic ionisation rate which includes Coulomb corrections from the CQSFA transition amplitude; and (c) the hybrid forward-boundary CQSFA (H-CQSFA), in which an ensemble of trajectories are launched from the tunnel exit, propagated until an asymptotic momentum value and subsequently used to solve a boundary problem as in (a). The boundary-type CQSFA is the least numerically demanding, but the least versatile because it requires initial guesses and pre-knowledge of the orbits' dynamics. Thus, it may miss whole classes of trajectories and it is not easily adaptable to arbitrary fields. The other two versions of the CQSFA are somewhat more expensive, from the numerical viewpoint, as one must numerically calculate an ensemble of classical trajectories for a large range of initial conditions. However, they are more easily adaptable if the shape of the driving field is modified, as they require no prior assumptions about the orbits' dynamics.  For details we refer to \cite{Lai2015,Rodriguez2023}.

\section{Symmetries used in this work}
\label{sec:symmetries}
Next, we briefly outline the dynamical symmetries used in our work. 
In what follows, we are concerned with the symmetries of a physical system and its time evolution (as opposed to a static geometric object). 
Depending on the theoretical framework, different mathematical objects may serve as the relevant invariant quantity under a transformation, such as the Hamiltonian, action or equations of motion. In laser–matter interaction, the system is explicitly time-dependent due to the presence of the driving field. Consequently, the most relevant symmetries are those that leave the equations of motion invariant (and hence preserve the dynamics of the system), even if the Hamiltonian itself is not invariant at each instant in time.  

We therefore identify physical symmetries as transformations that map dynamical trajectories onto symmetry-related trajectories. Such symmetries typically mix spatial and temporal transformations. 

\subsection{Field-related symmetries}
\label{sec:field_symm}
Below, we introduce the operators describing temporal and geometric transformations, which will be combined to describe the symmetries of the electric fields used in this work, as well as the corresponding vector potentials. They are: 
\label{subsec:operators}
\begin{enumerate}
    \item[$\mathcal{O}_1$:] The\textbf{ time-translation} operator $\hat{\mathcal{T}}_T(\tau_T)$ is defined as 
\begin{equation}
\hat{\mathcal{T}}_T(\tau_T): \quad t \rightarrow t + \tau_T ,
\label{eq:timetranslation}
\end{equation}
which shifts the electric field waveform along the time axis, $E(t) \rightarrow E(t+\tau_T)$. If the driving field is invariant under $\hat{\mathcal{T}}_T(\tau_T)$, the equations of motion are unchanged by this transformation, and solutions occur in symmetry-related `families' separated by the time shift $\tau_T$. 
\item[$\mathcal{O}_2$:] The\textbf{ time-reflection} operator $\hat{\mathcal{T}}_R(\tau_R)$ is defined as
\begin{equation}
\hat{\mathcal{T}}_R(\tau_R): \quad t \rightarrow -t + 2\tau_R ,
\end{equation}
which reflects the electric field waveform about a reference time $\tau_R$, $E(t) \rightarrow E(-t+2\tau_R)$. Time-reflection symmetry relates ionisation and recollision events occurring before and  after the reference time. 
\item [$\mathcal{O}_3$:] For linearly polarised fields, the \textbf{field-inversion} operator $\hat{\mathcal{F}}$ is defined as 
\begin{equation}
\hat{\mathcal{F}}: \quad E(t) \rightarrow -E(t), A(t) \rightarrow - A(t)
\end{equation}
which is an inversion of the electric field (or vector potential) along the temporal axis.

\item[$\mathcal{O}_4$:] For orthogonally polarised fields, the discrete rotation operator $\hat{R}_n$ plays a role, and is defined as,
\begin{equation}
\hat{R}_n: \quad \mathbf{E}(t) \rightarrow \mathcal{R}_{\theta_n}\mathbf{E}(t) ,
\end{equation}
where $\mathcal{R}_{\theta_n}$ denotes a rotation by an angle $\theta_n$. Note that while $\hat{R}_\theta$ is a generic rotation operator, rotational symmetry only exists for specific angles $\theta = \theta_n$. 
\end{enumerate}
Note that these are only a subset of the possible operators and combinations - see \cite{Neufeld2025} for a more thorough catalogue.

\subsubsection{Combined Operators}
Below are the physically relevant dynamical symmetries of the driving fields considered in this work, obtained from combining the above-stated operations. These are:
\begin{enumerate}
    \item[$\mathcal{S}_1$:] \textbf{Half-cycle symmetry}. The driving field is symmetric upon a  
 translation of half-cycle followed by a reflection along the time axis, i.e.,  $E(t)=\hat{\mathcal{F}}\,\hat{\mathcal{T}}_T(T/2)E(t)$. This symmetry is commonly expressed as $E(t\pm T/2) = -E(t)$. 
\item[$\mathcal{S}_2$:] \textbf{Time-reflection about extrema}. The driving field is reflection-symmetric with regard to extrema, i.e.,  $E(t)=\mathcal{T}_R (\tau_{ex})E(t)$ where $\tau_{ex}$ are the times where the extrema occur.
\item[$\mathcal{S}_3$:] \textbf{Time-reflection about zero crossings, followed by reflection along the time axis}. The driving field satisfies $E(t)=\mathcal{FT}_R(\tau_{cr})E(t)$ where $\tau_{cr}$ are the times of the zero crossings.
\item[$\mathcal{S}_4$:] \textbf{Rotation-time-translation}. This is time-translation followed by a discrete rotation in the polarisation plane $\hat R_n\,\hat{\mathcal T}_T(T/n)$. If the field is invariant under this operator then the system exhibits a discrete rotational symmetry of order $n$.
\end{enumerate}

\subsubsection{Field shapes}
\label{sec:fieldshapes}

Next, we discuss the importance of  the above-mentioned combined symmetries for the driving fields used in the remainder of this review. 
Symmetries $\mathcal{S}_1$ - $\mathcal{S}_3$ play an important role for  bi- or monochromatic linearly polarised fields

\begin{equation}
\textbf{E}(t) = [E_r \sin(r\omega t) + E_s \sin(s\omega t + \phi)] \mathbf{\hat{e}}_z,
\label{eq:bichromatic_field}
\end{equation}
where $E_r$, $E_s$ are the strengths of each component wave, $r$ and $s$ are assumed to be integer numbers (i.e., the frequencies are chosen to be commensurate), $\mathbf{\hat{e}}_z$ is the polarisation vector and $\phi$ is the relative phase between both driving waves. 
They were first discussed systematically in \cite{Rook2022}, and applied to photoelectron holography.~Subsequently, in our work, these symmetries were also explored in RESI \cite{Hashim2024b} and above-threshold ionisation with a bright squeezed vacuum \cite{Singh2026BSVAsymmetry}. 

The field given by Eq.~\eqref{eq:bichromatic_field} corresponds to the vector potential 
\begin{equation}
    \mathbf{A} (t) = \left[\frac{E_r}{r\omega}\cos(r\omega t) + \frac{E_s}{s\omega}\cos(s\omega t + \phi)\right]\mathbf{\hat{e}}_{z}   .
    \label{eq:vector_potential} 
\end{equation}
One should note that Eq.~\eqref{eq:bichromatic_field} is sufficiently general, as  all combinations $(r, s)$ are reducible either to the case in which r and s have opposite parity ($r+s$ odd) or to both r and s being odd ($r+s$ even). For a linearly polarised monochromatic field, $(E_s=0, r=1)$ the three symmetries $\mathcal{S}_1$ - $\mathcal{S}_3$ hold. These symmetries also hold for the vector potential, but with  $\tau_{cr}$ and  $\tau_{ex}$ swapped.

Introducing a second wave may or may not break such symmetries. If $r$ and $s$ have the same parity ($r+s=2n$), the half-cycle symmetry will be retained, but \textit{both} $\mathcal{S}_2$ and  $\mathcal{S}_3$ may or may not be broken, depending on the relative phase $\phi$ between the two waves, i.e., that is, the bichromatic field retains either the three symmetries or only the half-cycle symmetry. If $r$ and $s$ have opposite parities, the half-cycle symmetry \textit{and one of the other two symmetries} will be broken. This is due to the structure of the symmetry group governing this behavior. Intuitively, one can understand how the symmetries $\mathcal{S}_2$ and $\mathcal{S}_3$ hold or are broken by analyzing the field zero crossings or extrema of each constituent wave. If the zero crossings (extrema) are aligned, this means that the resulting bichromatic field is reflection-symmetric about its zeroes (extrema). In principle, this alignment can be generalised to an arbitrary number of constituent waves, but whether symmetries will be preserved will depend on the frequencies used. 

\begin{figure}[h]
    \centering
    \includegraphics[width=0.7\linewidth]{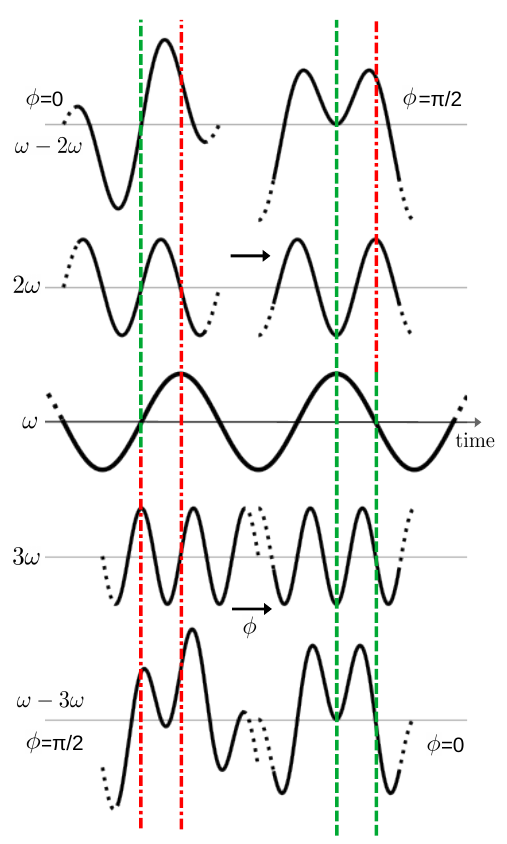}
    \caption{Schematic representation of bichromatic fields (up and bottom row) given by Eq.~\eqref{eq:bichromatic_field}, where $E_r=E_s$, together with their constituent fields (second to 4th row) as functions of time. The field frequencies are indicated on the left-hand side. The vertical green dashed lines indicate that either the extrema or the zero crossings are aligned, while the red dotted-dashed lines show that they are misaligned. The relative phases $\phi$ are given in left and right corners of the up and bottom rows, close to the time axis.  }
    \label{fig:bichromatic}
\end{figure}

This alignment is illustrated in Fig.~\ref{fig:bichromatic}, for $(\omega, 2\omega)$ and $(\omega, 3\omega)$ fields. If, for an $(\omega, 2\omega)$ field, the zero crossings are aligned, the symmetry $\mathcal{S}_3$ will hold. This happens, for instance, for $\phi=0$ in Eq.~\eqref{eq:bichromatic_field}. If, on the other hand, the extrema are aligned, such as for $\phi=\pi/2$ the resulting field will exhibit the reflection symmetry  $\mathcal{S}_2$. It is possible to show that, if the half-cycle symmetry $\mathcal{S}_1$ is broken, then either $\mathcal{S}_2$ or $\mathcal{S}_3$ is automatically broken.  Fig.~\ref{fig:bichromatic} also shows that, if the half-cycle symmetry is retained, this is no guarantee that the other two will hold. For instance, for $(\omega, 3\omega)$, $\phi=\pi/2$, the resulting field is half-cycle symmetric. However, the symmetries $\mathcal{S}_2$ and $\mathcal{S}_3$ are broken. In contrast, for $(\omega, 3\omega)$, $\phi=0$ $\mathcal{S}_1$ -- $\mathcal{S}_3$ hold. 

It is often convenient to write the field components in terms of the ponderomotive energy $U_p=\braket{\mathbf{A}^2(t)}_t/2$, where $\braket{\cdot}_t$ indicates a temporal average. This is the average kinetic energy acquired by the electron from the field.

For a linearly polarised monochromatic field this gives
\begin{equation}
    \mathbf{A}(t)=2\sqrt{U_p}\cos\omega t,
    \label{eq:Amonochromatic}
\end{equation}
with $U_p=A^2_r/4$ and $A_r=E_r/\omega$. For a bichromatic field,  Eq.~\eqref{eq:vector_potential} gives
\begin{equation}
    U_p=\frac{1}{2}\left(\frac{A^2_r}{r^2}+\frac{A_s^2}{s^2}\right).
\end{equation}

Subsequently, we consider linearly polarised $\sin^2$ few-cycle pulses, corresponding to the vector potential
\begin{equation}
\mathbf{A}(t) = 2\sqrt{U_p} \sin^2 \left( \frac{\omega t}{2N} \right) \sin(\omega t + \phi)\mathbf{\hat{e}}_{z}  ,
\label{eq:Apulse}
\end{equation}
where $N$ is the number of optical cycles and $\phi=\phi_1-\phi_{\mathrm{offset}}$ the carrier-envelope phase. 
The waveform is defined by the above expression over the interval $0 \leq t \leq NT$, with $T=2\pi/\omega$, and set to zero otherwise. One should note that, here, $N$ is not necessarily an integer. Throughout, we approximate the amplitude of the vector potential by
$A_0 = 2\sqrt{U_p}$. This is exact for a monochromatic linearly polarised field, but not for a few-cycle pulse. By varying the parameters in Eq.~\eqref{eq:Apulse}, we can influence the main rescattering and ionisation events.

These fields differ significantly from the monochromatic and bichromatic cases, in that the field envelope breaks their periodicity. Therefore, different field cycles are not equivalent, i.e.,  $E(t+T) \neq E(t)$.
They lose their time-translation symmetry, and with it the half-cycle symmetry. Time-reflection symmetry about the crossings and extrema are also generally broken, except in highly specific cases. 

To understand this issue, it is helpful to write the few-cycle pulse [Eq.~\eqref{eq:Apulse}] as a three-color field using trigonometric relations. This gives
\begin{equation}
 \mathbf{A}(t)= \sqrt{U_p}\left[ \mathbf{A}_0- \frac{1}{2}\left[\mathbf{A}_+(t)+\mathbf{A}_-(t) \right] \right],
\end{equation}
with
\begin{eqnarray}
    \mathbf{A}_0(t)&=&\sin(\omega t + \phi)\mathbf{\hat{e}}_{z},\\
    \mathbf{A}_{\pm}(t)&=&\sin\left( \omega \left( 1 \pm \frac{1}{N} \right) t + \phi \right)\mathbf{\hat{e}}_{z},
\end{eqnarray}
where all waves carry the carrier-envelope phase throughout. The frequencies $\omega$ and $\omega_{\pm}=(1\pm 1/N)\omega$ composing this three-color field are very specific, with $\omega_{\pm}$ being non-integer. Regardless of $N$, $\omega+\omega_{-}+\omega_{+}=3\omega$, which is odd. With these constraints upon the frequencies it is possible to make a limited number of extrema or zero crossings align, for specific cases. This is desirable to ensure, for instance, reflection symmetries about extrema, even if approximately.  Aligning the zero crossings of $\mathbf{A}(t)$ means aligning the extrema of the electric field $\mathbf{E}(t)$, which reads as 
\begin{eqnarray}
 \hspace*{-0.8cm}   \mathbf{E}(t)\hspace*{-0.3cm} &=\hspace*{-0.3cm} &-\omega \sqrt{U_p} \left[ \cos(\omega t + \phi) - \frac{1}{2} \left( \left( 1 + \frac{1}{N} \right) \times \right. \right. \\
\times&& \hspace*{-0.8cm}\left. \left. \cos\left[ \omega \left( 1 + \frac{1}{N} \right) t + \phi \right] + \left( 1 - \frac{1}{N} \right) \cos\left[ \omega \left( 1 - \frac{1}{N} \right) t + \phi \right] \right) \right] \mathbf{\hat{e}}_z.\nonumber
\end{eqnarray}
This can be achieved if each of the waves is dephased by half a cycle or a multiple thereof, which holds only for specific values of the CEP and pulse region.

In addition to linearly polarised fields, here we will bring examples of orthogonally polarised light, the simplest of which are elliptically polarised fields. Elliptically polarised fields introduce a transverse component of the electric field,
\begin{equation}
    \mathbf{E}_{\varepsilon}(t) = E_0 \left[ \sin(\omega t)\,\hat{\mathbf{e}}_z + \varepsilon \cos(\omega t)\,\hat{\mathbf{e}}_y \right].
    \label{eq:elliptical}
\end{equation}
where $\varepsilon \in [0,1]$ is the field ellipticity and $\hat{\textbf{e}}_z$ and $\hat{\textbf{e}}_y$ are the unit vectors along the major and minor polarisation axes respectively. This formulation ensures that for $\varepsilon=0$ we recover a linearly polarised monochromatic field, while $\varepsilon=1$ represents circular polarisation and has perfect continuous rotational symmetry which breaks with ellipticity. Half-cycle symmetry is retained regardless of ellipticity. While few-cycle pulses primarily change the dominance of events, elliptically polarised fields can change whether recollision happens at all. Traditionally, these types of fields have been used as a gate for several phenomena, as highlighted in Sec.~\ref{sec:shaped}.

We now take a brief foray into fields exhibiting discrete rotational symmetry, namely bicircular fields. These are formed by the superposition of two circularly polarised components with commensurate frequencies $r$ and $s$,
\begin{eqnarray}
   \textbf{ E}(t) &=&  E_1 [\cos(\omega t)\hat{\mathbf{e}}_z + \sigma_1\sin(\omega t)\hat{\mathbf{e}}_y] \nonumber \\ &+& E_2[ \cos(s \omega t + \phi) \hat{\mathbf{e}}_z + \sigma_2\sin(s\omega t + \phi)\hat{\mathbf{e}}_y],
\label{eq:bircular}
\end{eqnarray}
where we have taken $r=1$, $\sigma_{1,2} = \pm1$ is the helicity (left or right handed rotation), and $\sigma_1 = - \sigma_2$ denotes counter-rotating fields. Here $E_1$ and $E_2$ are the amplitudes of the two colours, the relative phase $\phi$ defines the alignment in time of the two colours similar to the bichromatic case.

By changing $s$ and $\phi$, one can tune these types of fields such that they remain invariant under time translation by $T/n$ and rotation of the polarisation plane by $2\pi/n$ where $n=(s+1)$ \cite{Orth2021, Madsen2016}.  Therefore, the field exhibits a discrete rotational symmetry of order $n=s+1$,
\begin{eqnarray}
\hat{R}_{n}\,\hat{\mathcal{T}}_{T}\left(\frac{T}{n}\right)\mathbf{E}(t)
&=&
E_{1}
\left(
\begin{array}{c}
\cos(\omega t)\\[2mm]
\sin(\omega t)
\end{array}
\right)
\nonumber\\&&+
E_{2}
\left(
\begin{array}{c}
\cos\left(n\omega t+\phi-\frac{2\pi}{n}\right)\\[2mm]
-\sin\left(n\omega t+\phi-\frac{2\pi}{n}\right)
\end{array}
\right),
\label{eq:bicircularInvariance}
\end{eqnarray}
where, for simplicity, we chose $\sigma_1=1$. 
The symmetry $\hat R_n\,\hat{\mathcal{T}}_T(T/n)\textbf{E}(t)  = \textbf{E}(t)$ holds if and only if $\phi \equiv \phi - 2\pi/n$ (mod $2\pi$), which is not satisfied for arbitrary $\phi$. 
This can be understood intuitively from the fact that the $\omega$ component rotates once per period while the $s\omega$ component rotates $s$ times per period, in the opposite direction. When the time translation and rotation is performed, for the discrete rotational symmetry to be retained, the total relative phase must be equal to $2\pi$. Note that the $(\omega, 2\omega, \phi=90^{\circ})$ bicircular field does not have true three-fold rotational symmetry as it does not satisfy the above constraint on $\phi$. Nonetheless, it can have time-reflection symmetry about the field extremum, meaning saddle point solutions may be linked by reflection symmetry but are not rotationally degenerate \cite{Habibovic2021}. 

\section{Mapping symmetries into observables}
\label{sec:mapping}

The observables employed in this work are photoelectron momentum distributions (PMDs). Here, we explore how the symmetries of the field are mapped onto PMDs. We also analyse the factors that shape this mapping, including the saddle-point equations, the biases introduced by the prefactors and, where applicable, the residual potential, and discuss how each of these influences the resulting distributions. This discussion provides a basis for predicting the shapes of electron momentum distributions, identifying the ionisation events whose contributions are dominant, and compensating for artefacts that may otherwise break symmetries that are physically present. It also discusses how symmetries can be broken and how this is reflected in the shapes of the PMDs. 

\subsection{Photoelectron momentum distributions and momentum components}
For ATI, the PMDs are plotted as functions of the electron components $(p_{\parallel},\mathbf{p}_{\perp})$ parallel and perpendicular to driving field polarisation, or, if the field is orthogonally polarised, parallel and perpendicular to the major polarisation axis. 
Explicitly, this gives the observable
\begin{equation}
\mathcal{P}(p_{f\parallel},\mathbf{p}_{f\perp})=|M_{\mu}(p_{\parallel},\mathbf{p}_{\perp})|^2,
\label{eq:distr1electron}
\end{equation}
where the subscript can be $\mu=\mathrm{dir}$, $\mu=\mathrm{resc}$ or $\mu=\mathrm{CQSFA}$. In the CQSFA plots, we employ $\mathbf{p}_f$ instead of $\mathbf{p}$ to emphasise that the electron momentum changes during the continuum propagation. For the outcome of the SFA, this distinction is not necessary due to the field-dressed momentum being conserved.

For RESI, the quantity of interest is the two-electron momentum distribution 
\begin{align}
	\mathcal{P}(p_{1\parallel},p_{2\parallel})= \int\int d^2 p_{1\perp}d^2 p_{2\perp}\mathcal{P}(\mathbf{p}_{1},\mathbf{p}_{2}), \label{Eq:Channels}
\end{align}
as a function of the electron momentum components $p_{n\parallel}$, $(n=1,2)$ parallel to the driving-field polarisation, 
where 
$\mathcal{P}(\mathbf{p}_{1},\mathbf{p}_{2})$ 
is the fully resolved two-electron momentum probability density, and the transverse momentum components have been integrated over. Here, we do not consider RESI in orthogonally polarised fields, but the above-stated expressions are also applicable in this case, considering that the subscript $\parallel$ and $\perp$ refer to the directions parallel and perpendicular to the major polarisation axis, respectively. The fully resolved probability density $\mathcal{P}(\mathbf{p}_{1},\mathbf{p}_{2})$  gives a six-dimensional  function, which may be difficult to visualise. However,  kinematically complete NSDI measurements do exist \cite{Weckenbrock2004}. 

The evaluation of Eq.~\eqref{Eq:Channels} requires several points to be accounted for. First, one must include electron exchange: since the two electrons are indistinguishable, the transition amplitude in Eq.~\eqref{eq:MRESI} has to be symmetrised under \(\mathbf{p}_1 \leftrightarrow \mathbf{p}_2\). Second, a laser pulse may contain several ionisation--excitation events, all of which contribute to the signal. Third, for a realistic target, different excitation channels may be available, each associated with its own transition amplitude. Therefore, within the saddle-point approximation, the total RESI amplitude involves coherent sums over: (i) exchange-related contributions arising from electron indistinguishability for each pair of excitation and ionisation times; (ii) the different events occurring within the pulse; and (iii) the transition amplitudes associated with distinct excitation channels.

The fully coherent sum over events, channels and symmetrisation reads
\begin{equation}
\mathcal{P}_{(\mathrm{ccc})}(\mathbf{p}_{1},\mathbf{p}_{2})=\left|\sum_{\varepsilon}\sum_{\mathcal{C}}M_{\varepsilon}^{(\mathcal{C})}(\mathbf{p}_1,\mathbf{p}_2)+M_{\varepsilon}^{(\mathcal{C})}(\mathbf{p}_2,\mathbf{p}_1)\right|^2. 
\label{eq:fullcoherent}
\end{equation}
In Eq.~\eqref{eq:fullcoherent}, subscripts \(\varepsilon\) and \(\mathcal{C}\) label the pulse event and excitation channel, respectively, while \(\mathcal{S}\) denotes the symmetrisation associated with electron exchange.  This symmetrization is implicitly incorporated in the dual sum of the right-hand side of Eq.~\eqref{eq:fullcoherent}, in which the momenta of each electron have been swapped. The composite subscript \((\mathcal{S}\mathcal{E}\mathcal{C})=(ccc)\), standing for coherent--coherent--coherent, indicates that full coherence is retained in all three contributions.~Throughout, whenever we need to specify which channels are summed over, we use the notation $\mathcal{P}^{(\mathcal{C}_1,\mathcal{C}_2,\ldots,\mathcal{C}_n)} _{(s\varepsilon\mathcal{C})}$,  where the three entries in the lower index specify, in order, how the sums over symmetrisation, pulse events, and excitation channels are performed.~Thus, $\mathcal{P}^{(\mathcal{C}_1,\mathcal{C}_2,\ldots,\mathcal{C}_n)}_{(ccc)} $ denotes a probability density in which the contributions from exchange symmetrisation, different pulse events, and the channels \(\mathcal{C}_1,\mathcal{C}_2,\ldots,\mathcal{C}_n\) are all summed coherently. Omitting the superscripts means that either all excitation channels are included, or, depending on the context, only a single channel is taken into consideration.

However, depending on the objective, it may be convenient to construct the two-electron probability density in several ways.
A summary of the possible different sums for the events, channel and symmetrisation is provided in Table \ref{tab:possiblesums}. For simplicity,  we focus on two-channel interference, although the sums may be generalised to an arbitrary number of excitation channels. 

\begin{table}[ht]
\centering
\begin{tabular}{cc}
\hline
\textbf{Process} $\mathcal{SEC}$ &  \textbf{$\mathcal{P_{SEC}}(p_{1}, p_{2})$} \\
\hline \hline
(a) ccc & $|\sum_{\varepsilon}\sum_{\mathcal{C}}M_{\varepsilon}^{(\mathcal{C})}(\mathbf{p}_1,\mathbf{p}_2)+M_{\varepsilon}^{(\mathcal{C})}(\mathbf{p}_2,\mathbf{p}_1)|^2$ \\
(b) cic & $\sum_{\varepsilon}\left|\sum_{\mathcal{C}}M_{\varepsilon}^{(\mathcal{C})}(\mathbf{p}_1,\mathbf{p}_2)+M_{\varepsilon}^{(\mathcal{C})}(\mathbf{p}_2,\mathbf{p}_1)\right|^2$ \\
(c) icc & $\left|\sum_{\varepsilon}\sum_{\mathcal{C}}M_{\varepsilon}^{(\mathcal{C})}(\mathbf{p}_1,\mathbf{p}_2)\right|^2+\left|\sum_{\varepsilon}\sum_{\mathcal{C}}M_{\varepsilon}^{(\mathcal{C})}(\mathbf{p}_2,\mathbf{p}_1)\right|^2$\\
(d) iic & $\sum_{\mathcal{\varepsilon}}|\sum_{\mathcal{C}}M_{\varepsilon}^{(\mathcal{C})}(\mathbf{p}_1,\mathbf{p}_2)|^2+|\sum_{\mathcal{C}}M_{\varepsilon}^{(\mathcal{C})}(\mathbf{p}_2,\mathbf{p}_1)|^2$\\
(e) cci & $\sum_{\mathcal{C}}\left|\sum_{\varepsilon}M_{\varepsilon}^{(\mathcal{C})}(\mathbf{p}_1,\mathbf{p}_2)+M_{\varepsilon}^{(\mathcal{C})}(\mathbf{p}_2,\mathbf{p}_1)\right|^2$ \\
(f) cii & $\sum_{\varepsilon}\sum_{\mathcal{C}}|M_{\varepsilon}^{(\mathcal{C})}(\mathbf{p}_1,\mathbf{p}_2)+M_{\varepsilon}^{(\mathcal{C})}(\mathbf{p}_2,\mathbf{p}_1)|^2$\\
(g) ici & $\sum_{\mathcal{C}}|\sum_{\varepsilon}M_{\varepsilon}^{(\mathcal{C})}(\mathbf{p}_1,\mathbf{p}_2)|^2+|\sum_{\varepsilon}M_{\varepsilon}^{(\mathcal{C})}(\mathbf{p}_2,\mathbf{p}_1)|^2$\\
(h) iii & $\sum_{\varepsilon}\sum_{\mathcal{C}}\left|M_{\varepsilon}^{(\mathcal{C})}(\mathbf{p}_1,\mathbf{p}_2)\right|^2+\left|M_{\varepsilon}^{(\mathcal{C})}(\mathbf{p}_2,\mathbf{p}_1)\right|^2$ \vspace*{0.1cm} \\ \hline \hline
\end{tabular}
\caption{ Possible ways of constructing probability densities for a pulse, in terms of events, channels and symmetrisation. The letters c and i indicate coherent and incoherent sums, respectively regarding symmetrisation (denoted by $\mathcal{S}$), event ($\mathcal{E}$) and channel ($\mathcal{C}$). \label{tab:possiblesums} }
\end{table}
\subsection{Saddle-point equations and kinematic constraints}

If written as functions of the electron momentum components in which we are interested, the saddle-point equations can be employed to determine kinematic constraints that shed light on the shapes of the PMDs. Although saddle-point solutions are not observables, within the present framework they help establish a link between classical trajectories and quantum pathways. Furthermore, they illustrate when symmetries are kept or broken, and when specific events cease to be relevant. For Coulomb-distorted theories such as the CQSFA, the constraints cease to be purely kinematic. Therefore, in this section, we will focus on the information one can extract from the SFA.

 \subsubsection{Above-threshold ionisation}
 For direct ATI, the ionisation equation \eqref{Eq:SPE} determines the system's dynamics. Below, this equation is re-written as 
 \begin{equation}
    \left[p_{\parallel}+A_z(t')\right]^2 +  \left[\mathbf{p}_{\perp}+\mathbf{A}_{\perp}(t')\right]^2 = -2I_p,\label{eq:directmapping}
 \end{equation}
 where we considered the major field polarisation component along $\hat{\mathbf{e}}_z$ following the convention of Sec.~\ref{sec:fieldshapes}. For generality, we assume that this field may have other components spanning the whole plane perpendicular to the major axis $\hat{\mathbf{e}}_z$, and thus be defined in all three dimensions.   This is the equation of a sphere of complex radius centred at $(p_{\perp_x},p_{\perp_y},p_{\parallel})=(-A_x(t'),-A_y(t'),-A_z(t'))$, with $\mathbf{p}_{\perp}=p_{\perp_x}\hat{\mathbf{e}}_x+p_{\perp_y}\hat{\mathbf{e}}_{y}$. In practice, however, we will consider a polarisation plane, so that we take $A_x(t')=0$. The remaining vector-potential components will serve to determine the centres and the symmetry of the PMDs.  Although there is not a rigid kinematic boundary, as the condition expressed by Eq.~\eqref{eq:directmapping} is classically forbidden throughout. 
 
 Inspection of Eq.~\eqref{eq:directmapping} provides information about what to expect, depending on the field geometry.  The equation is quadratic with regard to the field and momenta. Thus, for a half-cycle symmetric field  the PMD will be reflection-symmetric upon $p_{\parallel} \rightarrow -p_{\parallel} $ and $\mathbf{p}_{\perp} \rightarrow -\mathbf{p}_{\perp}$. This can be verified, for instance, for linearly and elliptically polarised fields. For circularly polarised fields, one can show that the PMD will be annular and rotation-invariant with regard to the origin \cite{Kim2022}. 
 
 If the field is linearly polarised, $\mathbf{A}_{\perp}(t')=\mathbf{0}$ so that Eq.~\eqref{eq:directmapping} reads
 \begin{equation}
    \left[p_{\parallel}+A_z(t')\right]^2   = -2I_p-\left[\mathbf{p}_{\perp}\right]^2.
 \end{equation}
In this case, the perpendicular momentum component effectively shifts the potential barrier through which the electron must tunnel and will locate the PMD around the $|\mathbf{p}_{\perp}|=0$ axis. The centre of the distribution will be such that $E_z(t')=-dA_z(t')/dt'$ is near its maximum. For a monochromatic field and few-cycle pulses, this implies that $A_z(t')\approx 0$, but for two-colour fields this will depend on the field-strength ratio. The half-cycle symmetry will be broken for few-cycle pulses and some two-colour fields, which implies that they will not be reflection-symmetric about the $p_{\parallel}=0$ axis. An example of this reflection symmetry is provided in the lower panels of Fig.~\ref{fig:sfa_inter_intra}, for linearly polarized monochromatic fields.    

Orthogonally polarised fields will move the PMDs away from the $\mathbf{p}_{\perp}=\mathbf{0}$ axis. For example, within the SFA, elliptically polarised fields will lead to two maxima at non-vanishing perpendicular momenta, that are reflection-symmetric about the $p_{\perp}=0$ axis. For fields with rotation symmetries, such as circularly polarised (continuous) or bicircular (discrete) fields, the PMD symmetry will reflect that dictated by the vector potential \cite{Habibovic2021,Chen2026}. The classical direct ATI cutoff is dictated from the maximal energy transferred to the electron by the field. For example, for a linearly polarised monochromatic field, the maximum of Eq.~\eqref{eq:Amonochromatic} can be used to obtain a cutoff energy of $2U_p$~\cite{Reiss1980,Paulus1994,Becker2018}.  

\begin{figure}
    \centering
    \includegraphics[width=0.8\linewidth]{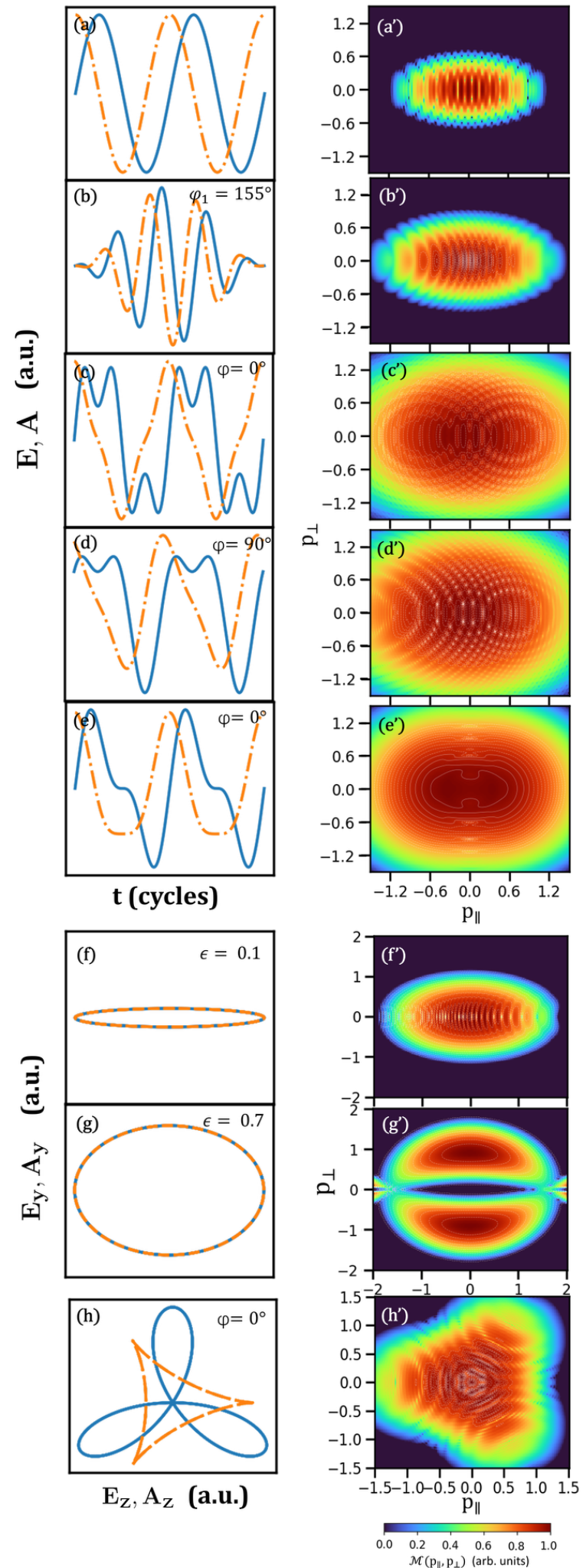}
   \caption{Examples of tailored driving fields and the corresponding direct above-threshold ionisation (ATI) photoelectron momentum distributions (PMDs). The blue solid curves show the electric field and the orange dash-dotted curves the corresponding vector potential. For the linearly polarised fields, these are displayed as functions of time, whereas the polarisation-plane trajectories are shown for the elliptically polarised and bicircular fields. From top to bottom, the rows correspond to: (a$'$) a monochromatic field; (b$'$) a few-cycle pulse with $N=4.3$ and carrier-envelope phase $\phi_1=155^\circ$; (c$'$) an $(\omega,3\omega)$ linearly polarised bichromatic field with $\phi=0$; (d$'$) an $(\omega,2\omega)$ field with $\phi=0$; (e$'$) an $(\omega,2\omega)$ field with $\phi=90^\circ$; (f$'$) an elliptically polarised field with $\varepsilon=0.1$; (g$'$) an elliptically polarised field with $\varepsilon=0.7$; and (h$'$) a counter-rotating $(\omega,2\omega)$ bicircular field with $\phi=0$, displaying threefold rotational symmetry. The bichromatic linearly polarised fields have a field-amplitude ratio $E_s/E_r=0.8$. The fundamental wavelength is $\lambda=800 \hspace*{0.2cm}\mathrm{nm}$, corresponding to $\omega=0.057$ a.u., and the PMDs are evaluated at $I=2\times10^{14}\mathrm{W/cm^2}$. This figure has been adapted from \cite{Chen2026}. 
   In contrast to the multi-cycle PMDs shown in the corresponding original figures, inter-cycle interference is omitted here. Consequently, the ATI-ring structure associated with interference between different optical cycles is absent, while the overall momentum-space envelopes, dynamical symmetries, and interference between sub-cycle ionisation events are retained.}
    \label{fig:DATI_Fields}
\end{figure}

This is shown in Fig.~\ref{fig:DATI_Fields}, where the field profiles are plotted together with the direct ATI photoelectron momentum distributions (left and right column, respectively). For clarity, we have omitted the ATI rings and plotted solely intra-cycle interference patterns. 
Fig.~\ref{fig:DATI_Fields}(a) shows a linearly polarised monochromatic field. The corresponding PMD, in Fig.~\ref{fig:DATI_Fields}(a'), exhibits the expected reflection symmetry $p_{\parallel} \rightarrow -p_{\parallel}$, as discussed above. The same  holds for the $(\omega, 3\omega)$ field with $\phi=0$  [see Figs.~\ref{fig:DATI_Fields}(c) and (c') for the field and the PMD, respectively], which is also predicted as this field exhibits the same symmetries as a monochromatic field. For the few cycle pulse in Fig.~\ref{fig:DATI_Fields}(b), the PMD is approximately reflection symmetric [see Fig.~\ref{fig:DATI_Fields}(b')]. However, this near symmetry is accidental, as the given few-cycle pulse does not satisfy $\mathcal{S}_1-\mathcal{S}_3$. For $(\omega, 2\omega)$ fields [Figs.~\ref{fig:DATI_Fields}(d) and (e)], the PMDs are asymmetric. However, the asymmetry is much more pronounced for $\phi=\pi/2$, with the PMD extending towards negative larger values of $p_{\parallel}$ [see Fig.~\ref{fig:DATI_Fields}(d')]. This is due to the fields associated with each half cycle being unequally strong, although the reflection symmetry about the maxima holds. In contrast, the PMD in Fig.~\ref{fig:DATI_Fields}(e'),  for $\phi=0$, is almost reflection symmetric about $p_{\parallel} \rightarrow -p_{\parallel}$, although a closer inspection shows that it is slightly skewed. This is due to the unequal gradients around the field extrema [see Fig.~\ref{fig:DATI_Fields}(e)].

If the field is taken to be ortogonally polarised [Figs.~\ref{fig:DATI_Fields}(f), (g) and (h)], the centres of the PMDs move away from the $p_{\perp}=0$ axis. For instance, for elliptically polarised fields the PMDs split, but are reflection-symmetric about the $p_{\perp}=0$ axis, as shown in Figs.~\ref{fig:DATI_Fields}(f'), (g'). For a bicircular field, the PMD essentially follows the shape and discrete rotational symmetry of the vector potential, as seen in Figs.~\ref{fig:DATI_Fields}(h'). The overall message is that the field symmetry is imprinted in direct ATI PMDs.

For rescattered ATI, symmetry information can be obtained from the return condition and the energy conservation upon rescattering. The return condition \eqref{eq:rescSP3} shows that the intermediate momentum is predominantly along the direction of the field, while Eq.~\eqref{eq:rescSP2} defines the shape of the PMD. If rewritten in terms of the momentum components, it reads as 
\begin{equation}
\left[p_{\parallel}+A_{z}(t)\right]^2+ \left[\mathbf{p}_{\perp}+\mathbf{A}_{\perp}(t)\right]^2
=
\left[\mathbf{k}+\mathbf{A}(t)\right]^2,    \label{eq:rescATImapping}
\end{equation}
which, once more, is the equation of a sphere in momentum space. However, there are two key differences. First, the right-hand side of Eq.~\eqref{eq:rescATImapping} is positive, which means that, in principle, this equation has a classical counterpart. However, because the intermediate momentum $\mathbf{k}$ is also a function of the ionisation time $t'$ given by Eq.~\eqref{eq:rescSP1}, the solutions of Eq.~\eqref{eq:rescATImapping} are complex.  Still, in the region for which rescattering can be described classically, the imaginary parts of the rescattering time $t$ is vanishingly small. Second, the center of the sphere is defined by the vector potential at the \textit{rescattering time}, $\mathbf{A}(t)$. Because the electron is most likely to return at a field zero crossing, the center of the sphere will be non-vanishing. The orbits for the returning electron obtained from Eq.~\eqref{eq:rescATImapping} occur in pairs which coalesce at the ATI cutoff: the long orbit, for which the electron left slightly earlier and returned after the field zero crossing, and the short orbit, for which the electron was freed slightly later and returned after the zero field crossing. The cutoff corresponds to the maximal classical kinetic energy the electron can obtain, upon rescattering. 
\begin{figure}
    \centering
    \includegraphics[width=0.99\linewidth]{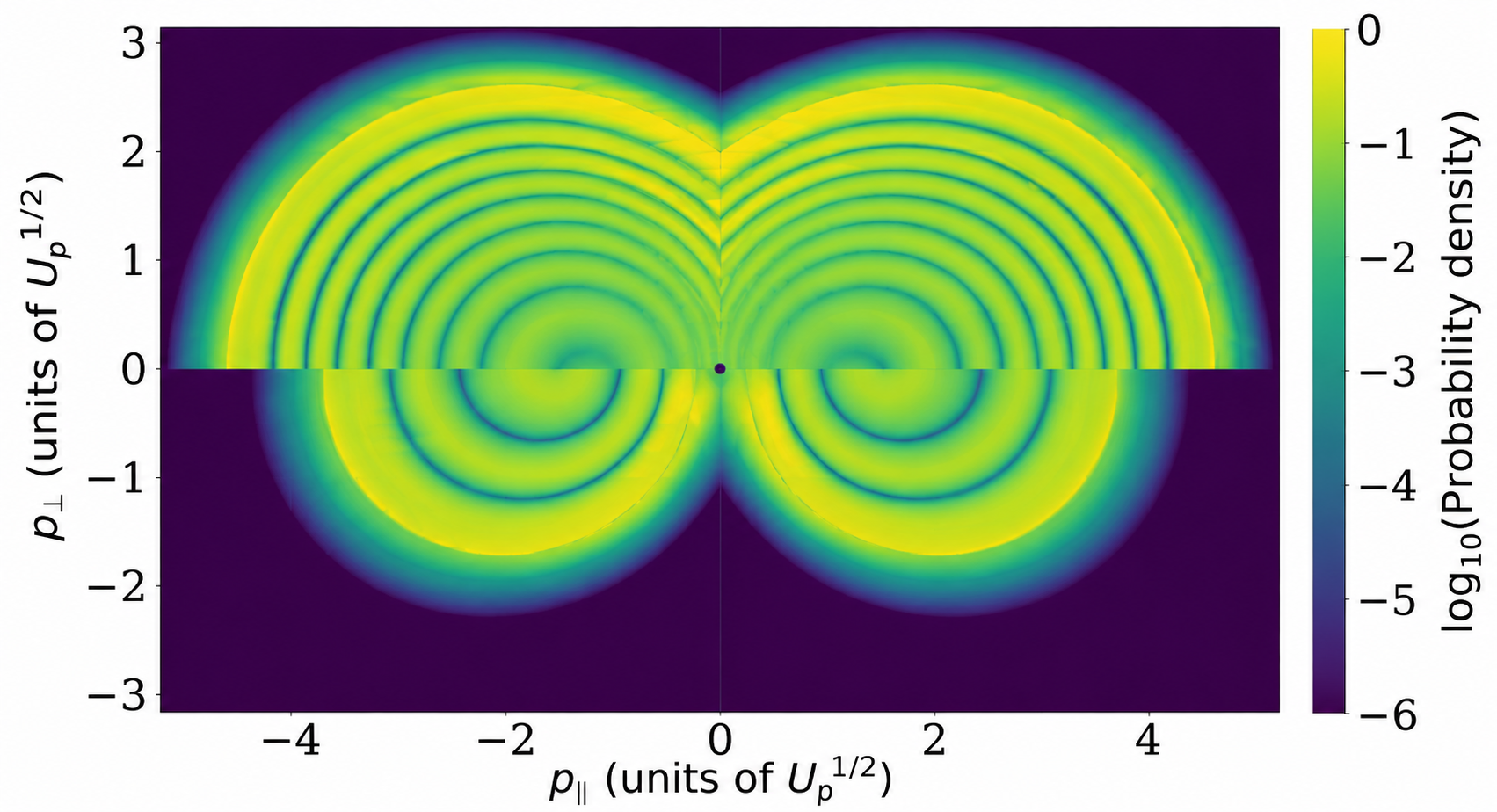}
    \caption{Adapted from Ref.~\cite{Rook2024a} to show the contribution of backward scattering trajectories in the SFA with rescattering. Here, we show the PMD due to the pair of trajectories which rescatter approximately (upper half plane) one cycle after ionisation and (lower half plane) one-and-a-half cycles after ionisation. The target is hydrogen with ionisation potential $I_p=0.5$~a.u., driven by a linearly polarised monochromatic field $E(t)=E_0\sin(\omega t)$ of intensity $I=2\times10^{14} \mathrm{W/cm^2}$ and wavelength $\lambda=800$~nm ($\omega\simeq0.057$~a.u.). The parallel and perpendicular photoelectron momenta are given in units of $U_p^{1/2}$, and the probability density is displayed on a logarithmic scale.}
    \label{fig:ridge}
\end{figure}
For each pair, the maximal kinetic energy will define a rescattering ridge, whose radius is given by the RHS of Eq.~\eqref{eq:rescATImapping}.  Similarly to what happens to direct ATI, half-cycle symmetry will imply symmetry reflection upon $p_{\parallel} \rightarrow -p_{\parallel} $ and $\mathbf{p}_{\perp} \rightarrow -\mathbf{p}_{\perp}$. 

If the field is linearly polarised, Eq.~\eqref{eq:rescATImapping} reduces to 
\begin{equation}
\left[p_{\parallel}+A_{z}(t)\right]^2+ \left[\mathbf{p}_{\perp}\right]^2
=
\left[k+A_z(t)\right]^2,    \label{eq:rescATImapping_rw}
\end{equation}
which means that the rescattering ridges will be centered at $ (p_{\parallel},\mathbf{p}_{\perp})=(-A(t),\mathbf{0})$. For orthogonally polarised fields, the centre of the rescattering ridge  will move away from the major polarisation axis. For a detailed discussion for bicircular fields see, e.g., \cite{Milos2016}.

For linearly polarised monochromatic light, it has been shown that the rescattering ridge has an energy of approximately $10U_p$ for the shortest orbit pair~\cite{Becker2018}. In general, the rescattered ATI cutoff energy is far above that obtained for direct ATI, and longer orbit pairs will coalesce at other maxima of the electron's kinetic energy at $t$. Beyond the cutoff, the ATI signal will be exponentially decaying. Although, in the SFA framework, this constraint is purely kinematic, adding a residual binding potential in the continuum propagation will blur this distinction (see \cite{Rook2024} for details). 

Fig.~\ref{fig:ridge} shows the rescattering ridge computed for the SFA in a monochromatic field. The upper (lower) panel shows the structure obtained for the shortest (second shortest) pair of orbits. The concentric interference fringes are associated with the long and the short orbits in each pair.   

\subsubsection{Non-sequential double ionisation  (RESI)}
\label{sec:NSDIconstraints}

 Similarly to what has been done in the previous section, one may employ the saddle-point equations to determine momentum constraints for the PMDs for laser-induced nonsequential double ionisation. Here, we will focus on the RESI mechanism, the saddle-point equations in Sec.~\ref{sec:saddlepoints} and linearly polarised fields, although the present line of argument can also be used for electron-impact ionisation and/or orthogonally polarised fields. We will partly rely on the discussion of the previous section, as RESI can be interpreted as two time-ordered ATI processes.  The constraints outlined below have been derived in previous publications \cite{Shaaran2010}, but they are illustrative of how the field symmetries map into the two-electron PMDs $\mathcal{P}(p_{1\parallel},p_{2\parallel})$. One should note that the times employed here have a slightly different notation, namely $t''$ and $t'$ are the ionisation and rescattering time of the first electron, and $t$ is the ionisation time for the second electron. 
 
 The first electron undergoes a rescattered ATI-like process, with the difference that, upon return, it must give some kinetic energy to the core. Rewriting Eq.~\eqref{eq:RESIsp3} in terms of the momentum components gives
 \begin{equation}\label{eq:RESIsp3_rw}
[p_{1\parallel}+A_z(t')]^2 +\mathbf{p}^2_{1\perp}= [k+A(t')]^2 - 2(E^{(\mathcal{C})}_{2g}-E^{(\mathcal{C})}_{2e}),
\end{equation}
which shows again the equation of a sphere centered at $(p_{1\parallel},\mathbf{p}_{1\perp})=(-A_z(t'),\mathbf{0})$. However, the kinetic energy of the returning electron must be larger than the energy gap $\Delta E^{(\mathcal{C})} =(E^{(\mathcal{C})}_{2g}-E^{(\mathcal{C})}_{2e})$ between the ground and excited state of the singly ionised target in order for the right-hand side to be real. Shifting $\mathbf{p}^2_{1\perp}$ to the right-hand side shows that this term effectively shifts $\Delta E^{(\mathcal{C})}$, so that setting $\mathbf{p}^2_{1\perp}=0$ gives an upper bound for the range of $p_{1\parallel}$ that should be occupied by the PMD. Thus, $-A_z(t')-\sqrt{[k+A(t')]^2-2\Delta E^{(\mathcal{C})}}\leq p_{1\parallel}\leq -A_z(t')+\sqrt{[k+A(t')]^2-2\Delta E^{(\mathcal{C})}} $. A similar reasoning may be employed for the second electron, using the saddle-point equation \eqref{eq:RESIsp4}. In this case, $-2\sqrt{U_p}\leq p_{2\parallel}\leq 2\sqrt{U_p}$, where we used the direct ATI cutoff $2U_p$ for an approximate estimate. These two conditions give approximate bounds in the $p_{1\parallel}p_{2\parallel}$ plane. Considering that one may not distinguish between the first and the second electron means that both conditions must be transposed. 

\begin{figure}
    \centering
    \includegraphics[width=1.1\linewidth]{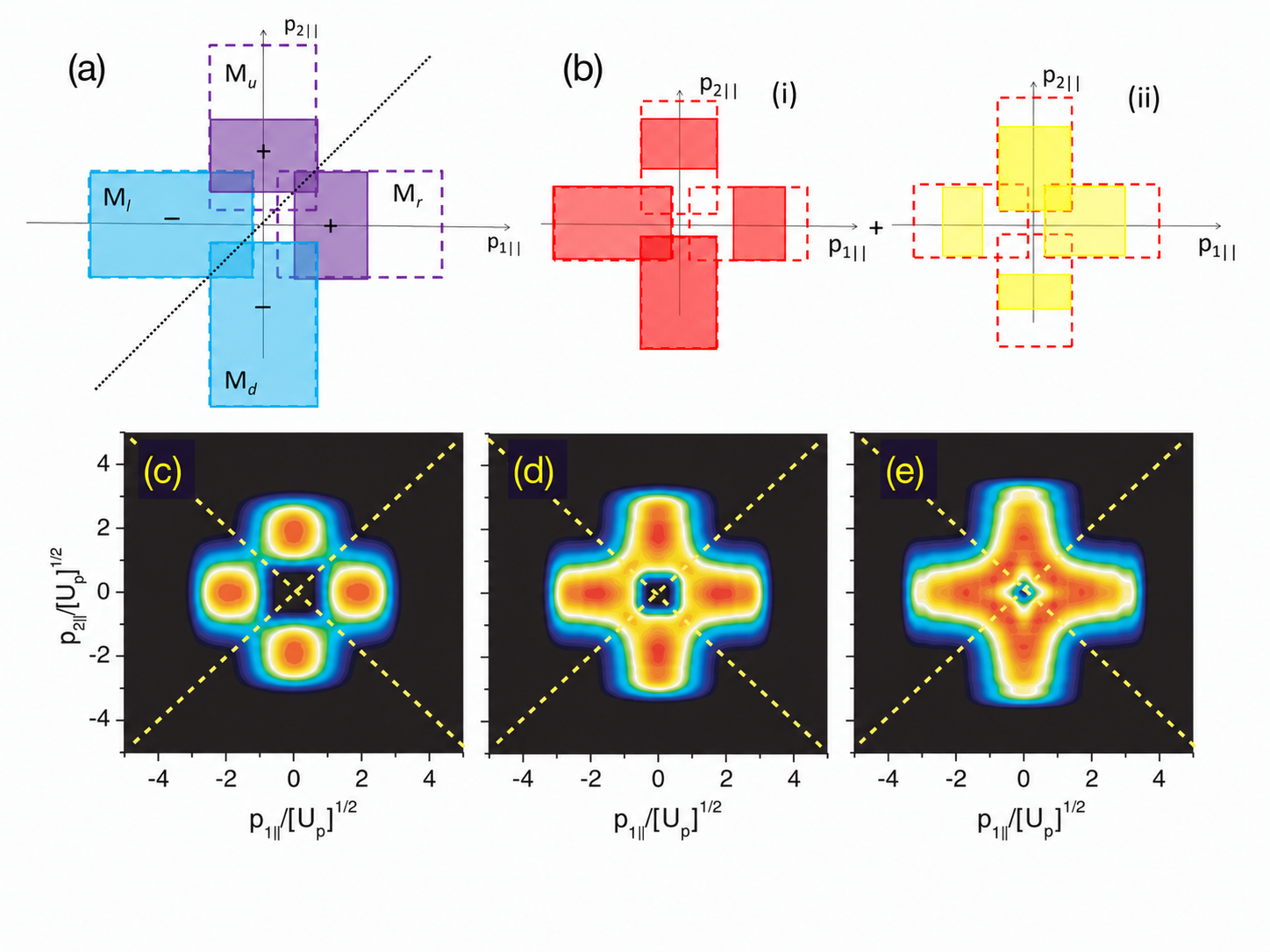}
    \caption{Momentum-space constraints and correlated two-electron distributions in recollision-excitation with subsequent ionisation (RESI). Panel (a) shows the schematic momentum-space regions associated with the RESI transition amplitudes $M_l$, $M_u$, $M_r$, and $M_d$, where the subscript is associated with the direction they occupy in the $p_{1\parallel}p_{2\parallel}$ plane, i.e., $l=$ left, $u=$ up, $r=$ right and $d=$ down. Dashed rectangles show the momentum constraints for a driving field with half cycle symmetry and $A(t)=0$, while the shaded regions indicate the corresponding ranges for fields in which this symmetry is broken. Their overlap marks momentum regions in which interference between different contributions may become significant. The amplitudes $M_l$ and $M_d$ are shown in blue, and $M_u$ and $M_r$ in purple, reflecting the exchange relations $M_l(\mathbf{p}_1,\mathbf{p}_2)=M_d(\mathbf{p}_2,\mathbf{p}_1)$ and $M_r(\mathbf{p}_1,\mathbf{p}_2)=M_u(\mathbf{p}_2,\mathbf{p}_1)$. These relations lead to reflection symmetry about the diagonal $p_{1\parallel}=p_{2\parallel}$, indicated by the black dotted line. The minus and plus signs identify the most probable momentum regions for the blue and purple contributions, respectively. 
  Panel (b) gives the schematic momentum-space regions associated with different ionisation–recollision events within the pulse. Each subpanel represents contributions from events occurring within a single optical cycle, with pairs of events separated by half a cycle and subsequently combined coherently. As in panel (a), dashed rectangles denote the monochromatic-field momentum constraints, while the shaded regions show their few-cycle counterparts. Overlap between these regions indicates where interference between different events is expected to be important. Panels (c)–(e) display the correlated electron momentum distributions for helium with $E_{1g}=0.97$ a.u., $E_{2g}=2$ a.u., and $E_{2e}=0.5$ a.u., calculated for a linearly polarised monochromatic field of frequency $\omega=0.057$ a.u. and constant prefactors. The panels correspond to driving-field intensities $I=2.16\times10^{14}\mathrm{W/cm^2}$, $I=2.5\times10^{14}\mathrm{W/cm^2}$, and $I=3.0\times10^{14}\mathrm{W/cm^2}$, respectively. Each distribution is normalised to its maximum value, and the probabilities associated with each transition amplitude have been summed incoherently. The dashed diagonals indicate the reflection-symmetry axes of the correlated momentum distributions. Figures adpated from \cite{Shaaran2010} and \cite{Hashim2024}.
}
    \label{fig:NSDISchematic}
\end{figure}

These estimates are shown in Fig.~\ref{fig:NSDISchematic}. The reflection symmetry about the diagonal $p_{1\parallel}=p_{2\parallel}$ is always present, while the symmetry about the anti-diagonal $p_{1\parallel}=-p_{2\parallel}$ only holds for fields that are half-cycle symmetric [Fig.~\ref{fig:NSDISchematic}(a)].  The regions occupied will be different for different channels because of different energy gaps $\Delta E^{(\mathcal{C})}$. In Fig.~\ref{fig:NSDISchematic}(b), we also provide schematic representations of regions occupied for different events within a  pulse: depending on the pulse shape and/or channel, two generic events (i) and (ii) will occupy different regions in the $p_{1\parallel}p_{2\parallel}$ plane, and both need to be considered in the overall two-electron PMD. Finally, Figs.~\ref{fig:NSDISchematic}(c) to (e) display the RESI PMDs calculated for a linearly polarised monochromatic field of increasing intensity and summed incoherently according to Table \ref{tab:possiblesums}. The figure shows that the width of the distributions, obtained from the direct ATI cutoff, remains the same, while its length increases. This is expected, as the latter constraint will depend on the difference between the kinetic energy of the returning first electron and the energy gap $\Delta E^{(\mathcal{C})}$. The RESI distributions are also symmetric with regard to $p_{1\parallel}=\pm p_{2\parallel}$.

The constraints plotted in Fig.~\ref{fig:NSDISchematic} give the impression that the half-cycle symmetry $\mathcal{S}_1$, together with the electron indistinguishability, is sufficient for the RESI distributions to be fourfold symmetric. However, fourfold symmetry only holds if, in addition, the reflection symmetries $\mathcal{S}_2$ and $\mathcal{S}_3$ hold. This is because these symmetries are necessary in order for the momentum distributions not to be skewed about their peak values, as they guarantee symmetric gradients about these points.

\begin{figure*}
    \centering
    \includegraphics[width=\linewidth]{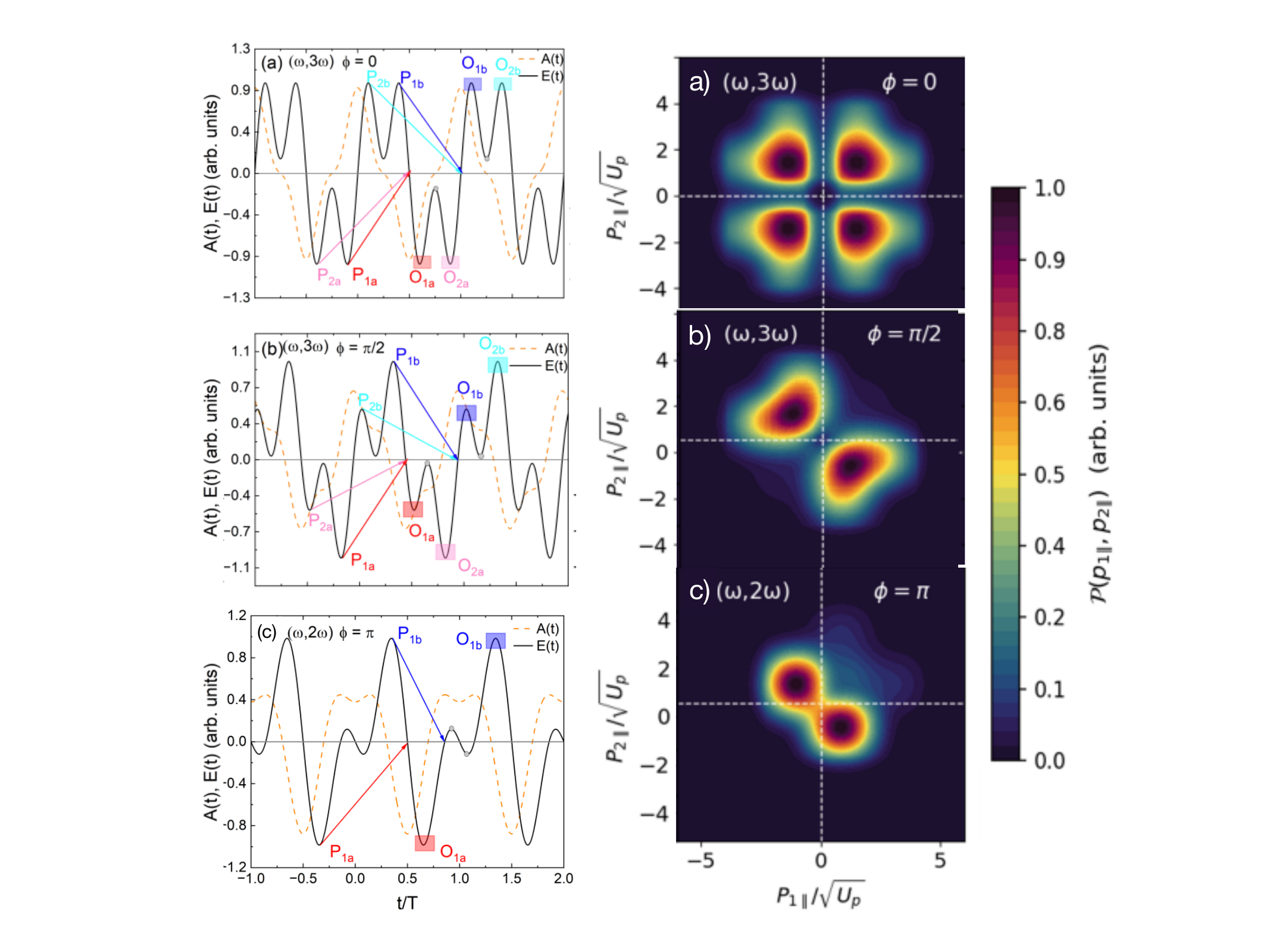}
    \caption{Dominant RESI events and the corresponding correlated electron momentum distributions for bichromatic linearly polarised fields. Panels (a) and (b) correspond to $(\omega,3\omega)$ fields with relative phases $\phi=0$ and $\phi=\pi/2$, respectively, while panel (c) shows an $(\omega,2\omega)$ field with $\phi=\pi$. In the left-hand subpanels, the electric field $E(t)$ and vector potential $A(t)$ are shown by the black solid and orange dashed curves, respectively. The arrows indicate approximate real ionisation and rescattering times of the first electron for the saddle-point pairs giving the dominant contributions, while the shaded rectangles mark the corresponding ionisation times of the second electron. The pairs $P_{n\mu}$ are ordered by increasing continuum excursion time, with $n=1,2$, while $\mu=a,b$ distinguishes events originating in the two consecutive half-cycles considered. Red and pink arrows denote rescattering events leading predominantly to positive first-electron parallel momenta, whereas blue and cyan indicate negative momenta. The colors of the shaded regions associate the second-electron orbits $O_{j\mu}$ with the corresponding first-electron pairs and do not specify the momentum region populated by the second electron; $j=1,2$ distinguishes the ionisation events according to their temporal proximity to rescattering. Gray dots indicate negligible ionisation events. The electric fields are normalised to their maximum amplitude in each panel. The right-hand subpanels show the corresponding incoherent two-electron RESI momentum distributions $\mathcal{P}(p_{1\parallel},p_{2\parallel})$, calculated for argon ($E_{1g}=0.58$ a.u., $E_{2g}=1.016$ a.u. and $E_{2e}=0.52 $ a.u.) considering the $3s \rightarrow 3p$ excitation channel without prefactors. For all panels, the field-amplitude ratio is $E_s/E_r=0.8$, the intensity of the fundamental component is $E_\omega^2=6\times10^{13}\mathrm{W/cm^2}$, and the fundamental wavelength is $\lambda = 800\mathrm{nm}$.}
    \label{fig:NSDIdominant}
\end{figure*}

In Fig.~\ref{fig:NSDIdominant}, we provide an example, using bichromatic fields, that illustrates this issue, and also sheds light on the dominant orbits. The left and the right columns show the field and the resulting PMD, respectively. An important feature is that the instantaneous vector potential corresponding to the field extrema is no longer vanishing. This implies that, for the second electron, the centres of the distributions will be non-vanishing.  When mapped to the $p_{1\parallel}p_{2\parallel}$ plane, this translates into moving the axes of the distributions away from the $p_{n\parallel}$ axes. The approximate center can be read from considering the field extrema associated with orbits $O_n$ of the second electron, inspecting the vector potential (dashed line) and performing the mapping $p_{2\parallel}=-A(t)$.

The field in Fig.~\ref{fig:NSDIdominant}(a), namely $(\omega, 3\omega)$ with $\phi=0$, exhibits the symmetries $\mathcal{S}_1$-$\mathcal{S}_3$ also present for the monochromatic field. There are two rescattering pairs $P_{1\nu}$, $P_{2\nu}$, $\nu=a,b$, for the first electron per half cycle contributing to the PMD, indicated by two arrows converging to the same zero crossing. Subsequently, the excited second electron leaves at the times $O_{1\nu}$, $O_{2\nu}$, $\nu=a,b$, for which the most probable momentum will be non-vanishing and have opposite directions. This leads to the fourfold-symmetric distributions on the right panel, which is no longer cente
red at the axes but at $(p_{1\parallel}, p_{2\parallel})=(\mp A(t'),\mp A(t))$ and its transpose. 

Changing the relative phase to $\phi=\pi/2$ [Fig.~\ref{fig:NSDIdominant}(b)] breaks the symmetries $\mathcal{S}_2$ and $\mathcal{S}_3$ for the $(\omega, 3\omega)$ field, but keeps the half-cycle symmetry $\mathcal{S}_1$. As a direct consequence, the two field extrema within a half-cycle become unequal, with the ionisation events associated with $P_{1\nu}$, and $O_{2\nu}$, with $\nu=a,b$, becoming more prominent. This moves the RESI distributions to the second and fourth quadrants. One should note that the distribution is still reflection symmetric about the diagonal and the antidiagonal. However, it is no longer fourfold symmetric. 

Finally, considering an $(\omega, 2\omega)$ field [Fig.~\ref{fig:NSDIdominant}(c)] will break the reflection symmetry of the RESI distribution with regard to the anti-diagonal $p_{1\parallel}=-p_{2\parallel}$. However, the distributions will still be reflection symmetric about  $p_{1\parallel}=p_{2\parallel}$ due to electron indistinguishability. For the specific value of $\phi$ chosen, there is only one relevant event per half cycle, which will mainly populate the second and fourth quadrant of the $p_{1\parallel}p_{2\parallel}$ plane.

\subsection{Biases introduced by the target}
In addition to the momentum constraints discussed in the previous sections, the target will introduce additional biases, which will influence the single and two-electron PMDs. In the context of the SFA, this will happen via the prefactors, while for the CQSFA the potential will also influence the continuum propagation by accelerating the electron in the continuum, thus affecting its intermediate momentum and the phases that will be picked up.
The ionisation equations given by the saddle-point approximation also will lead to singularities in the prefactors $V_{\mathbf{p}0}$, $V_{\mathbf{k}0}$ and $V_{\mathbf{p}_2e}$ if the length-gauge is considered. This problem can be overcome by considering a Gaussian basis \cite{Augstein2010,Hashim2024}, or exponentialising the prefactor and incorporating it in the action \cite{Faria2005}. 
In the CQSFA, the bound-state singularity can be overcome by other means, as discussed in \cite{Maxwell2017}.

In the specific case of RESI, for few cycle pulses or monochromatic fields, the prefactor given by Eq.\eqref{eq:Vp2e} is particularly important, as it determines the shape of the PMD. Thus, care must be taken when it is calculated. The vector potential is approximately vanishing for the most probable tunneling times of the second electron. In those cases, one may use the velocity gauge and obtain similar results \cite{Shaaran2010}. Within the context of the SFA, the gauge transformation will cause a momentum translation that will cancel out the dressing in the prefactors, so that one must omit $\mathbf{A}(t)$ in Eqs. \eqref{eq:directprefactor}, \eqref{eq:atipref1}, \eqref{eq:Vp2e}. However, care must be taken in the case of bichromatic fields with comparable intensities, as $\mathbf{A}(t)$ can be quite large for the electrons' ionisation times. Typically, $s-$state prefactors will lead to cross shaped distributions, but may reduce the
$p$ state prefactors lead to nodes at the $p_{n\parallel}$ axes, and $d-$state prefactors will lead to maxima in the diagonals and axes for the RESI PMDs. However, in practice there may be variations, and extra biases induced by the excitation prefactor $V$ (see \cite{Hashim2024,Hashim2024b,Hashim2025} for details). 

\section{Photoelectron holography}
\label{sec:holography}

\begin{figure}
    \centering
    \includegraphics[width=0.99\linewidth]{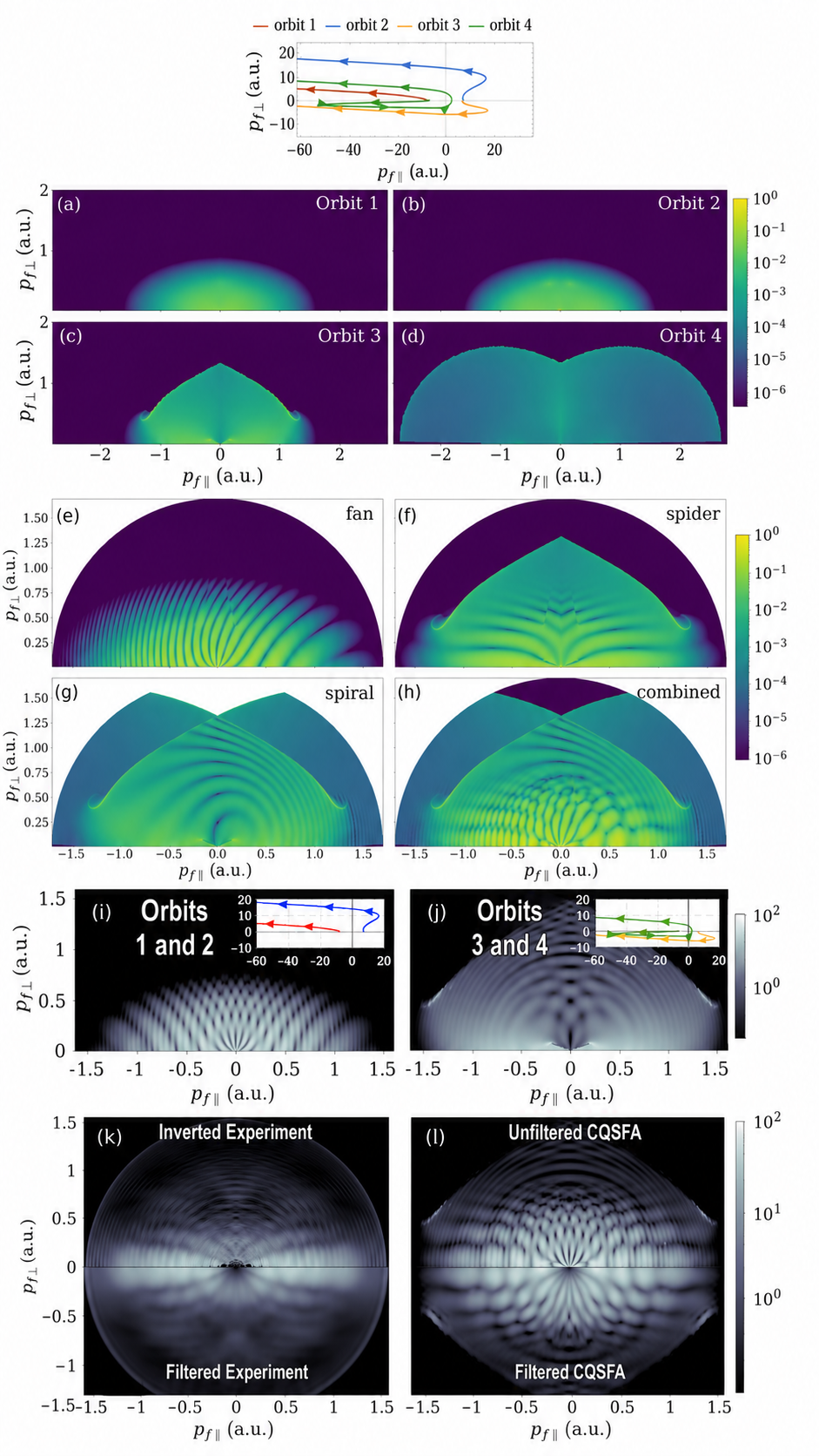}
    \caption{Adapted from (a)-(h) Ref.~\cite{Rook2024thesis} and (i)-(l) Ref.~\cite{Werby2021}. The trajectories labelled by orbits 1-4 are shown in the top panel. Panels (a)-(d) shows the single orbit distribution for each of these orbit types, computed using the CQSFA for hydrogen monochromatic field of frequency $\omega = 0.057$ a.u and intensity 1.5 $\times 10^{14}\mathrm{W/cm}^2$ . (e) Fan pattern due to combination of orbits 1 and 2. (f) Spider pattern due to combination of orbits 2 and 3. (g) Spiral pattern due to combination of orbits 3 and 4. (h) Pattern due to all orbits combined, computed for the same target and field parameters as in panels (a)-(d). (i) and (j) Unit cell averaged fan and spiral recovers the expected symmetry of the PMD, computed using the CQSFA for argon field of wavelength $\lambda = 800$ nm and intensity 2 $\times 10^{14}\mathrm{W/cm}^2$. (k) and (l) Comparison of CQSFA with experiment, for the same parameters as in (i) and (j).  }
    \label{fig:PMDCQSFA1}
\end{figure}

\subsection{Key structures and orbits}

Here, we will focus on the CQSFA, discussing the main structures observed and linking them to the interference between the types of orbits. For simplicity, we will consider a linearly polarised monochromatic field, which is a good approximation for long enough pulses. There are basically four types of orbits in the CQSFA, whose classification is more fluid than ``direct'' or ``rescattered'' due to the presence of the Coulomb potential. Their schematic representation is shown in the top panel of Fig.~\ref{fig:PMDCQSFA1}. Orbit 1 leaves the tunnel exit, and is decelerated by the potential, behaving like the SFA short orbit in direct ATI. Orbits 2 and 3 are field-dressed hyperbolae for which the electron is freed half a cycle earlier or later, on the opposite side of the detector. Both types of orbit are deflected by the joint action of the field and the potential, eventually turning and reaching the detector. In the absence of the residual binding potential, orbits 2 and 3 are degenerate and become the SFA long orbit for direct ATI \cite{Lai2015}. These two orbits differ as follows: for orbit 2, the  final momentum component of the electron perpendicular to the field does not change sign, with regard to its value at the tunnel exit, while it does for orbit 3. Thus, orbit 2 can be loosely associated with the long SFA orbit for direct ATI, while orbit 3 is of hybrid character, but sometimes behaves like the SFA forward-scattered trajectories. Orbit 4 starts on the side of the detector, but it is driven around the core before it reaches the detector, thus behaving like a backscattered SFA trajectory. One should note, however, that there is no one-to-one correspondence between the saddle-point solutions obtained with these two theories, due to them being structurally different. Several studies have shown how far this analogy can be pushed  \cite{Lai2015,Maxwell2017a,Maxwell2018,Rook2024,Rook2024a}, with good agreement for orbit 4 and high photoelectron momenta, but varying agreement for the remaining orbits. 

The above classification was first introduced in \cite{Yan2010}, and is based on the product $\Pi_{\parallel}=z_0p_{f\parallel}$ of the tunnel exit and the final momentum component parallel to the driving-field polarisation, and the product $\Pi_{\perp}=p_{0\perp}p_{f\perp}$ of the initial and final momentum components perpendicular to the driving-field polarisation. A positive (negative) product indicates that the orbit has retained (changed) its direction. For orbits 1 and 4, $\Pi_{\parallel}>0$, while the signs of $\Pi_{\perp}$ differ, being positive for orbit 1 and negative for orbit 4. Both orbits 2 and 3 start on the ``wrong'' side, so that $\Pi_{\parallel}<0$, but along the continuum propagation their behavior differs: orbits 2 are only slightly deflected, so that  $\Pi_{\perp}>0$, while for orbits 3 $\Pi_{\perp}<0$, which indicates a stronger deflection by the core. One should note, however, that these criteria allow a broader range of orbits to be classified under the umbrella of a specific orbit. In fact, previous work by us employing these criteria and spatial filters to single out direct, deflected or recolliding orbits has shown that, for instance, hard-colliding, backscattered orbits may also satisfy the criteria for orbit 1 or 2, or  non-colliding/lightly deflected orbits may satisfy the criteria for orbit 4 \cite{Rodriguez2023}.  These subsets of orbits, however, are rarer, so that to first approximation the classification gives the expected dynamics.

Figs.~\ref{fig:PMDCQSFA1}(a) to (d) show PMDs calculated using single orbits, according to the pre-specified dynamics. The PMDs associated with orbits 1 and 2 [Figs.~\ref{fig:PMDCQSFA1}(a) and (b)] are centered at the origin and roughly extend towards the direct ATI cutoff, confirming that they can be approximately  interpreted as direct orbits, while those obtained from orbits 3 and 4 [Figs.~\ref{fig:PMDCQSFA1}(c) and (d)] exhibit the expected behavior from forward-scattered and backscattered orbits, respectively. The PMD from orbit 3 is located in a low-momentum region, with a central caustic, and orbit 4 gives rise to a PMD with a clear scattering ridge centered at ($p_{f\parallel},p_{f\perp})=(\mp 2 \sqrt{U_p}, 0)$, as expected from backscattered trajectories. All those single-orbit PMDs are reflection-symmetric with regard to the axis $p_{f\parallel}=0$, as expected from a half-cycle symmetric field. They are also centered at and reflection-symmetric about the $p_{f\perp}=0$ axis, as expected from a linearly polarised half-cycle symmetric field. 

Figs.~\ref{fig:PMDCQSFA1}(e) to (h) show the fan, the spider, the spiral and the resulting patterns calculated with the standard CQSFA. The fan [Fig.~\ref{fig:PMDCQSFA1}(e)] results from the interference of 1 and 2, the spider stems [Fig.~\ref{fig:PMDCQSFA1}(f)] from the interference of orbits 2 and 3, and the spiral [Fig.~\ref{fig:PMDCQSFA1}(g)] is due to the interference of orbits 3 and 4. All these contributions result in an intricate pattern, shown in Fig.~\ref{fig:PMDCQSFA1}(h). Interestingly, from these figures only the spider is reflection-symmetric about $p_{f\parallel}=0$. This is due to a numerical artefact, that arises from the necessity of limiting the ionisation times to a finite range. The limited range introduces boundary effects associated with the unit-cell choice that break the field symmetry \cite{Werby2021}. Because the spider stems from orbits starting at the same half cycle, it is not affected by this choice. These unphysical effects are eliminated if one includes contributions over many cycles, which would correspond to a more realistic setting. However, this also makes the ATI rings resulting from inter-cycle interference overpowering. A solution is to average incoherently over unit cells starting at different instances within a cycle, which will filter out the ATI rings and, yet, eliminate this arbitrariness. 
For a general polychromatic linearly polarised electric field
\begin{equation}
    E(t)=\sum_nE_nf_n(t),
\end{equation}
of amplitudes $E_n$ and time profiles $f_n(t)$, shifting the unit cell is equivalent to taking $f_n(t) \rightarrow f_n(t+\varphi/\omega)$ in the above equation, and performing an incoherent sum over PMDs calculated for different phases  $\varphi$. 
This averaging has been employed in \cite{Werby2021,Werby2022b}, in good agreement with experiments.
Unit-cell averaged fan and spiral are shown in Fig.~\ref{fig:PMDCQSFA1}(i) and (j), respectively. The resulting patterns are reflection-symmetric about $p_{f\parallel}=0$, as expected for a monochromatic driving field. Finally,  Fig.~\ref{fig:PMDCQSFA1}(k) and Fig.~\ref{fig:PMDCQSFA1}(l) show some of the experimental results in \cite{Werby2021}, compared with the CQSFA results. One should note that, additionally to the unit-cell averaging in the CQSFA, a filter has been applied to theory and experiment to eliminate spurious high-frequency oscillations. 

\subsubsection{Breaking symmetries in photoelectron holography}

In this section, we illustrate how symmetry may be broken by appropriate choices of driving fields, binding potentials, or the interplay of both.  Resulting from broken symmetries, holographic patterns and features such as caustics may be skewed, blurred or suppressed by employing fields in which some of the symmetries $\mathcal{S}_1-\mathcal{S}_3$ are broken. Examples are bichromatic fields, in which the relative phases between the two driving waves are varied in such a way as to induce different gradients around maxima and minima, or maxima of different amplitudes at each half cycle, elliptically polarised fields, which will lead to broken symmetries in PMDs due to the Coulomb potential, and differences in rescattering ridges due to hidden potential symmetries. 

\begin{figure}
    \centering
    \includegraphics[width=0.99\linewidth]{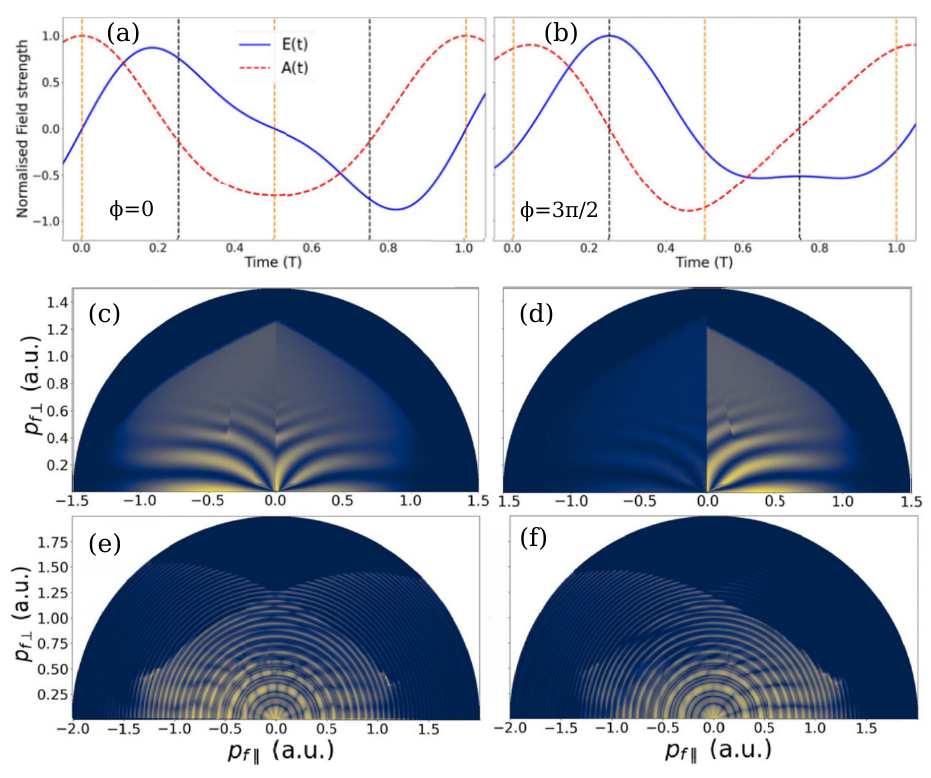}
    \caption{Adapted from Ref.~\cite{Rook2022} to show how the [(a) and (b)] field symmetry propagates through  to [(c)-(f)] symmetry in the CQSFA PMDs, for a linearly polarised bichromatic field with $\phi=0$ and $\phi=3\pi/2$.  Panels (c) and (d) plot the spider and panels (e) and (f) show a CQSFA computation for all patterns, taking the bichromatic fields in the upper panels. The target was hydrogen, and the fundamental  of the bichromatic field is $\omega=0.057$ a.u.  The bichromatic field intensity is set such that the \textit{total} ponderomotive energy is the same as that of a monochromatic field of intensity $1.4~ \times 10^{14}\mathrm{W/cm}^2$. The ratio between the intensities of the two constituent waves is $I_{\omega}/I_{2\omega}=100$.      }
    \label{fig:PMDCQSFA2}
\end{figure}

Fig.~\ref{fig:PMDCQSFA2} exemplifies what kinds of imprints broken symmetries leave on the spider and on the full PMDs, for a bichromatic $\omega - 2\omega$ field of different phases.  Because, the half-cycle symmetry is broken, the PMDs will no longer be symmetric about $p_{f\parallel}=0$, but breaking the other symmetry ($\mathcal{S}_2$  or $\mathcal{S}_3$) will influence the holographic structures differently.  
For $\phi=0$ [Fig.~\ref{fig:PMDCQSFA2}(a)], both the half-cycle symmetry $\mathcal{S}_1$ and the time reflection $\mathcal{S}_2$ about the extrema are broken, but the symmetry  $\mathcal{S}_3$ about the crossing remains. Consequently, the field gradients about the extrema are unequal which will skew the spider [see Fig.~\ref{fig:PMDCQSFA2}(c)], shifting the fringes unequally for positive and negative $p_{f\parallel}$.
In contrast, for $\phi=3\pi/2$ [Fig.~\ref{fig:PMDCQSFA2}(b)], $\mathcal{S}_3$ is broken, but $\mathcal{S}_2$ is retained. This leads to the maximum field strength being  weaker for one of the half cycles ($0.5T\leq t\leq T$), although it roughly plateaus. Therefore, for orbits 2 and 3 starting at that specific half cycle, ionisation is strongly suppressed. This will result in the spider being strongly suppressed, but not skewed as in the previous case, as shown in Fig.~\ref{fig:PMDCQSFA2}(d).  Bringing the contributions of all orbits together in a PMD [Figs.~\ref{fig:PMDCQSFA2}(e) and (f)] shows that, overall, this effect is subtle. However, while in  Fig.~\ref{fig:PMDCQSFA2}(e) the spider is asymmetric and somewhat skewed, in Fig.~\ref{fig:PMDCQSFA2}(f) it has almost completely vanished for $p_{f\parallel}<0$. 

\begin{figure*}
    \centering
    \includegraphics[width=0.9\linewidth]{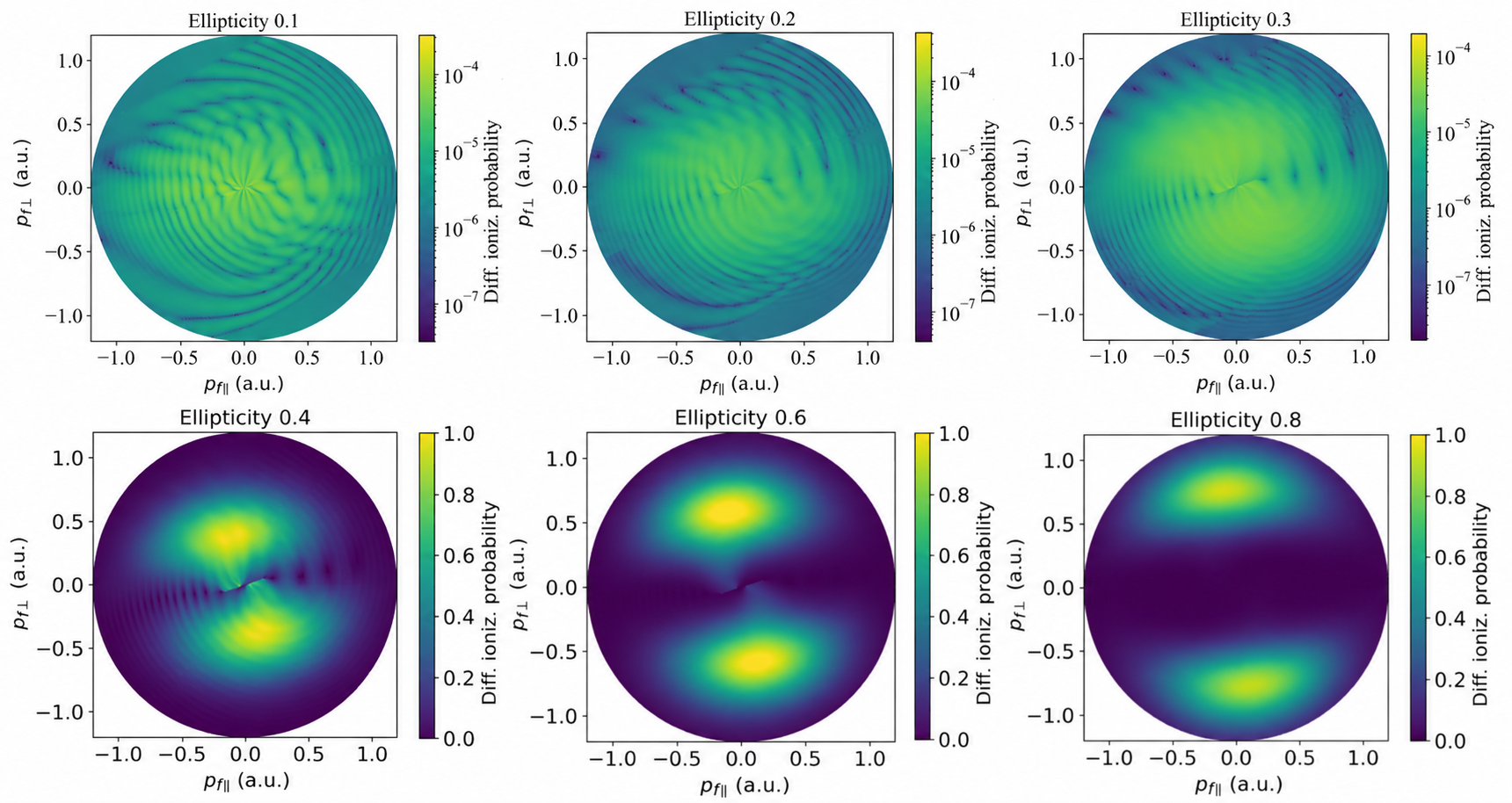}
    \caption{Photoelectron momentum distributions for helium in an elliptically polarised driving field, shown for ellipticities $\epsilon=0.1,\,0.2,\,0.3,\,0.4,\,0.6,$ and $0.8$. The horizontal and vertical axes give the final electron momentum components $p_{f\parallel}$ and $p_{f\perp}$, respectively, in atomic units. The upper row highlights the low-ellipticity regime, in which twisted holographic interference structures are still visible, although their contrast decreases as $\epsilon$ increases. The lower row illustrates the transition to the high-ellipticity regime, where the distributions evolve into split, attoclock-like lobes and the intra-cycle interference patterns become strongly suppressed. The calculations are performed for helium within the single-active-electron approximation using a Coulomb-type binding potential, for a laser field of intensity $I=2.5\times10^{14}\,\mathrm{W/cm^2}$ and wavelength $\lambda=735\,\mathrm{nm}$. The low-ellipticity panels are displayed on a logarithmic scale in order to reveal weak interference features, whereas the high-ellipticity panels are shown on a linear normalised scale; in the single-cycle results, inter-cycle ATI rings are absent. Adapted from Ref.~\cite{Kim2022}.}
    \label{fig:elliptical}
\end{figure*}

 Fig.~\ref{fig:elliptical} shows how the  Coulomb potential  breaks the reflection symmetries identified for the ATI photoelectron momentum distributions in an elliptically polarised field, with regard to the parallel and perpendicular momentum axis. For comparison, we refer to Fig.~\ref{fig:DATI_Fields} for direct ATI. Instead of being symmetric about  $p_{f\parallel} \rightarrow -p_{f\parallel}$ and $p_{f\perp} \rightarrow -p_{f\perp}$ reflections,  the PMDs exhibit angular offsets that cause a twist in the holographic patterns. For higher ellipticities, these offsets have been employed in the attosecond angular streaking ``attoclock" technique to extract tunneling times from sickle shaped PMDs \cite{Eckle2008}. In particular, it is possible to map specific times to specific offset angles, and there has been substantial discussion of whether they stem from sub-barrier or continuum propagation \cite{Landsman2015,Hofmann2019,Hofmann2021}. 
 The upper panel of Fig.~\ref{fig:elliptical}  displays the intra-cycle, holographic interference patterns calculated with the CQSFA, and how they become increasingly twisted and blurred as the driving-field ellipticity increases. The cause of the effect is the same as that in the ``atto clock", with the difference that there are many types of interfering orbits, with distinct, orbit-dependent  offsets. Eventually, for higher ellipticities, as shown in the lower row of the figure, the PMDs become sickle-shaped, the holographic interference vanishes and the maxima are increasingly further apart. 

\begin{figure}
    \centering
    \includegraphics[width=0.99\linewidth]{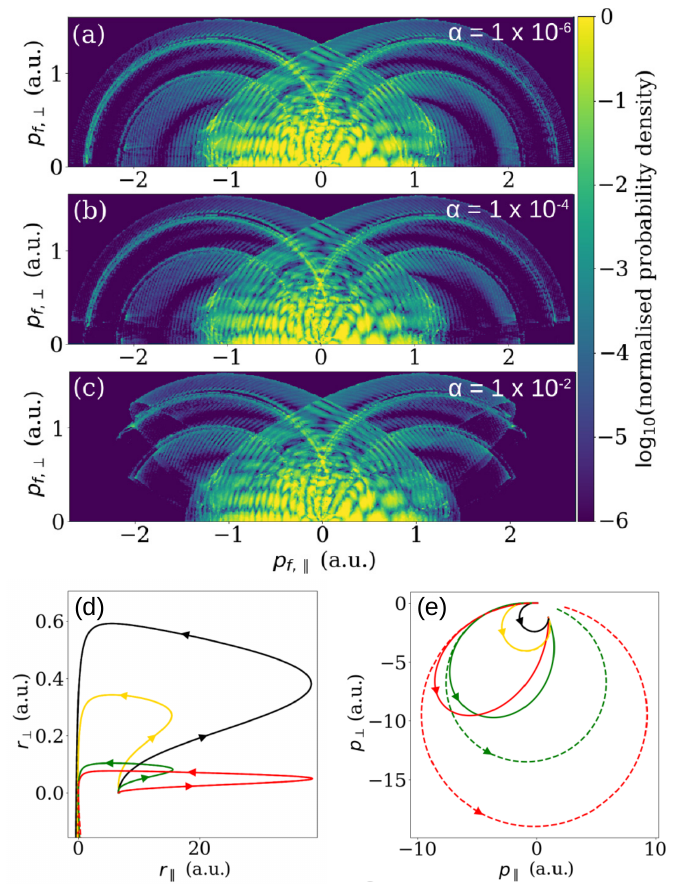}
    \caption{
    Photoelectron momentum distributions and representative quantum trajectories for atomic hydrogen ($I_p=0.5$ a.u.) driven by a linearly polarized monochromatic field with wavelength $800$ nm and intensity $1.5\times10^{14}\,\mathrm{W\,cm^{-2}}$, calculated using the H-CQSFA. Panels (a)--(c) show the distributions as functions of the final momentum components $p_{f\parallel}$ and $p_{f\perp}$, parallel and perpendicular to the laser-field polarization, for Coulomb-softening parameters $\alpha=10^{-6}$, $10^{-4}$, and $10^{-2}$, respectively. Panels (d) and (e) display representative trajectories in position and momentum space, respectively. For the trajectory comparison, the softened potential with $\alpha=10^{-2}$ is shown by solid curves, while the nearly hard-Coulomb case with $\alpha=10^{-6}$ is indicated by dashed curves. Adapted from Ref.~\cite{Rook2024}}.
    \label{fig:softening}
\end{figure}

Fig.~\ref{fig:softening} is an example of how some of the kinematic constraints determined by the SFA are in fact dynamic constraints and how they can be influenced by the symmetry of the residual binding potential. The figure shows PMDs obtained with the CQSFA for soft-core potentials 
\begin{equation}
    V(r)=\frac{-1}{\sqrt{r^2+\alpha^2}}
\end{equation}
of increasing softening parameter $\alpha$. Soft-core potentials have been used for around four decades in the modelling of strong-field phenomena as they do not exhibit the singularities associated with Coulomb-type potentials. However, much more attention has been given to their spectral properties and the associated tunnel dynamics, than to how it behaves upon rescattering. 

Figs.~\ref{fig:softening}(a)-(c) show different behaviors of the rescattering ridge near $10U_p$, which is associated with orbits type 4. For vanishingly small values of the softening parameter [Fig.~\ref{fig:softening}(a)], the ridge essentially closes, reaching up to the $p_{f\parallel}$ axis. For the remaining softening parameters [Fig.~\ref{fig:softening}(b) and (c)], the ridge folds around a non-vanishing angle, which increases with $\alpha$. Two additional solutions arise for the soft-core potential, which coalesce at the ridge and are the result of a symmetry that existed for the Coulomb potential being broken in the soft-core case. Physically, this implies that the rescattering angle for the Coulomb potential can go up to $\pi$, while for the soft-core potential, after a maximal angle the electron will just pass through. 

In Figs.~\ref{fig:softening}(d) and (e), we analyze the orbits leading to the ridges in position and momentum space, respectively.  The orbits plotted in red and green are closer to the driving-field polarisation axis than those in yellow and black, but, essentially, the position-space orbits exhibit a similar behavior, being scattered by the potential and coalescing towards similar values. In contrast, their momentum-space behavior is markedly different: the yellow and black orbits are practically circular and practically indistinguishable from those obtained for the Coulomb potential, while the green and red orbits are elliptical. Their counterparts for the Coulomb potential, plotted as dashed lines, are circular and qualitatively very different. 

These patterns are associated with them being unequally affected by a broken symmetry for the soft-core potential, which is present for the Coulomb potential: the conservation of the Laplace-Runge-Lenz (LRL) vector.  This additional symmetry will lead to circular orbits in  momentum space, with a radius given by $1/L$, where $L$ is the angular momentum, and a center defined by the LRL vector \cite{Bander1966,Rogers1973,Cariglia2013}. 
The black and yellow orbit types interact less strongly with the core, so that their dynamics are very close to those scattered by a Coulomb potential, while non-Coulomb effects are more prevalent for the red and green orbits.  A full classification of type 4 orbits, and more details about the folding of the rescattering ridge are provided in \cite{Rook2024}.

\subsection{Extracting Target Information: Parity and Structure}

\begin{figure}
\hspace*{-0.5cm}    \includegraphics[width=1.1\linewidth]{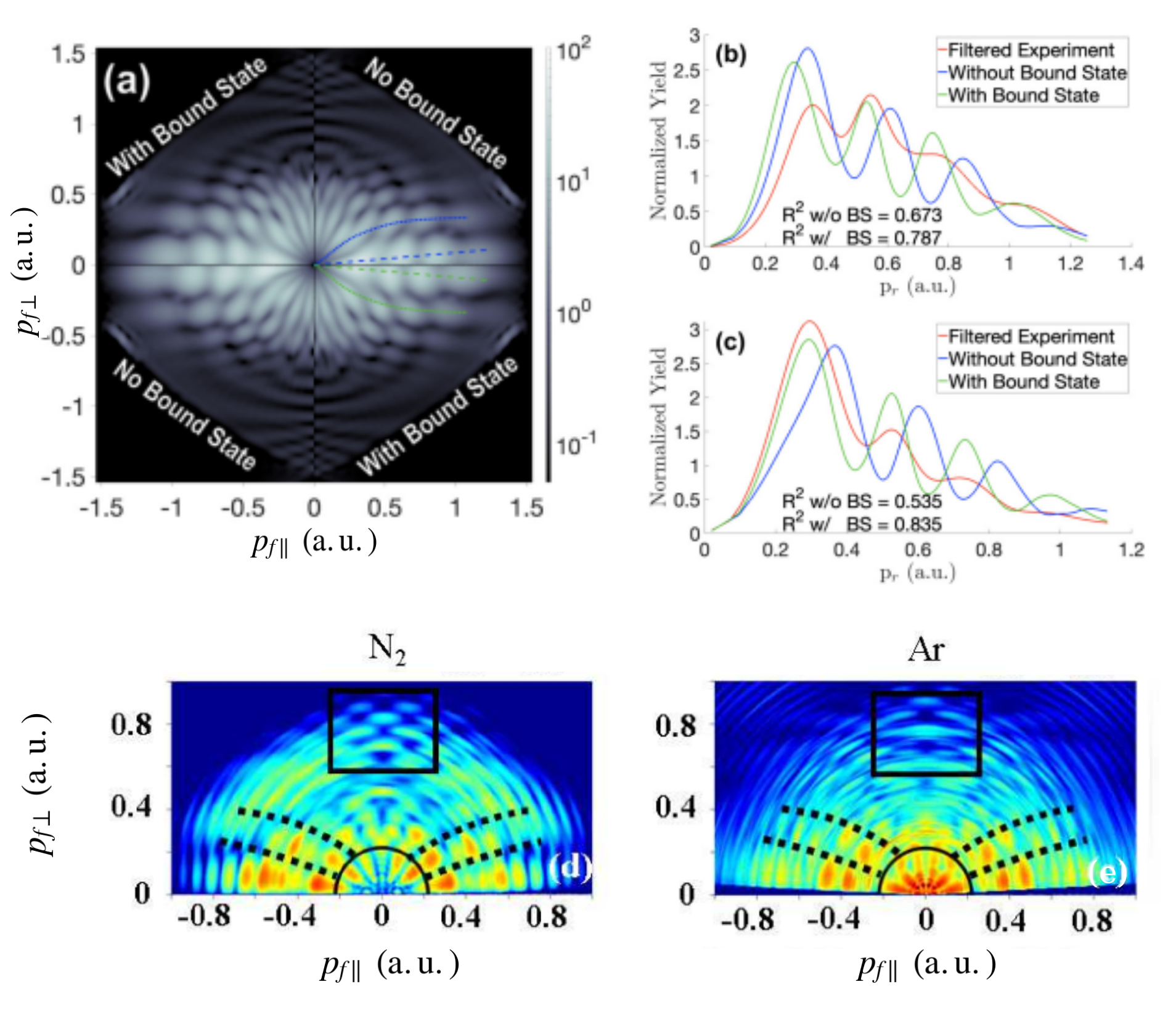}
    \caption{Adapted from [(a)-(c)] Ref.~\cite{Werby2021} and [(d) and (e)] Ref.~\cite{Kang2020}.  Panels (a) to (c) exemplify multi-path interference in PMDs computed with the CQSFA, for argon in a driving field of  intensity $ 2 \times 10 ^{14} \mathrm{W/cm}^2 $ and wavelength $\lambda = 800$ nm. Panel (a) displays CQSFA distributions with and without the bound-state prefactor, shown in alternating quadrants to highlight the resulting phase shifts along both momentum axes. High-frequency structures have been filtered from the CQSFA distributions. In panels (b) and (c), we plot electron-yield lineouts taken close to the longitudinal polarisation axis and along the first spider leg, respectively. The corresponding filtered experimental lineouts in \cite{Werby2021} included for comparison, together with the $R^2$ goodness-of-fit values. Panels (d) and (e) show TDSE photoelectron momentum distributions for ionisation of Ar and randomly aligned $\mathrm{N}_2$, respectively, by a linearly polarised laser field with a central wavelength of $788\,\mathrm{nm}$ and peak intensity of $6.5\times10^{13}\mathrm{W/cm^2}$, matching the experimental parameters in \cite{Kang2020}. The focal-volume effect is included, and the $\mathrm{N}_2$ distribution is averaged over the random orientation of the molecular axis. }
    \label{fig:parity}
\end{figure}

In this section, using Fig.~\ref{fig:parity}, we will exemplify how one can use specific holographic patterns, or multi-path interference, to extract information about a specific target. The work discussed here is focused on parity, and what holographic patterns are sensitive to it. Parity is the symmetry associated with spatial inversion $\mathbf{r} \rightarrow -\mathbf{r}$. A state has even (odd) parity if $\psi(-\mathbf{r})=\psi(\mathbf{r})$($\psi(-\mathbf{r})=-\psi(\mathbf{r})$). Holographic patterns are insensitive to the parity of the initial bound state when the interfering electron trajectories are released within the same half cycle, as in the spider. By contrast, for structures such as the fan or spiral, the interfering trajectories originate from different half cycles and are therefore released from opposite sides of the target. Neglecting subcycle laser-induced distortions of the bound state, this leads to an additional phase shift of $\pi$ for an initial state of odd parity, whereas no such shift occurs for an even-parity state.

Fig.~\ref{fig:parity} illustrates two ways in which this can be employed. In Figs.~\ref{fig:parity}(d) and (e), we compare holographic patterns from $N_2$ molecules and their companion atoms $Ar$. These two targets have very similar ionisation potentials, but highest occupied orbitals of different parities: while the HOMO (highest occupied molecular orbital) of $N_2$, i.e., $3\sigma_g$ has even parity, for $Ar$ the electron is initially bound in $3p$, whose parity is odd. The similar ionisation potentials ensure that additional phase shifts are not picked up during the continuum propagation, but, rather, are associated with the initial bound states as stated above. For our comparison, we chose the spiral as, (i) due to its high photoelectron energy, it is not obfuscated by the fan or the spider; (ii) since it stems from the interference of orbits 3, and 4, which are separated by half a cycle, a $\pi$ phase shift should be present. Comparing these figures shows that this is indeed the case. 

An alternative to measuring parity that does not rely on a companion target, shown in Figs.~\ref{fig:parity}(a) to (c), is to employ parity-sensitive modulations caused by a third type of orbit on a prominent holographic structure that is insensitive to parity, i.e., the spider. The figure shows CQSFA computations in which the bound-state prefactor is included or excluded, and emphasises the modulations in the legs of the spider. These modulations are sensitive to parity and caused by type 1 orbits. Omitting or including the prefactor associated with the $3p$ initial bound states causes phase shifts in these modulations. 

From the application perspective, one should note that the holographic pattern to be used will be chosen depending on the property one is willing to investigate. For example, the spider is insensitive to parity, but is strongly influenced by nodal planes \cite{Meckel2014} or the asymmetry in a heteronuclear molecule \cite{Haertelt2026}. Alternatively, one may consider, instead of standard PMDs, photoelectron spin textures. Theoretical studies comparing  $\mathrm{He}_+$ and $\mathrm{Xe}$ using the CQSFA show drastic differences in this specific observable, although the PMDs remain similar \cite{Chen2026spin}. In contrast, the spiral is a good measure for parity, as discussed in this section.  One should note that changes in holographic patterns also be use to probe nuclear dynamics \cite{Walt2017}.

Overall, one important complication is that, often, the momentum-space regions occupied by different holographic patterns overlap. This can be seen in the intricate patterns obtained in the experiments and calculations discussed in this section. One possibility for disentangling these contributions is to employ an additional orthogonally polarised field, which can deflect electron trajectories associated with different ionisation times and thereby separate otherwise overlapping contributions in momentum space \cite{Li2016BackwardHologram,He2016Holography,Henkel2015Trajectories,Xie2017Disentangling}, but this strategy may change the problem's dynamics \cite{Li2016BackwardHologram,He2016Holography,Xie2017Disentangling} and thus should be approached with care. 

\section{Nonsequential double ionisation}
\label{sec:NSDI}

The main objective of this section is to revise how quantum interference breaks symmetries in NSDI. We will focus on the RESI mechanism, for which interference is robust, in the context of the SFA.  An important nuance of NSDI is that electron-electron correlation is, to some extent, generated and shaped by the laser field, the recollision dynamics and, if included, the post-collision Coulomb interaction. Thus, employing tailored fields can shape quantum interference in two-electron momentum distributions.

\subsection{Symmetry and interference in NSDI}

For any linearly polarised fields satisfying the symmetries $\mathcal{S}_1 - \mathcal{S}_3$ exactly, such as monochromatic fields and $(\omega, 3\omega)$, $\phi=0$ fields, or approximately, such as long-enough pulses or CEP-averaged few-cycle pulses, the RESI distributions are expected to be fourfold symmetric in the $p_{1\parallel}p_{2\parallel}$ plane.  This symmetry may either be broken by breaking the field symmetry directly, as discussed in the previous sections, or by quantum interference, as we will illustrate next. The first theoretical observations that quantum interference was prominent in RESI was provided in \cite{Hao2014}, in which the main question was how to move from the fourfold distributions obtained with the SFA towards the anti-correlated distributions observed in experiments. Theoretical studies by us not only confirmed that this symmetry could be broken, by producing a myriad of shapes for RESI by adding different quantum pathways coherently, but also identified different types of quantum interference in RESI \cite{Maxwell2015,Maxwell2016}. These types of interference were associated with a single or several excitation channels, and the former type of interference supports an analytical description. 

\begin{figure}
    \centering
    \includegraphics[width=\linewidth]{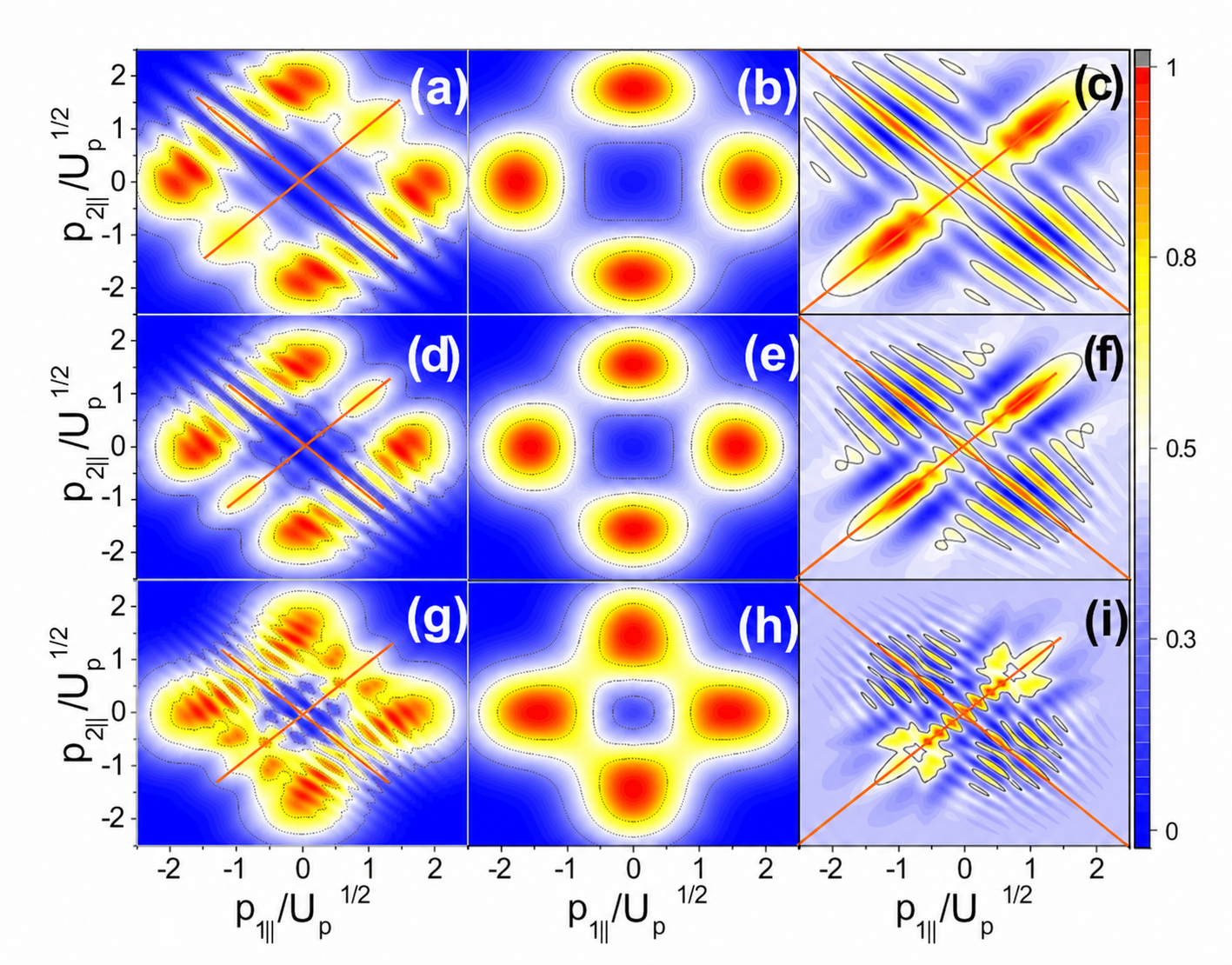}
    \caption{Quantum interference between symmetry-related RESI events in correlated two-electron momentum distributions. From left to right, the columns show the fully coherent sum of the four transition amplitudes [panels (a), (d), and (g)], the corresponding incoherent sum [panels (b), (e), and (h)], and the difference between the two [panels (c), (f), and (i)]. The perpendicular momentum components have been integrated out and the bound-state prefactors are neglected in order to isolate interference between the different events. The rows correspond, from top to bottom, to laser intensities $I=2.28\times10^{13}$, $4.56\times10^{13}$, and $6.84\times10^{13}\mathrm{W/cm^2}$, for which the ponderomotive energies are $U_p=0.05$, $0.10$, and $0.15$ a.u., respectively. A linearly polarised monochromatic field with angular frequency $\omega=0.057$ a.u. is used. The target is Ar, with $E_{1g}=0.58$ a.u. and $E_{2g}=1.02$ a.u., and the RESI channel considered corresponds to excitation of the second electron from $3s$ to $3p$, with $E_{2e}=0.52$ a.u. The orange lines mark the diagonal and anti-diagonal $p_{1\parallel}=\pm p_{2\parallel}$, along which interference is particularly pronounced. Each panel is normalised to its maximum value. Adapted from Ref.~\cite{Maxwell2015}.}
    \label{fig:NSDIInterf1}
\end{figure}

Fig.~\ref{fig:NSDIInterf1} shows how this fourfold symmetry is broken for a monochromatic, linearly polarised field even if a single excitation channel is taken into account, with fringes along the diagonal and anti-diagonal whose spacing decrease with the driving-field intensity.
  
\begin{figure}
    \centering
    \includegraphics[width=\linewidth]{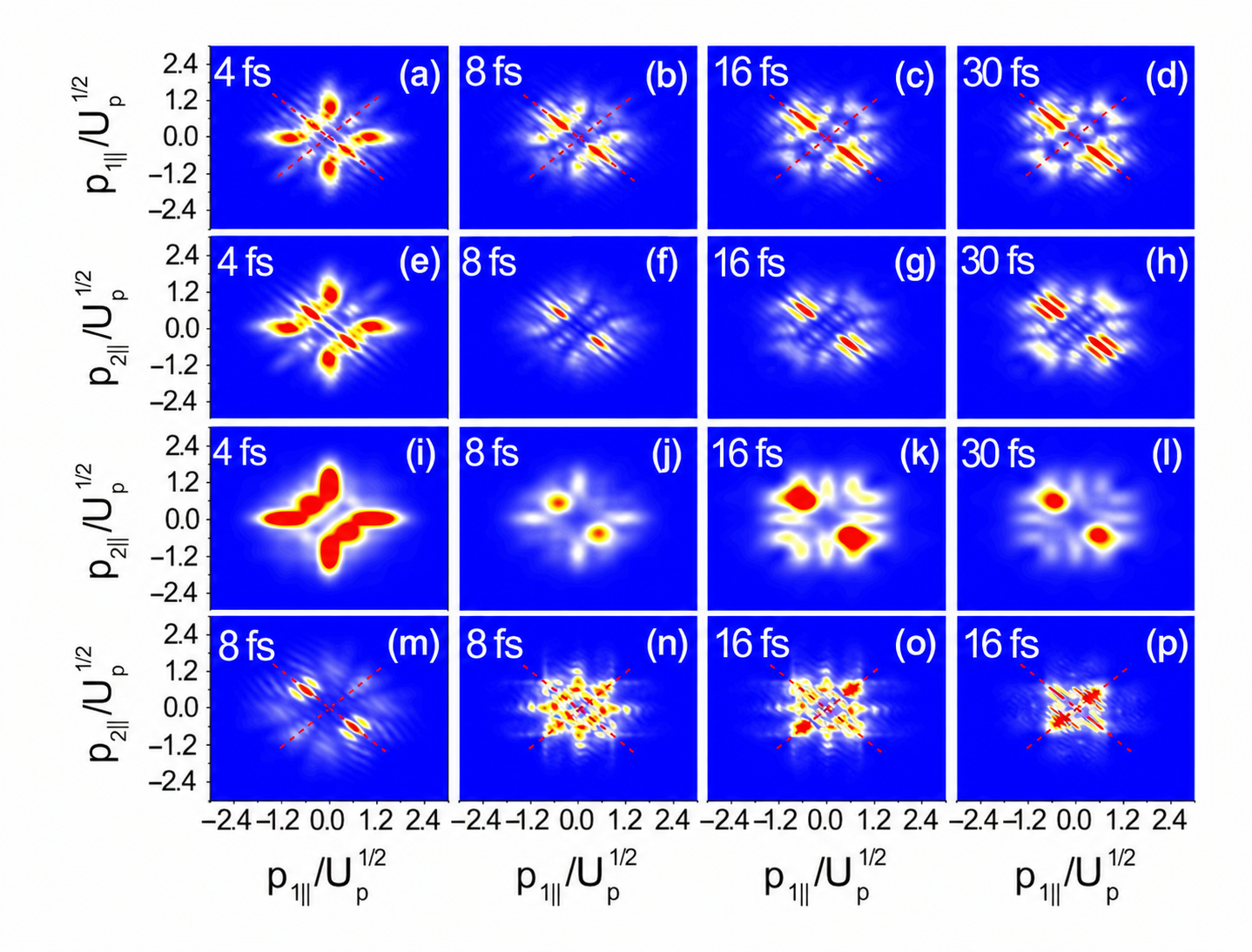}
    \caption{Quantum-interference control of below-threshold nonsequential double ionisation in Ar under experimentally relevant conditions. Shown are correlated two-electron momentum distributions as functions of the momentum components $p_{1\parallel}$ and $p_{2\parallel}$ parallel to the laser polarisation, scaled by $\sqrt{U_p}$. The calculations employ the RESI mechanism and coherent superpositions of several excitation channels of Ar$^+$, with target energies $E_{1g}=0.58$ a.u. and $E_{2g}=1.02$ a.u. The first row, panels (a)--(d), shows coherent distributions for a monochromatic field with $\omega=0.057$ a.u. $\left(\lambda\simeq800\,\mathrm{nm}\right)$, peak intensity $I=1.0\times10^{14}\,\mathrm{W/cm^2}$, and $U_p=0.22$ a.u. Panels (e)--(h) include focal-volume averaging for the same peak field parameters, while panels (i)--(l) additionally suppress interference between symmetry-related RESI events by summing these events incoherently. Their comparison shows that event interference remains visible after focal averaging and integration over the transverse electron momenta. In the last row, calculated without focal averaging, the peak intensities are $I=0.8\times10^{14}\,\mathrm{W/cm^2}$ for (m), $1.2\times10^{14}\,\mathrm{W/cm^2}$ for (n) and (o), and $1.4\times10^{14}\,\mathrm{W/cm^2}$ for (p), corresponding to $U_p=0.18$, $0.26$, and $0.30$ a.u., respectively. The pulse durations indicated inside the panels refer to the experimental Ar distributions that the calculations were designed to reproduce; those measurements employed near-transform-limited, CEP-averaged pulses centred around $790\,\mathrm{nm}$ with durations from $4$ to $30$ fs. The dashed lines indicate $p_{1\parallel}=\pm p_{2\parallel}$, where event-interference structures are particularly prominent. All field intensities lie below the direct electron-impact ionisation threshold, so that RESI remains the dominant NSDI pathway. The persistence of these structures under focal averaging and experimentally accessible momentum integration illustrates that quantum interference can strongly shape two-electron correlations under realistic strong-field conditions. Adapted from Ref.~\cite{Maxwell2016}.}
    \label{fig:NSDIInterf2}
\end{figure}

Fig.~\ref{fig:NSDIInterf2} exemplifies the myriad shapes the RESI distributions may acquire, with computations done for incorporating prefactors, intra-and inter-channel interference and focal averaging. The coherent superpositions were adjusted to mimic the  frequency widths of a CEP averaged few-cycle pulse. The figure shows a myriad of shapes, such as the maxima at the diagonal and hyprebolic fringes. 
Next, we will examine some of these features closely, starting from the types of quantum interference that may occur.  The discussions will be based on our recent work \cite{Hashim2024,Hashim2025,Hashim2026}, in which this interference has been explored more systematically. 
\subsection{Types of quantum interference}

Quantum interference in NSDI arises whenever multiple quantum pathways lead to the same final two-electron momentum state. NSDI, and RESI in particular, provides a rich setting in which to investigate such interference phenomena.  This is because the process involves two electrons, intermediate excited states and multiple possible ionisation/rescattering events with a single laser cycle leading to a large number of competing quantum pathways. While the detailed manifestation of interference depends on the driving field and target, its physical origin can be traced back to degeneracies in the semiclassical action associated with these competing pathways.

\begin{figure*}
    \centering
    \includegraphics[width=1\textwidth, keepaspectratio]{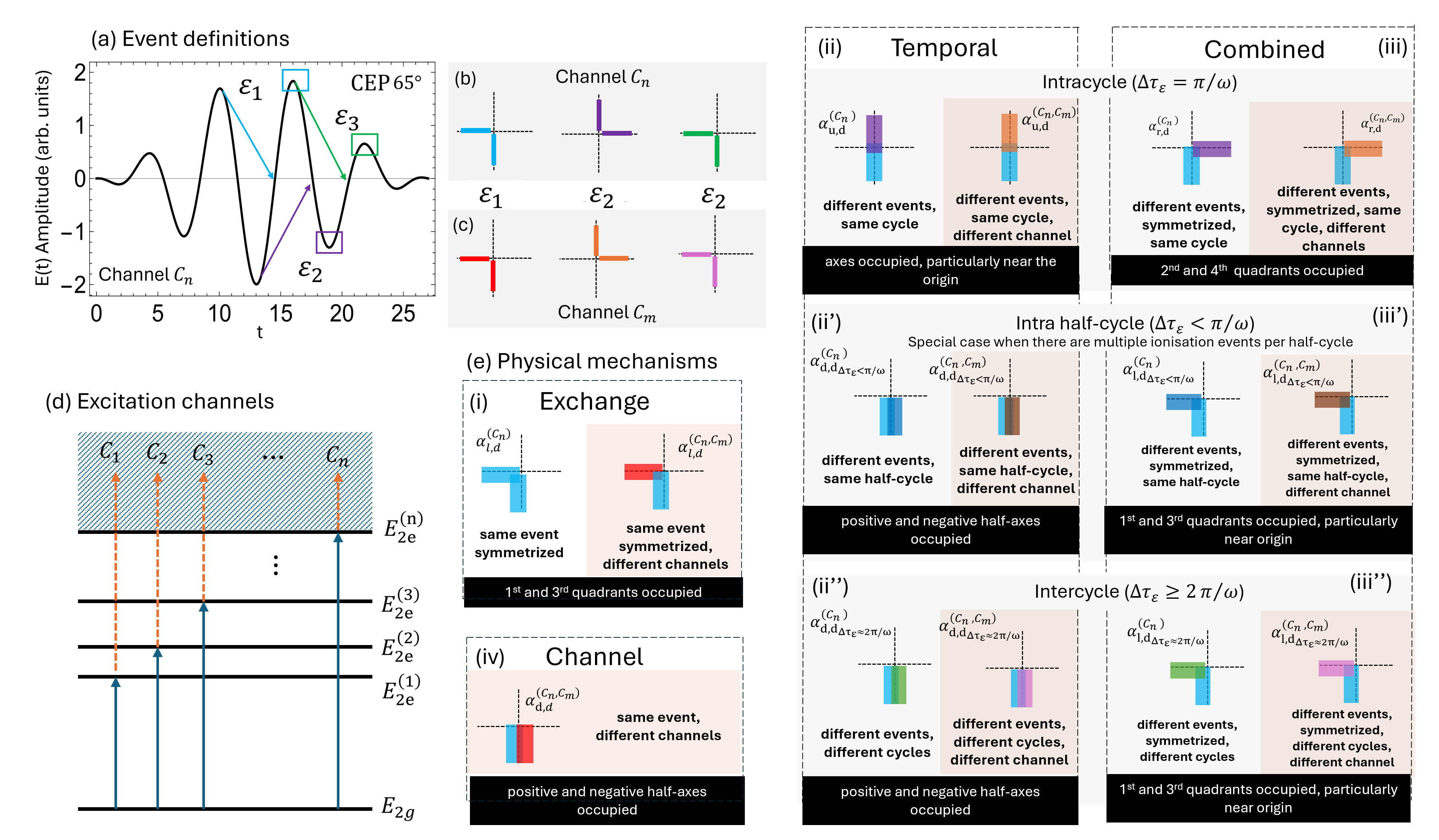}
    \caption{Panels (a)-(c) show a representative few-cycle pulse (peak intensity $I = 1.5 \times 10^{14}\text{W/cm}^2$, $\lambda = 800$nm ($\omega = 0.057$a.u.), $N = 4.3$, and CEP $65^{\circ}$) and two excitation channels solely to illustrate the events $\varepsilon_n$ and their momentum-space mapping. Only three most dominant events have been taken ($n=1,2,3$). The arrows indicate the classical ionisation and return times for the first electron and the rectangles show the most relevant ionisation times for the second electron. Panel (d) displays a schematic of the channels by which the second electron can be excited and then ionised into the continuum. Panel (e) illustrates the momentum mappings associated with different physical mechanisms: (i) exchange interference, (ii, iii) intracycle temporal and combined temporal-exchange interference, (ii', iii') intra half-cycle temporal and combined interference where the momentum mapping of events within a half-cycle in each channel are indicated by a darker and a lighter shade, (ii'', iii'') intercycle temporal and combined interference, and (iv) channel interference. The interchannel interference schematics have orange backgrounds to distinguish them from intrachannel interferences. The regions of the momentum space occupied are given in the black boxes. Phase differences associated with the specific schematics shown are provided - note that this is not exhaustive. }
    \label{fig:giantInterf}
\end{figure*}

The probability distribution for two coherently interfering transition amplitudes can be written as, 
\begin{align}
\Omega 
  &= |M_1 + M_2|^2 \nonumber \\
  &= |M_1|^2 + |M_2|^2 
     + 2|M_1||M_2| 
       \cos\!\big(\mathrm{Re}[\Delta S_{12}]\big)
       e^{-\mathrm{Im}[\Delta S_{12}]}
\label{eq:generalinterf}
\end{align}
where $\Delta S_{12} = S_1 - S_2$. Therefore the interference maxima and minima are determined by $\text{Re}[\Delta S_{12}] = 2n\pi$ and $\text{Re}[\Delta S_{12}] = (2n+1)\pi$ respectively where $n$ is an integer. The imaginary parts of the actions and the prefactors control the visibility and contrast of the fringes. In practice, observable interference requires the contributing trajectories to have comparable weights, i.e. similar imaginary parts of the action. In realistic strong-field processes, the total transition amplitude generally involves coherent sums over many trajectories, events, and channels. Hence, it is useful to classify interference phenomena according to which pathways are interfering. Figure~\ref{fig:giantInterf} provides a unified overview of the interference mechanisms discussed throughout this section.

When considering a single excitation channel, interference can arise from three main mechanisms: electron exchange, temporal displacement between RESI events and a combination of both electron exchange and temporal displacement of events. When more than one excitation channel is considered, one must also consider interference between events in different channels in conjunction with the aforementioned three mechanisms.

\subsubsection{Exchange Interference}

Because RESI is a two-electron process, the indistinguishability of the electrons must be taken into account. Since the final state does not label which detected electron corresponds to which emission route (i.e. rescattered ATI-like process or direct ATI-like process), the physical probability distribution requires symmetrisation under exchange of the electron momenta. The coherent superposition of amplitudes related by $\textbf{p}_1 \rightarrow \textbf{p}_2$ gives rise to ``exchange interference'' and has no direct analogue in single-electron strong-field ionisation. Note that in some studies \cite{Maxwell2015,Maxwell2016}, this type of interference was called ``event interference''. Exchange interference, illustrated schematically in Fig.~\ref{fig:giantInterf}(e)(i) predominantly occupies the first and third quadrants of the $p_{1\parallel}p_{2\parallel}$ plane and produces characteristic structures aligned with the diagonal $p_{1\parallel}=p_{2\parallel}$.

\subsubsection{Temporal Interference}
A second class of interference arises from distinct RESI events occurring at different times within the laser field. The RESI transition amplitude is expressed as an asymptotic expansion over saddle-point solutions, each corresponding to a physically distinct ionisation–recollision event. Whenever such two temporally-displaced events populate overlapping momentum regions, they may interfere leading to ``event'' or ``temporal'' interference. In general, the first relevant recollision return, which occurs approximately three quarters of an optical cycle after ionisation is the biggest contributor to this type of interference given that later returns are exponentially suppressed due to wave-packet spreading and reduced tunnelling probability. 
Temporal interference can be classified according to the relative displacement $$ \Delta\tau=(\Delta t'',\Delta t',\Delta t) $$ where $t''$, $t'$ and $t$ denote the ionisation, recollision and subsequent ionisation times of the RESI process.
``Intracycle (subcycle) interference'' refers to interference between pathways whose relevant saddle-point times lie within the same optical cycle and is depicted in Fig.~\ref{fig:giantInterf}(e)(ii). Such interference typically generates structures aligned with the momentum axes.

More generally, tailored driving fields may enable multiple ionisation events within a single half-cycle. In this case, a further class of temporal interference becomes possible, namely ``intra-half-cycle interference'', illustrated in Fig.~\ref{fig:giantInterf}(e)(ii'). This type of interference is absent in monochromatic fields and emerges naturally in bichromatic \cite{Hashim2025} and more structured driving fields.

Note that another contribution to intracycle interference arises from ``orbit interference''. Even for a fixed event within a single cycle, the saddle-point equations often lead to multiple solutions known as the long and the short orbits, which differ in excursion time. In RESI, the rescattering saddle points of the first electron may coalesce and so this type of orbit interference primarily affects the first electron during its excursion prior to recollision. For the second electron, the relevant saddle points are well separated in the complex-time plane. This is naturally included in the present formalism since the first electrons' orbits are treated in pairs using the uniform approximation \cite{Faria2002}. However, previous studies have shown that their signatures may be strongly reduced after transverse momentum integration. Orbit interference effects are expected to become more pronounced when Coulomb corrections to the continuum dynamics are included.
Finally, ``intercycle interference'' involves pathways separated by one or more field periods [Fig.~\ref{fig:giantInterf}(e)(ii'')]. For single-electron processes like ATI with monochromatic fields, intercycle interference leads to the well-known concentric ring structure. For RESI, this type of interference mainly occupies the positive or negative half-axes of the momentum plane, although the exact interference structures are strongly dependent on the overlap of the momentum regions occupied by each of the interfering events \cite{Hashim2024}.

\subsubsection{Combined Interference}
There also exists interference arising from a combination of temporal and exchange interference referred to as ``temporal-exchange'' or ``combined interference''. Depending on the time delay between the interfering events, this can be intra or intercycle combined interference [see Fig.~\ref{fig:giantInterf}(e)(iii)-(iii'')], occupying the diagonals in the latter case, and the anti-diagonals in the former. 
The intra-half-cycle combined interference special case occurs in the first and third quadrants and thus is dominant along the diagonals. 

Together, exchange, temporal and temporal-exchange interference constitute the three fundamental classes of ``intrachannel'' interference, as they arise from pathways belonging to the same excitation channel.

\subsubsection{Interchannel Interference}
In RESI, the return of the first electron causes excitation of the second electron into different bound excited states of the parent ion for most atomic targets. Each excitation pathway constitutes a distinct channel leading to the same final electron momenta. ``Channel interference'' arises when transition amplitudes associated with different excited states are summed coherently as shown in Fig.~\ref{fig:giantInterf}(d). This type of interference is expected to reflect target-specific structure through the bound and excited state prefactors. Pure channel interference arises from interference of the same event in different channels [Fig.~\ref{fig:giantInterf}(e)(iv)]. Channel interference can also occur in combination with the intrachannel interferences, as highlighted in orange in all panels of Figs.~\ref{fig:giantInterf}(e). This is collectively known as ``interchannel interference'', and in principle occupies the same momentum regions as the intrachannel interference counterparts albeit with distinct structures. In practice, the excited state plays a large role in the shape of the PMD through the prefactor, meaning the same events in different channels may not occupy similar momentum regions, leading to little overlap and hence faint or vanishing interchannel interferences. 

\subsection{Building Blocks of Interference and Analytical phase shifts}

Although the interference mechanisms discussed above arise from physically distinct pathways, their observable signatures can ultimately be traced back to a small number of recurring phase ``building blocks''. These building blocks arise from the semi-classical actions and provide a useful framework for understanding the interference patterns in the parallel momentum plane. 
In general, the phase difference associated with two interfering pathways may be written as
\begin{equation}
\alpha_{\mu,\nu}^{(m,n)} = S_\mu ^{(m)}-S_{\nu}^{(n)},
\end{equation}
where $\mu,\nu$ are $l,r,u$ or $d$ (left, right, up or down) denoting the regions in momentum space occupied by the event to which the semi-classical action is associated, and $m, n$ refer to the specific event in question. The latter notation is useful in cases where intercycle, and intra-half-cycle interferences are investigated.
The condition for constructive interference is
$\alpha = 2n\pi $
where $n$ is an integer. The geometry of the resulting interference pattern is therefore determined by how the phase difference depends upon the electron momenta and the relevant ionisation and rescattering times.

Exchange interference originates from the symmetrisation of the two electrons and is therefore present even when a single RESI event is considered. For an event and its symmetrised counterpart (hence $m=n$), the phase difference can be decomposed into two contributions,
\begin{equation}
\alpha_{\mu,\nu}^{(m,m)} = \alpha^{(\mathrm{exch})}_{\mathbf p_1,\mathbf p_2} (t^, t')
+
\alpha^{}_{\mathbf p_1,\mathbf p_2}(t, t'),
\end{equation}
where 
\begin{equation}
\alpha^{(\mathrm{exch})}_{\mathbf{p}_1,\mathbf{p}_2})(t,t')=\frac{1}{2}  \left(\mathbf{p}_{2 }^2 -\mathbf{p}_{1}^2\right)(t-t'),
\label{eq:alphaexch}
\end{equation}
\begin{equation}
\alpha_{\mathbf{p}_1,\mathbf{p}_2}(t,t')=  (\mathbf{p}_2-\mathbf{p}_1) \cdot \left[\mathbf{F}_{A}(t')-\mathbf{F}_{A}(t)\right]
  \label{eq:alphaexch2}
\end{equation}
and $\textbf{F}_A (t)$ is the temporal integral of the vector potential $\int^t \mathbf A(\tau),d\tau.$

These two contributions generate the characteristic structures associated with exchange interference. The field-independent term gives rise to families of hyperbolic fringes whose asymptotes lie along the diagonals of the parallel momentum plane. The field-dependent contribution produces spine-like structures parallel to the diagonal $p_{1\parallel}=p_{2\parallel}$. Exchange interference structures are expected to be the most robust interference pattern, surviving even focal averaging \cite{Maxwell2016}.

The other phase difference solely related to electron momentum exchange is 
\begin{equation}
  \alpha_{ru}=S_r-S_u=  -\alpha^{(\mathrm{exch})}_{\mathbf{p}_1,\mathbf{p}_2}+ \alpha_{\mathbf{p}_1,\mathbf{p}_2}(t+\frac{\pi}{\omega},t'+\frac{\pi}{\omega}),
\label{eq:phaserightup}
\end{equation}
which, if made to vanish, also leads to hyperbolae and a bright fringe at the main diagonal for linearly polarised fields. One should note, however, that the temporal shifts in the second term of Eq.~\eqref{eq:phaserightup} imply that $|\alpha_{ru}| \neq |\alpha_{ld}|$ unless there is half-cycle symmetry, in which case $\alpha_{ru} = - \alpha_{ld}$. If half-cycle symmetry is broken, for instance for a driving field consisting of a wave and its second harmonic, or for the few-cycle pulses, the hyperbolae symmetric about the main diagonal will be asymmetric about $(p_{1\parallel},p_{2\parallel}) \leftrightarrow (-p_{1\parallel},-p_{2\parallel})$, that is, in the first and third quadrants of the $p_{1\parallel}p_{2\parallel}$ plane.

For temporal interference, we must consider the actions $S_r$ ($S_u$) which give events shifted from those corresponding to $S_l$($S_d$) by half a cycle, but no momentum exchange between the first and second electron takes place. This means that $S_l(\mathbf{p}_1,\mathbf{p}_2,t,t',t'')=S_r(\mathbf{p}_1,\mathbf{p}_2,t+\pi/\omega,t'+\pi/\omega,t''+\pi/\omega)$ and that $S_r(\mathbf{p}_1,\mathbf{p}_2,t,t',t'')=S_l(\mathbf{p}_1,\mathbf{p}_2,t+\pi/\omega,t'+\pi/\omega,t''+\pi/\omega)$. For $\alpha_{lr}$, this gives 
\begin{align}
     \alpha_{lr}=S_l-S_r&=-\frac{1}{2(t'-t'')}\alpha^{(A^2)}_{\pi/\omega}(t',t'')- \frac{1}{2}\alpha^{(\mathrm{pond})}_{\pi/\omega}(t,t'')\nonumber \\-&\alpha^{(\mathrm{ene})}_{\pi/\omega}-\alpha^{(\mathbf{p}_1,\mathbf{p}_2)}_{\pi/\omega}(t,t'), 
     \label{eq:alphalr}
\end{align}
where the purely temporal phase shifts are defined as 
\begin{equation}
  \alpha^{(A^2)}_{\Delta\tau}(t',t'')=\left[\mathbf{F}_A(t'+\Delta\tau)-\mathbf{F}_A(t''+\Delta\tau)\right]^2 -\left[\mathbf{F}_A(t')-\mathbf{F}_A(t'')\right]^2,
    \label{eq:alphaA2}  
\end{equation}

\begin{align}
    \alpha^{(\mathrm{pond})}_{\Delta \tau}(t,t'')=F_{A^2}(t+\Delta\tau)+F_{A^2}(t''+\Delta\tau)-F_{A^2}(t)-F_{A^2}(t''),
     \label{eq:alphapond}
\end{align}
and
\begin{equation}
 \alpha^{(\mathrm{ene})}_{\Delta\tau}=    \frac{\Delta\tau}{ 2 }  \left(2 \text{E}_{2 g}+2 \text{E}_{1 g}+\mathbf{p}_{1 }^2+\mathbf{p}_{2 }^2\right), 
 \label{eq:alphaene}
\end{equation}
with $F_{A^2}$ as defined as $\int^t \mathbf A^2(\tau),d\tau.$

One should note that Eqs.~\eqref{eq:alphaA2}, \eqref{eq:alphapond} have temporal arguments, while \eqref{eq:alphaene} depends only on the time difference $\Delta\tau$. The phase difference in Eq.~\eqref{eq:alphapond} is expected to give ponderomotive terms, which will add to that in Eq.~\eqref{eq:alphaene}. The latter is linear in time and contains the bound-state and kinetic energies. 

The phase difference $\alpha^{(\mathbf{p}_1,\mathbf{p}_2)}_{\pi/\omega}(t,t')$ is a particular case of 
  \begin{align}
\alpha^{(\mathbf{p}_1,\mathbf{p}_2)}_{\Delta \tau}(t,t')=\mathbf{p}_1\cdot\left[\mathbf{F}_A(t'+\Delta \tau) - \mathbf{F}_A(t') \right]  &\\+ \mathbf{p}_2\cdot \left[\mathbf{F}_A(t+\Delta \tau) - \mathbf{F}_A(t) \right],
   \label{eq:alphap1p2tau}  
  \end{align}
which also incorporates the electron momenta. Similarly, 
\begin{align}
     \alpha_{du}=S_d-S_u&=-\frac{1}{2(t'-t'')}\alpha^{(A^2)}_{\pi/\omega}(t',t'')- \frac{1}{2}\alpha^{(\mathrm{pond})}_{\pi/\omega}(t,t'') \\&-\alpha^{(\mathrm{ene})}_{\pi/\omega}-\alpha^{(\mathbf{p}_2,\mathbf{p}_1)}_{\pi/\omega}(t,t'), 
     \label{eq:alphadu}
\end{align}
where $\alpha^{(\mathbf{p}_2,\mathbf{p}_1)}_{\pi/\omega}$ indicates that the electron momenta have been swapped in Eq.~\eqref{eq:alphap1p2tau}, with $\Delta \tau=\pi/\omega$. For a monochromatic wave, we have found that the effect of these terms is minimal \cite{Shaaran2010,Maxwell2015}.

Computing the phase shifts $\alpha_{lu}$ and $\alpha_{rd}$ means that, in addition to exchanging $\mathbf{p}_1$ and $\mathbf{p}_2$, we must consider that one of the interfering events will be displaced by half a cycle. Specifically,  we must note that $S_u(\mathbf{p}_1,\mathbf{p}_2,t,t',t'')=S_l(\mathbf{p}_2,\mathbf{p}_1,t+\pi/\omega,t'+\pi/\omega,t''+\pi/\omega)$ and that $S_r(\mathbf{p}_1,\mathbf{p}_2,t,t',t'')=S_d(\mathbf{p}_2,\mathbf{p}_1,t+\pi/\omega,t'+\pi/\omega,t''+\pi/\omega)$. These phase differences will mainly lead to fringes in the second and fourth quadrant of the $p_{1\parallel}p_{2\parallel}$ plane.  

Explicitly, this gives
\begin{align}
    \alpha_{lu}=S_l-S_u&=  -\alpha^{(\mathrm{exch})}_{\mathbf{p}_1,\mathbf{p}_2} -\frac{1}{2(t'-t'')}\alpha^{(A^2)}_{\pi/\omega}(t',t'')- \frac{1}{2}\alpha^{(\mathrm{pond})}_{\pi/\omega}(t,t'')\\&-\alpha^{(\mathrm{ene})}_{\pi/\omega}-\alpha^{(\mathbf{p}_1 \leftrightarrow \mathbf{p}_2)}_{\pi/\omega} (t,t'),
    \label{eq:phaseupleft}
\end{align}
where the first four phase differences are given by Eqs.~\eqref{eq:alphaexch}, \eqref{eq:alphaA2}, \eqref{eq:alphapond} and \eqref{eq:alphaene}. 
The term
\begin{align}
 \alpha^{(\mathbf{p}_1 \leftrightarrow \mathbf{p}_2)}_{\Delta \tau}&(t,t')=\mathbf{p}_2\cdot \mathbf{F}_A(t'+\Delta \tau) + \mathbf{p}_1\cdot \mathbf{F}_A(t+\Delta \tau)  \\&-\mathbf{p}_2\cdot \mathbf{F}_A(t)-\mathbf{p}_1\cdot \mathbf{F}_A(t'),
   \label{eq:alphaexctau}
\end{align}
contains a momentum dependence and the double arrow in the superscript indicates that the momentum in the second row of Eq.~\eqref{eq:alphaexctau} are exchanged with regard to the temporal arguments  $t$ and $t'$ of $\mathbf{F}_A$. We have verified that $\alpha_{lu}=-\alpha_{rd}$ regardless of the field shape. This guarantees symmetry of these contributions in the second and fourth quadrant of the $p_{1\parallel}p_{2\parallel}$ plane. 

Eq.~\eqref{eq:phaseupleft} shows that, in addition to $\alpha^{(\mathrm{exch})}_{\mathbf{p}_1,\mathbf{p}_2}$, which is also present for $\alpha_{ld}$ and leads to hyperbolic interference fringes, the phase shift  $\alpha_{lu}$ contains several terms which come from the temporal shifts. The effects of these equations are more difficult to visualise. For linearly polarised fields, the scalar products in Eq.~\eqref{eq:alphaexctau} will reduce to the products of $p_{2\parallel}$ and $p_{1\parallel}$ with field-dependent terms. Imposing that $\alpha^{(\mathbf{p}_1 \leftrightarrow \mathbf{p}_2)}_{\Delta \tau}(t,t')=0$ will lead to conditions that depend on the field symmetry, although a first inspection may give the impression that there should be fringes located at the axes $p_{n\parallel}=0$. 

For instance, one may verify that, for fields satisfying the half cycle symmetry, the condition $\mathbf{F}_A(t)=-\mathbf{F}_A(t + \pi/\omega)$ leads to $ \alpha^{(A^2)}_{\pi/\omega}(t',t'')=0$ and 
\begin{equation}
   \alpha^{(\mathbf{p}_1 \leftrightarrow \mathbf{p}_2)}_{\pi/\omega}(t,t')=-(\mathbf{p}_2+\mathbf{p}_1) \cdot \left[\mathbf{F}_A(t')+\mathbf{F}_A(t)\right].
     \label{eq:alphaulc2}
\end{equation}
For linearly polarised fields satisfying the half-cycle symmetry, the scalar product stated above becomes
$(p_{2\parallel}+p_{1\parallel})\left[F_A(t')+F_A(t)\right]$, and the requirement that $ \alpha^{(\mathbf{p}_1 \leftrightarrow \mathbf{p}_2)}_{\pi/\omega}$ vanishes for all times would imply that $p_{2\parallel}=-p_{1\parallel}$.

If, further to that, the field is taken to be monochromatic and linearly polarised, one can show that
\begin{align}
       \alpha^{(\mathrm{mono})}_{lu}=&-\frac{\pi}{2 \omega} \left[4U_p+2E_{1g}+2E_{2g}+\mathbf{p}_{1 }^2+\mathbf{p}_{2 }^2\right]
      \\&-\frac{1}{2}  \left(\mathbf{p}_{2 }^2 -\mathbf{p}_{1}^2\right)(t-t')  +\frac{\sqrt{U_p}}{\omega} (p_{2\parallel}+p_{1\parallel})  \left[\sin \omega t'+\sin \omega t\right],
      \label{eq:phaseleftupmono}
\end{align}
where the extra term in $U_p$ comes from $\alpha^{(\mathrm{pond})}_{\pi/\omega}(t,t'')= 4U_p \pi / \omega$ in this case. 

Due to several field- and energy-dependent terms in $\alpha_{lu}$, it is not straightforward to derive simple analytic expressions for constructive interference between $M_u$ and $M_l$. There is, however, a term in $\mathbf{p}^2_{2}-\mathbf{p}^2_{1}$, which, if required to vanish, leads to hyperbolic fringes. For fields in which the half-cycle symmetry is broken, we cannot use the argument that imposing that the last term in Eq.~\eqref{eq:phaseleftupmono} vanishes leads to a fringe along the anti-diagonal $p_{1\parallel}=-p_{2\parallel}$ \cite{Maxwell2015}. 

Furthermore, in case we are considering more than one cycle of the field, we must also assess what happens for contributions displaced by at least a full cycle ($\Delta \tau \geq 2 \pi/\omega$), without and with momentum exchange. The contributions from events separated by a full number of cycles overlap considerably in momentum space, which potentially makes their interference substantial. This is particularly important for fields in which the field cycles are not equivalent, such as for few-cycle pulses.

Shifting the action associated with the left peak in a full cycle gives $S_{l,T}(\mathbf{p}_1,\mathbf{p}_2,t,t',t'')=S_l(\mathbf{p}_1,\mathbf{p}_2,t+2\pi/\omega,t'+2\pi/\omega,t''+2\pi/\omega)$, where we have defined $T=2\pi/\omega$. 
Similarly, for the downward peak one obtains $S_{d,T}(\mathbf{p}_1,\mathbf{p}_2,t,t',t'')=S_d(\mathbf{p}_1,\mathbf{p}_2,t+2\pi/\omega,t'+2\pi/\omega,t''+2\pi/\omega)$. The phase difference  $\alpha_{lT,l}$ then reads
\begin{align}
\alpha_{lT,l}=S_{l,T}-S_{l}&=\frac{1}{2(t'-t'')}\alpha^{(A^2)}_{2\pi/\omega}(t',t'')= 
+\frac{1}{2}\alpha^{(\mathrm{pond})}_{2\pi/\omega}(t,t'')\\&+\alpha^{(\mathrm{ene})}_{2\pi/\omega}+\alpha^{(\mathbf{p}_1,\mathbf{p}_2)}_{2\pi/\omega}(t,t'), 
\label{eq:phaseperiodl}
\end{align}
where the phase differences have been defined in Eqs.~\eqref{eq:alphaA2},\eqref{eq:alphapond}, \eqref{eq:alphaene} and \eqref{eq:alphap1p2tau}. 
 The phase difference $\alpha_{dT,d}=S_{d,T}-S_{d}$ resembles Eq.~\eqref{eq:phaseperiodl}, but with the momenta swapped in the last building block, i.e., $\alpha^{(\mathbf{p}_2,\mathbf{p}_1)}_{2\pi/\omega}(t,t')$.  
The phase differences $\alpha_{rT,r}=S_{r,T}-S_{r}$ and  $\alpha_{uT,u}=S_{u,T}-S_{u}$ are calculated similarly, but the arguments of the functions $F_A$ and $F_{A^2}$ must be shifted in half a cycle with regard to those used in Eq.~\eqref{eq:phaseperiodl}. 

For fields of period  $T=2\pi/\omega$,  $\alpha^{(A^2)}_{2\pi/\omega}$ and $\alpha^{(\mathbf{p}_i,\mathbf{p}_j)}_{2\pi/\omega}$, with $i \neq j$, will vanish, while $ \alpha^{(\mathrm{pond})}_{2\pi/\omega}$ will give ponderomotive shifts which will add to $\alpha^{(\mathrm{ene})}_{2\pi/\omega}$. This will happen regardless of whether the half-cycle symmetry is broken. An example of this specific scenario is a monochromatic field, for which 
\begin{equation}
\alpha^{\mathrm{mono}}_{lT,l}= \alpha^{\mathrm{mono}}_{dT,d} =\frac{\pi}{\omega} \left[4U_p+2E_{1g}+2E_{2g}+\mathbf{p}_{1 }^2+\mathbf{p}_{2 }^2\right]
\end{equation}
is the equation of a six-dimensional hypersphere in momentum space.  
Re-writing the condition for interference maxima $\alpha^{\mathrm{mono}}_{lT,l}=2n\pi$ in terms of the momentum components parallel to the driving-field polarisation yields
\begin{equation}
p_{1\parallel}^2+p_{2\parallel}^2 =2n\omega-(4U_p+2E_{1g}+2E_{2g}+\mathbf{p}^2_{1\perp}+\mathbf{p}^2_{2\perp}), 
\end{equation}
whose projection in the $p_{1\parallel}p_{2\parallel}$ plane gives circular fringes if the integer $n$ is sufficiently large so that the right-hand side is positive. 
Interference fringes taking into account electron exchange and events displaced by a full cycle can be obtained by considering, for instance, $\alpha_{lT,d}=S_{l,T}-S_d=\alpha_{lT,l}+\alpha_{ld}$, or  $\alpha_{uT,l}=S_{u,T}-S_l=\alpha_{uT,u}+\alpha_{ul}$. 

One must also account for cases in which there are multiple relevant ionisation events per half-cycle, which typically occurs for bichromatic fields - see \cite{Hashim2025} for a detailed discussion on this issue. This leads to additional phase differences associated with temporal and temporal-exchange interferences.
For events still in different half-cycles but with second electron events separated by less than \( 0.5T \), the interference is associated with the phase difference \( \alpha_{\mu,\nu}^{(m,n)} \), where $\mu,\nu$ refer to the regions of the momentum space that are occupied and $m,n$ distinguish the events.
Notably, $\alpha_{\mu,\nu}^{(m,n)} \neq \alpha_{\mu,\nu}^{(n,m)}$ as these phase differences represent distinct combinations of events. 
When both interfering events occur within the same half-cycle, but the second electron events are separated in time, interference is described by phase differences \( \alpha_{\mu,\nu}^{(m,n)} \), where \( \mu,\nu \) = \( r,r \); \( l,l \); \( d,d \); or \( u,u \) depending on the event combination. This type of interference is unique to fields with multiple ionisation events per half-cycle, such as bichromatic fields.

\subsection{Statistical measures and inter-channel interference}

In RESI, the second electron is able to excite to a number of bound states in the target before being ionised by the driving field. These `excitation channels' provide yet more pathways by which the RESI process can occur, and between which quantum interference can be investigated. Therefore, extending the analysis of interference patterns from intrachannel to interchannel requires a  different strategy. For a single excitation channel, the relevant phase differences can be organised according to electron exchange and the temporal displacement between ionisation events, so that individual interference structures may be related fairly directly to specific phase contributions. Once different intermediate excited states are included, however, an additional channel index enters and the number of interfering pathways increases rapidly. Even for only two channels and three dominant events, there are six channel-only, six channel-exchange, twelve channel-temporal, and twelve combined channel-exchange-temporal phase differences. Moreover, the corresponding structures may occupy the same regions of momentum space and therefore overlap, distort or wash out one another. The problem is also more strongly target dependent than the intrachannel case, because different excitation channels have different energies, which will directly affect ionisation, and bound-state geometries, which will influence what regions in the $p_{1\parallel}p_{2\parallel}$ plane will be predominantly occupied. The simplest case is two-channel interference. 

Whether two channels can interfere appreciably is controlled by several competing factors. First, their individual momentum distributions must be comparable, so that the resulting PMD is influenced by both channels. In case this does not hold, one of the channels will prevail and obfuscate the contributions of the `weaker' channel. Second,  the excited states should have sufficient spatial overlap. Third, the dominant events associated with the two channels must map onto overlapping regions of the two-electron momentum plane. Finally, the difference between the excitation energies also enters the interchannel phase and changes the relative time shifts between otherwise equivalent events in the two channels. Thus, small energy differences generally favour comparable contributions, but this does not necessarily translate into stronger or clearer fringes: a large energy difference can enhance particular phase variations, while strong mixing can itself lead to several superimposed structures and hence to blurring. In the examples studied, the relative channel intensity and excitation-energy difference correlate most strongly with the degree of channel mixing, while the radial overlap plays a secondary but still physically important role. Thus, an exhaustive analysis of all possible phase differences becomes both cumbersome and, in many cases, unnecessary: if one channel overwhelmingly dominates a particular pair, most genuinely interchannel contributions will be weak.

A useful strategy, used by us in \cite{Hashim2026}, is therefore to determine first which channels are sufficiently comparable to interfere. Thus, in addition to deriving analytic phase conditions in terms of the same elementary building blocks that occur for intrachannel interference, we employed statistical measures, widespread in research areas such as \cite{Rubner2000,Zhang2020emd}, particle physics \cite{Komiske2019}, biology \cite{Orlova2016} and chemistry \cite{Wang2023emd}, to quantify whether the PMDs associated with each channel in a pair are comparable. Specifically, we have introduced an equal-mix metric (EMM) based on the Earth mover's distance and examine its relation to the relative channel strengths, excitation-energy separation and spatial overlap of the excited states. The resulting metric provides a means of screening the large number of possible channel combinations and identifying those for which interchannel interference is likely to be relevant. Having identified such combinations, the interference itself may then be exposed by comparing coherent and incoherent channel sums.

\subsection{Mapping field symmetry to correlated PMDs}

\begin{figure}
    \centering
    \includegraphics[width=\linewidth]{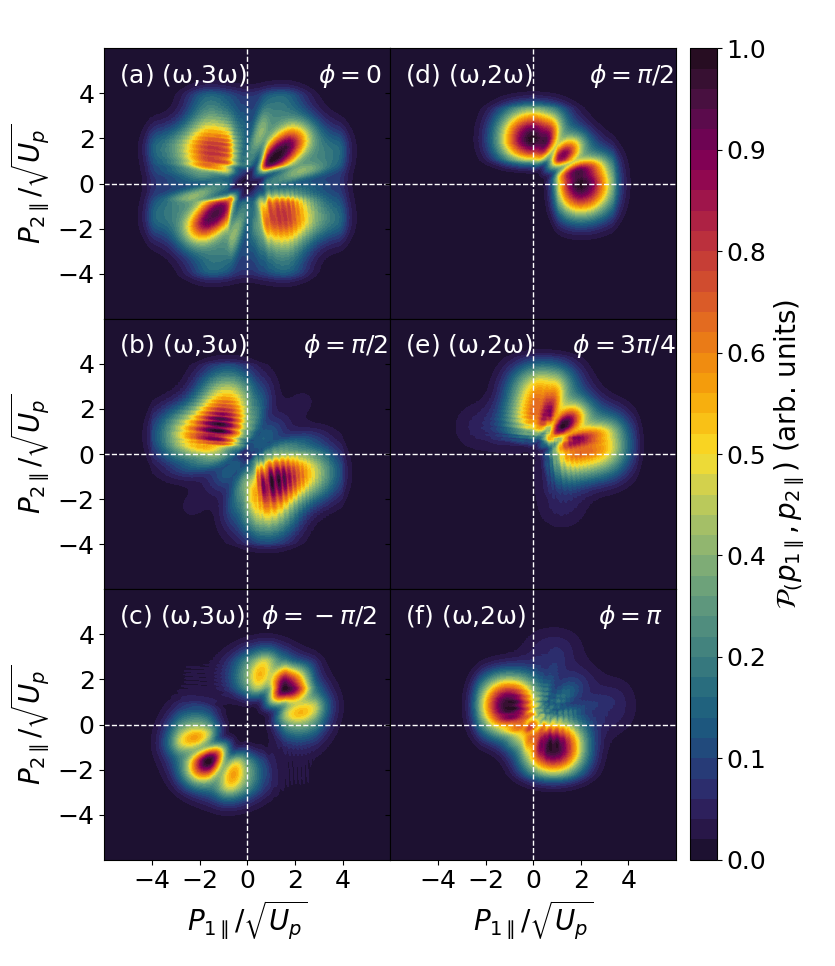}
    \caption{Fully coherent two-electron momentum distributions for RESI in Ar, calculated within the SFA for linearly polarised bichromatic fields and without bound-state prefactors. Panels (a)--(c) correspond to $(\omega,3\omega)$ fields with relative phases $\phi=0$, $\phi=\pi/2$, and $\phi=-\pi/2$, respectively, while panels (d)--(f) show $(\omega,2\omega)$ fields with $\phi=\pi/2$, $\phi=3\pi/4$, and $\phi=\pi$, respectively. The distributions are functions of the electron momentum components $p_{1\parallel}$ and $p_{2\parallel}$ parallel to the field polarisation, expressed in units of $\sqrt{U_p}$. The intensity of the fundamental component is $E_{\omega}^{2}=6\times10^{13}\,\mathrm{W/cm^2}$, its wavelength is $\lambda=800\,\mathrm{nm}$, and the field-amplitude ratio is $E_s/E_r=0.8$. The target parameters are $E_{1g}=0.58$ a.u., $E_{2g}=1.016$ a.u., and $E_{2e}=0.52$ a.u., corresponding to the $3s\rightarrow3p$ excitation channel of Ar. All relevant events and their electron-exchange counterparts are summed coherently. Adapted from Ref.~\cite{Hashim2025}.}
    \label{fig:RESIinterf3}
\end{figure}
Fig.~\ref{fig:RESIinterf3} shows the two-electron PMDs for several bichromatic fields considering a single excitation channel, but summing all amplitudes in Table \ref{tab:possiblesums} coherently. Some of the panels in the figure can be directly compared with those in Fig.~\ref{fig:NSDIdominant}, for which incoherent sums have been performed. A striking feature is that the fourfold symmetry of the RESI PMD present in Fig.~\ref{fig:NSDIdominant}(a) has been broken for the $(\omega, 3\omega, \phi=0)$ case [Fig.~\ref{fig:RESIinterf3} (a)], with spine-like fringes in the first and third quadrant associated with exchange interference, and a feather-like interference structure in the second and fourth quadrant which can be traced back to combined temporal-exchange interference. The latter fringes also break the reflection symmetry about the anti-diagonal $p_{1\parallel}=-p_{2\parallel}$. However, the exchange-only fringes are reflection-symmetric with regard to the diagonal  $p_{1\parallel}=p_{2\parallel}$. Choosing $(\omega, 3\omega, \phi=\pi/2)$ as a driving field concentrates the PMD in the second and fourth quadrant by suppresses the events $O_{1}$ [Fig.~\ref{fig:NSDIdominant}(b)]. In this case, for the coherent sum shown in Fig.~\ref{fig:RESIinterf3}(b) only the feather-like fringes related to the combined interference are visible. Once more, the reflection-symmetry about the anti-diagonal that is expected for half-cycle symmetric fields has been broken.  If, on the other hand, $\phi$ is chosen to suppress the probability density on the second and fourth quadrants is suppressed, as in Fig.~\ref{fig:RESIinterf3}(c), the spine-like fringes associated with exchange interference are very visible. For the remaining panels in Fig.~\ref{fig:RESIinterf3}, computed for $\omega, 2\omega$ fields of different phases, are mainly concentrated in the first quadrant spilling towards the second and the fourth [Figs.~\ref{fig:RESIinterf3}(d) and (e)], or are mainly located in the second and fourth quadrants [Figs.~\ref{fig:RESIinterf3}(f)]. These panels also 
show the following features: the spine-like fringes associated with exchange-type interference are present in the first quadrant, and, whenever there is significant signal in the second and fourth quadrants, there are visible temporal-exchange feather-shaped fringes. It is as if the dominance achieved by the field knobs highlights a specific momentum region, so that specific types of interference can be shown.  

\begin{figure}
\hspace*{-0.5cm}    \includegraphics[width=1.1\linewidth]{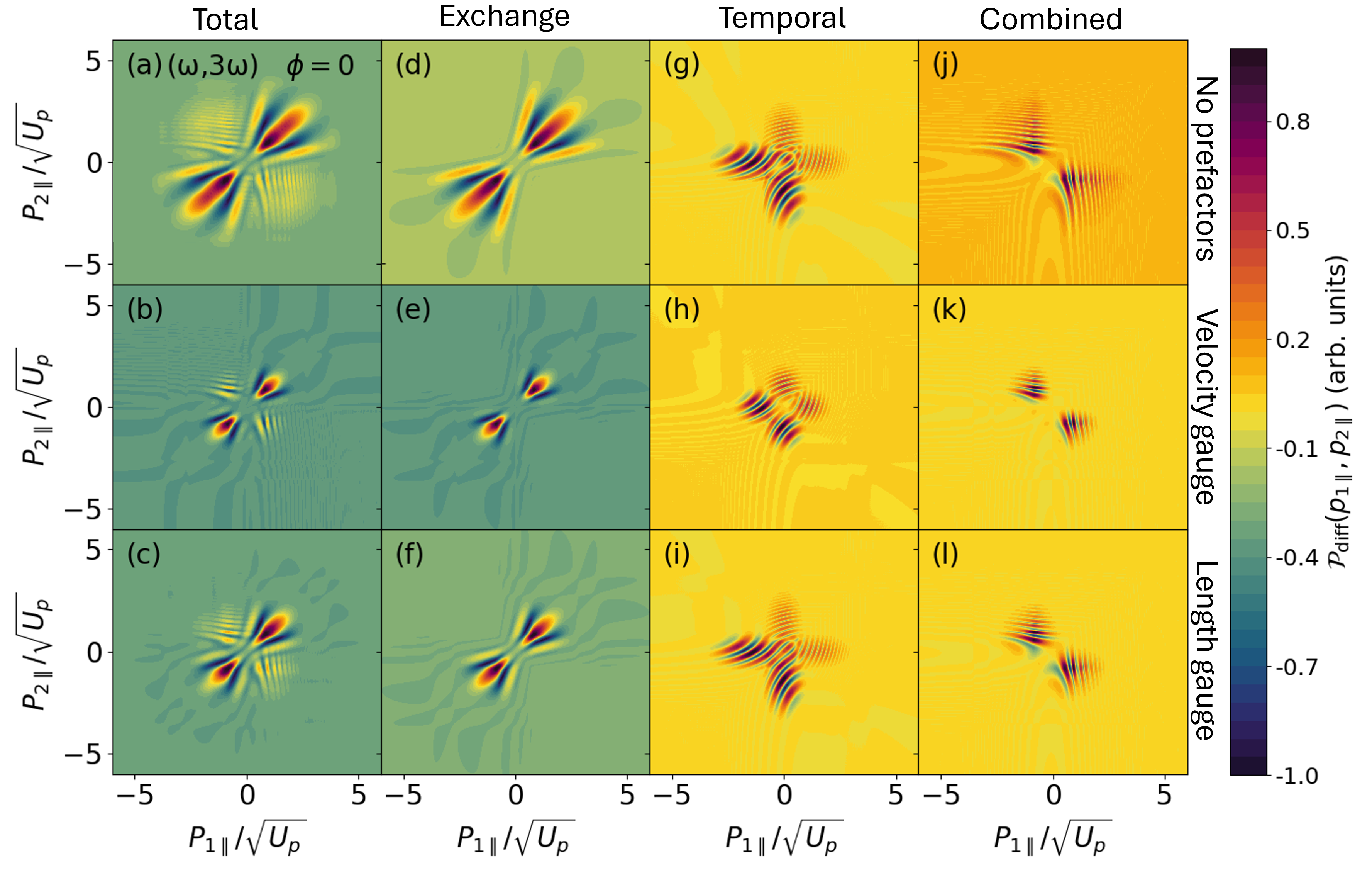}
    \caption{Interference contributions to RESI for the $3p\rightarrow 4s$ excitation channel in Ar, for a linearly polarised bichromatic $(\omega,3\omega)$ driving field with relative phase $\phi=0$. The rows show results obtained without prefactors [panels (a), (d), (g), (j)], with prefactors in the velocity gauge [panels (b), (e), (h), (k)], and with prefactors in the length gauge [panels (c), (f), (i), (l)]. From left to right, the columns display the difference distributions $\mathcal{P}_{\mathrm{diff}}(p_{1\parallel},p_{2\parallel})$ associated with the total interference, exchange-only interference, temporal-only interference, and combined exchange-temporal interference, respectively, obtained by comparing the corresponding coherent RESI distributions with their fully incoherent counterparts. The momentum components $p_{1\parallel}$ and $p_{2\parallel}$ are given in units of $\sqrt{U_p}$, and each panel is normalised to its maximum value. The fundamental component has wavelength $\lambda=800\,\mathrm{nm}$ and intensity $E_{\omega}^{2}=6\times10^{13}\,\mathrm{W/cm^2}$, with field-amplitude ratio $E_s/E_r=0.8$.  Adapted from Ref.~\cite{Hashim2025}.}
    \label{fig:RESIinterf5}
\end{figure}
Subtracting the coherent sums from their incoherent counterparts provide more details and also sheds some light into the temporal-only interference, which is subtler. These plots are presented in Fig.~\ref{fig:RESIinterf5}, in which the different types of interference have been detangled.  Apart from the fringes associated with exchange and combined interference, there are fringes concentrated along the axes associated with temporal phase shifts only [see third column in Fig.~\ref{fig:RESIinterf5}]. Fig.~\ref{fig:RESIinterf5} also illustrates the effect of incorporating prefactors in the calculations in the velocity and length-gauge (middle and lower row, respectively). The fringes and their shapes are retained, but they occupy momentum regions much more localised around the origin, compared to the no-prefactor
interference. Further to the three types of features discussed here, other, subtler structures associated with intra-channel interference have been reported in our original publications \cite{Hashim2024,Hashim2025}.

The next two figures illustrate inter-channel interference for RESI in argon, where we consider the channels given in Table \ref{tab:channels}. The letters on the right-hand top corner give the excited states for each pair of channels. For example, $3p4p$ indicates that the $3s\rightarrow3p$ and $3p\rightarrow4p$ excitations have been combined pairwise.
\begin{table}[] 
\begin{tabular}{cccc}
\hline \hline
 Channel & Excited-state configuration  &  $E^{(\mathcal{C})}_{2e}$ (a.u.)& Maxima   \vspace*{0.1cm} \\ \hline
 1& $3s3p^6 (3s \rightarrow 3p)$ &  0.52 & $d$   \\
 2& $3p^53d (3p \rightarrow 3d)$ &  0.41 & $a,d$  \\
 3& $3p^54d (3p \rightarrow 4d)$ &  0.18 & $ a,d$  \\
 4&  $3p^54s (3p \rightarrow 4s)$  &  0.40&  $a$ \\
 5& $3p^54p (3p \rightarrow 4p) $& 0.31 &  $d$ \\
 6& $3p^55s (3p \rightarrow 5s)$ & 0.19 &   $a$ \vspace*{0.1cm} \\ \hline \hline
\end{tabular}
\caption{Relevant excitation channels of $\mathrm{Ar}^+$, ordered by increasing principal and orbital quantum numbers. From left to right, the columns list the channel label, electronic configuration and corresponding excitation pathway, excited-state energy in atomic units, and the location of the dominant maxima in the resulting momentum distributions. In the final column, $a$ and $d$ denote maxima along the axes and diagonals, respectively. \label{tab:channels}}
\end{table}

\begin{figure}
    \centering
    \includegraphics[width=\linewidth]{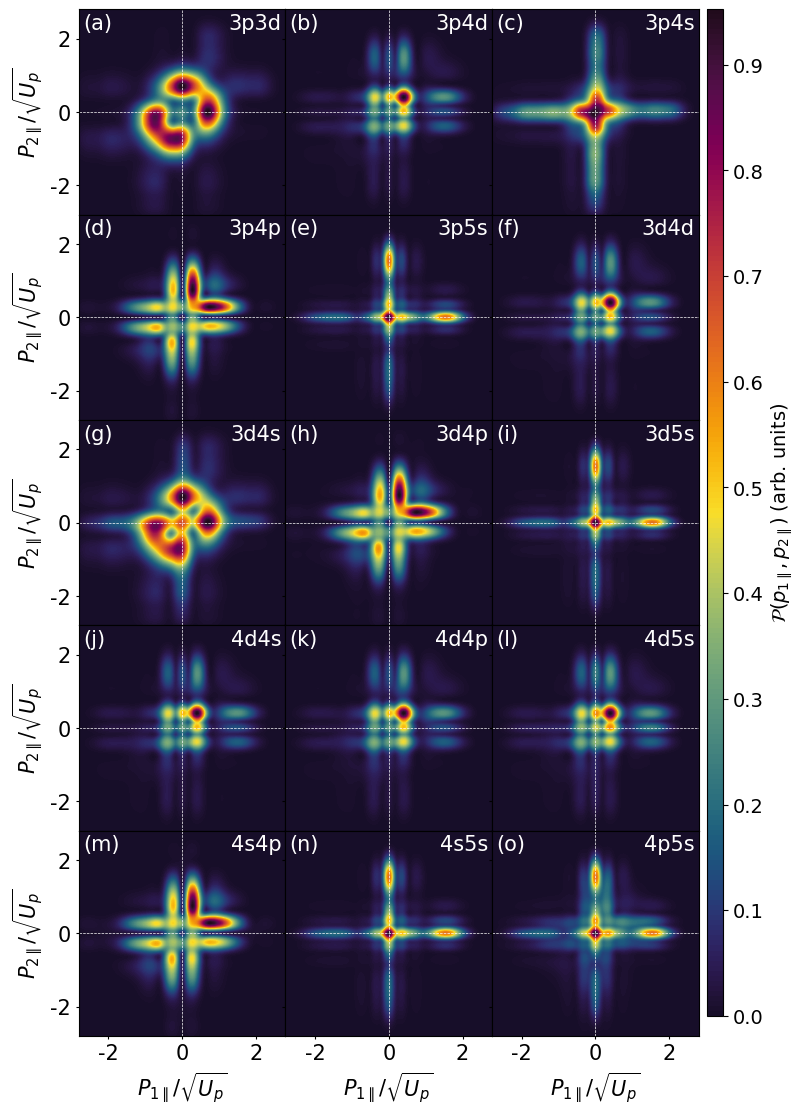}
    \caption{Fully incoherent two-channel momentum distributions, $\mathcal{P}{iii}(p{1\parallel},p_{2\parallel})$, for all channel pairs listed in Table~\ref{tab:channels}. The calculations are performed for Ar using a linearly polarised few-cycle pulse with peak intensity $I=1.5\times10^{14}\,\mathrm{W/cm}^2$, wavelength $\lambda=800$~nm ($\omega=0.057$~a.u.), $N=4.3$, and carrier-envelope phase $\phi_1=65^\circ$, including the three dominant events. The excited states associated with each channel pair are indicated in the upper-right corner of each panel. The signal in each panel has been normalised to its maximum. From \cite{Hashim2026}.}
    \label{fig:Multichannel1}
\end{figure}

Fig.~\ref{fig:Multichannel1}, which shows the incoherent sums of all pairwise channel combinations, illustrates qualitatively if two channels are or not comparable. While many distributions retain predominantly the characteristic structure of one of their constituent channels, others exhibit clear hybrid features, signalling that both channels contribute appreciably. The incoherent two-channel PMDs show that comparable channel contributions are relatively uncommon. The clearest mixtures occur for the pairs with the smallest excitation-energy separation, $3d4s$ and $4d5s$, whose distributions combine characteristic d- and s-state features. Most other channel pairs are dominated by a single orbital, producing maxima along the axes for s states, along the diagonals for p states, or in both regions for d states. This dominance is especially expected when both channels have the same orbital angular momentum, since their bound-state geometries are mapped similarly into momentum space. An exception is $4p5s$, which, despite a larger energy difference, also exhibits substantial contributions from both channels, indicating that energy separation alone does not determine the degree of channel mixing.
The strongest channel mixing occurs for $3d4s$ and $4p5s$, for which the equal-mix metric exceeds $0.7$, while $3p4s$, $4d5s$, and $3d4p$ exhibit intermediate levels of mixing. Notably, $4d5s$ appears strongly hybridised in Fig.~5 because of its very small excitation-energy difference, yet its EMM remains only moderate ($0.549$). By contrast, $4p5s$ yields the largest EMM despite a substantially larger energy separation. This shows that excitation-energy proximity alone is not sufficient to determine the degree of channel mixing, which also depends on the relative strengths and momentum-space overlap of the contributing channels.

\begin{figure}
    \centering
    \includegraphics[width=\linewidth]{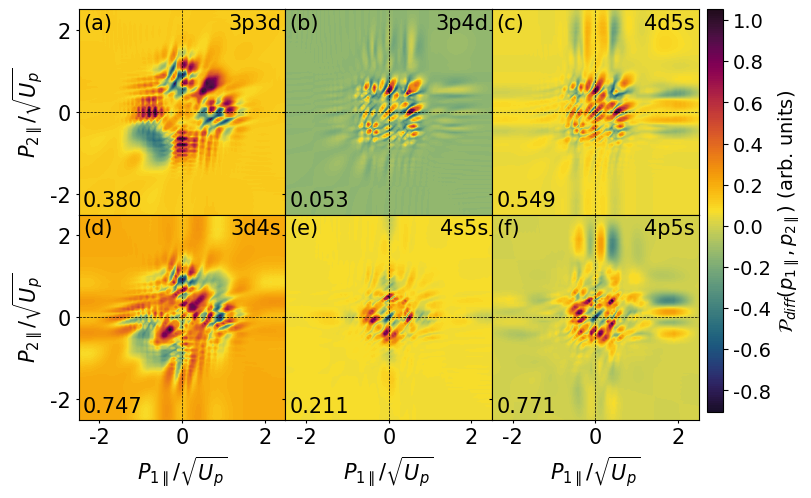}
    \caption{Difference between the fully coherent and fully incoherent RESI momentum distributions, $\mathcal{P}_{(ccc)}(p_{1\parallel},p_{2\parallel})-\mathcal{P}_{(iii)}(p_{1\parallel},p_{2\parallel})$, for selected channel pairs, calculated using the same driving-field parameters as in Fig.~\ref{fig:Multichannel1}. The excited states associated with each channel pair are labelled in the upper-right corner of each panel, while the black dashed lines mark the momentum axes. The corresponding equal-mix metric value is given in the lower-left corner. Each distribution is normalised to its maximum value. From \cite{Hashim2026}.}
    \label{fig:Multichannel2}
\end{figure}

Figure \ref{fig:Multichannel2} shows the difference between the coherent and incoherent distributions for representative pairs ranging from weakly to strongly mixed channels. The comparison reveals that stronger channel mixing generally increases the role of interchannel pathways, but does not necessarily yield sharper interference: different phase contributions may overlap and partially obscure the familiar intrachannel structures.

\section{Summary and Outlook}
\label{sec:conclusions}

Throughout this review, we have seen the rich structure that photoelectron momentum distributions may exhibit, allowing them to be understood as probability distributions encoding information about the driving field, the target, and the electronic quantum pathways underlying the ionisation process.~From this perspective, quantum interference provides the mechanism through which phase information becomes accessible, while the underlying symmetries constrain how such information is distributed.~Strong-field ionisation therefore constitutes a versatile platform for extracting information with accuracy that is beyond what is accessible through ionisation rates or integrated yields alone, including the sensitive characterisation of driving laser parameters close to the quantum efficiency limit~\cite{Maxwell2021a}.

Looking further ahead, this picture holds a particular promise in the emerging field of strong-field quantum optics~\cite{Stammer2025}, where non-classical states of light are used to drive strong-field processes~\cite{Rasputnyi2024,Heimerl2025NeedleTipsBSV,Lemieux2025PhotonBunching,tzur_attosecond-resolved_2025}.~Alongside the opportunities offered by such fields, significant challenges also arise due to the intrinsically strong-field regime in which they operate, as many standard quantum optical methods (e.g.~photon counting, homodyne detection or optical tomography) become difficult to implement under these conditions.~In this context, a natural route forward is to exploit approaches that are naturally suited to extreme-field environments.~Strong-field ionisation is especially well positioned in this regard, as the use of non-classical light is expected to substantially modify the ionisation dynamics~\cite{EvenTzur2023,RiveraDean2025}, significantly alter the ionisation response~\cite{Lyu2025PhotonStatisticsATI} which inherits the driving field's symmetries~\cite{Singh2026BSVAsymmetry}, and even enable the engineering of non-classical states whose properties are shaped by the underlying quantum pathways responsible for their generation~\cite{RiveraDean2026}.~Combined with appropriate tools from classical and quantum information theory, these developments open exciting prospects for future research, with the potential to introduce methodologies    from the strong-field community in regimes largely unexplored with conventional quantum optics.

\noindent \textbf{Acknowledgements}. J. R.-D. and S. H. were funded by the UK Engineering and Physical Sciences Research Council (EPSRC) Grant UKRI2300 - Attosecond Photoelectron Imaging with Quantum Light (APIQuL). 
\bibliographystyle{elsarticle-num}

\end{document}